\documentclass[11pt]{article}

\usepackage[no-natbib-sort]{jheppub}
\usepackage[T1]{fontenc} 

\usepackage{amssymb}
\usepackage{amsfonts}
\usepackage{amsbsy}
\usepackage{amsmath}
\usepackage{amsthm}
\usepackage{graphicx}
\usepackage{epstopdf}
\usepackage{array,booktabs}
\usepackage{color}

\newcommand{\baselink}[1]{\hyperref[eq:A#1]{$\widehat{B}_{#1}$}}

\def\F{\mathbb{F}}

\def\C{\mathbb{C}}

\def\P{\mathbb{P}}

\def\n3a{t}

\def\gsu{{\mathfrak{su}}}

\def\gg{{\mathfrak{g}}}

\def\person{Persson}

\def\ho{h^{1, 1}}
\def\htt{h^{2, 1}}
\newcommand{\eq}[1]{(\ref{#1})}

\def\ov{\over}
\def\eq#1{(\ref{#1})}

\def\fu{\mathfrak{u}}
\def\MW{\widehat{MW}}

\newcommand{\be}{\begin{equation}}
\newcommand{\ee}{\end{equation}}
\newcommand{\bea}{\begin{equation} \begin{aligned}}
\newcommand{\eea}{\end{aligned} \end{equation}}

\newcommand{\bi}{\begin{itemize}}
\newcommand{\ei}{\end{itemize}}
\newcommand{\ben}{\begin{enumerate}}
\newcommand{\een}{\end{enumerate}}

\title{Non-Higgsable 
abelian gauge symmetry
and \\
F-theory on fiber products of
rational elliptic surfaces}

\author[1]{David R.  Morrison}
\author[2,3]{Daniel S. Park}
\author[]{and}
\author[4]{Washington Taylor}

\affiliation[1]{Departments of Mathematics and  
Physics\\ University of California, Santa Barbara\\ Santa Barbara, CA 93106, 
USA}
\affiliation[2]{Simons Center for Geometry and Physics \\ 
State University of New York \\ 
Stony Brook, NY 11794, USA}
\affiliation[3]{
NHETC and Department of Physics and Astronomy\\
Rutgers University\\
Piscataway, NJ 08855, USA}
\affiliation[4]{Center for Theoretical Physics\\
Department of Physics\\
Massachusetts Institute of Technology\\
77 Massachusetts Avenue\\
Cambridge, MA 02139, USA}

\emailAdd{{\tt drm} {\rm at} {\tt math.ucsb.edu}}
\emailAdd{{\tt dspark} {\rm at} {\tt physics.rutgers.edu}}
\emailAdd{{\tt wati} {\rm at} {\tt mit.edu}}

\preprint{UCSB Math 2016-17, MIT-CTP-4839}

\abstract{We construct a general class of Calabi--Yau threefolds from
  fiber products of rational elliptic surfaces with section,
  generalizing a construction of Schoen to include all Kodaira fiber
  types.  The resulting threefolds each have two elliptic fibrations
  with section over rational elliptic surfaces and blowups thereof.
  These elliptic fibrations generally have nonzero Mordell--Weil rank.
  Each of the elliptic fibrations has a physical interpretation in
  terms of a six-dimensional F-theory model with one or more
  non-Higgsable abelian gauge fields.  Many of the models in this
  class have mild singularities that do not admit a Calabi--Yau
  resolution; this does not seem to compromise the physical integrity
  of the theory and can be associated in some cases with massless
  hypermultiplets localized at the singular loci. In some of these
  constructions, 
  however, we find examples of abelian gauge fields that cannot be
  ``unHiggsed'' to a nonabelian gauge field without producing
  unphysical singularities that cannot be resolved.  The models
  studied here can also be used to exhibit T-duality for a class of
  little string theories.}

\begin{document}
\maketitle

\flushbottom

\section{Introduction}
\label{sec:intro}

Recently, the phenomenon of gauge symmetries that cannot
be Higgsed in low-energy supersymmetric  theories
has been studied intensively in the context of F-theory
\cite{clusters,Grassi:2014zxa,Anderson-Taylor,4D-NHC,Halverson:2015jua,Taylor-Wang-mc,Halverson-nh}.
These works have focused on nonabelian gauge symmetries that may arise
through singular Kodaira fiber types that are present at
generic points in the complex structure moduli space of a Calabi--Yau
threefold or fourfold that is elliptically fibered
over a given base geometry.  Such
``non-Higgsable'' nonabelian gauge groups seem to arise in very
generic classes of F-theory vacua.  In this paper we consider a class
of models that give rise to analogous non-Higgsable {\it abelian}
gauge factors.

An interesting class of elliptic Calabi--Yau threefolds was constructed
by Schoen \cite{Schoen} using fiber products of rational elliptic
surfaces with section\footnote{There exist rational elliptic surfaces
with no section (see {\it e.g.}\ \cite[Chap.\ V, \S6]{MR986969}
or \cite{MR1053907}) but they will not concern us here.} that admit small resolutions.  These manifolds are of
interest for a number of reasons, including the fact that manifolds with
Mordell--Weil (MW) groups having relatively large rank $r$ are easily
constructed in this fashion, $r = 9$ being the largest out of the examples
appearing in this work.
Schoen's construction was generalized by Kapustka and Kapustka
\cite{Kapustka} to include a broader range of singular fibers in the
component rational elliptic surfaces.  In this paper we generalize
this construction further, by considering the full range of Kodaira
fiber types in the component rational elliptic surfaces.  We give a
systematic classification of the pairs of fiber types 
which, when combined into a fiber product, have at worst canonical
singularities in the total space.  
Following Miranda \cite{MR690264} and Grassi \cite{Grassi91},
for each of the projections from the fiber product to one of the factors,
there is a blowup of the base and a partial resolution of the total
space that yields a flat elliptic fibration ({\it i.e.}, one in which all fibers
are one-dimensional) such that the canonical bundle of the total space
is trivial.
Combining this information with Persson's list
\cite{Persson, Miranda} of the  combinations of singular fibers that can
appear in a rational elliptic surface gives in principle a
complete classification of all elliptic Calabi--Yau threefolds that can
be constructed in this fashion. 

Elliptic Calabi--Yau threefolds of this type are  of interest in
physics, where they can be used as compactification spaces for
F-theory \cite{Vafa-F-theory, Morrison-Vafa1,Morrison-Vafa2} to give six-dimensional
supergravity theories.  The role of abelian gauge fields in F-theory,
associated with elliptically fibered Calabi--Yau manifolds having
nonzero Mordell--Weil rank, has been of substantial recent interest
(see for example 
\cite{GrimmWeigand,DudasPalti,Marsano,DMSS,MSS,GKPW,%
Morrison-Park,%
Mayrhofer:2012zy, Braun:2013yti, Cvetic:2013nia, Braun:2013nqa,
Cvetic:2013uta,%
Borchmann:2013hta,%
Cvetic:2013qsa,
mt-sections,%
Antoniadis:2014jma, Kuntzler:2014ila, Esole:2014dea,
Lawrie:2015hia,Cvetic:2015uq, Krippendorf:2015kta,
Cvetic:2015uwu, Morrison-Park-2}.)
Some of
the threefolds that we consider here have  been constructed in this
related physics context.
  In \cite{Martini-WT}, following methods developed in
\cite{clusters, toric},
a
complete classification was given of complex surfaces that admit a
$\C^*$-action 
and can act as
bases for an elliptically fibered Calabi--Yau threefold.  Of the more
than 100,000 such bases, 13 have the property that the associated
elliptically fibered Calabi--Yau threefolds have Mordell--Weil groups of
nonzero rank everywhere in the moduli space.  We show in
\S\ref{sec:application} that all 13 of these models are special cases
of the construction  presented here.

These manifolds  shed light on the extent to which fibrations with
nonzero Mordell--Weil rank can be analyzed through deformations of
local Kodaira singularity structures.  In \cite{Morrison-Park,
  mt-sections}, we showed that in a very general class of situations,
sections in the free part of the Mordell--Weil group (associated with
abelian gauge fields) can be converted to divisors associated with
Kodaira singularities that are local in the base (associated with
nonabelian gauge fields) without changing $h^{1, 1}$ of the threefold.
This shows that in many cases, threefolds with nonzero Mordell--Weil
rank can be constructed from deformations of threefolds with vanishing
Mordell--Weil rank.  In some cases
the transition from ``horizontal''
divisors associated with sections to ``vertical'' divisors associated
with Kodaira singularities gives rise to singularities   that require
further treatment.  As discussed in \cite{mt-sections}, some of these
situations involve singularities that require a blowup in the base
manifold.  In other situations, however,
we find that the singularity cannot be
resolved (while keeping the canonical bundle trivial),  and there is no blowup of the base that leads to an
acceptable F-theory background. The threefolds
constructed here provide a variety of
concrete examples of such situations where the full Mordell--Weil rank
cannot be realized in this way without producing singularities that
lie at infinite distance from the interior of the moduli space of the manifold,
corresponding in physics terms to theories with abelian
gauge fields that cannot be ``unHiggsed'' or ``enhanced.''%
\footnote{Recently, another scenario has been suggested
in which a non-enhanceable abelian gauge symmetry
in an F-theory background can arise.
In \cite{Morrison-Park-2}, it was pointed out that
F-theory models with abelian gauge symmetry that simply
do not have the moduli that can be tuned to enhance the
abelian gauge algebra might exist.}

Another interesting aspect of the manifolds we study is that
they admit two distinct elliptic fibration structures. Depending
on which elliptic fibration structure we choose to preserve,
we arrive at two different resolutions, and two different
six-dimensional physical models in the F-theory limit. 
Calabi--Yau manifolds with distinct elliptic fibration structures were also
recently discussed in a related context in \cite{Anderson:2016cdu}.
Upon compactifying the two distinct
six-dimensional theories to five dimensions along a circle,
it can be shown that the two five-dimensional theories are actually
dual to each other. This duality, which we call $A$-$B$ duality,
can be understood in terms of flops relating the two resolutions.
It is also closely related to ``little string theories:''
choosing one singular fiber in each of the two fibrations and restricting
to neighborhoods of the chosen fibers, there is a scaling limit
that produces a little string theory from the F-theory compactification
\cite{LST, LST2}. Our $A$-$B$ duality
exhibits the expected T-duality of the corresponding little string theories.

Because the results in this paper may be of interest both to
mathematicians and to physicists, we have tried to arrange the
presentation in a way that makes the relevant points accessible to both
types of readers.  After a review of canonical singularities
in \S\ref{sec: canonical}, the construction is described in mathematical terms
in \S\ref{sec:products}, and the examples encountered in the physical
context are described in \S\ref{sec:application}.
We discuss the abelian sector of the F-theory vacua in \S\ref{sec:abelian},
while \S\ref{sec:AB} is devoted to studying $A$-$B$ duality.
Some concluding remarks are given in \S\ref{sec:conclusions}.

\section{Canonical singularities of algebraic threefolds}
\label{sec: canonical}

One common approach to studying F-theory compactifications is
to begin with M-theory compactified on a smooth
 elliptically fibered Calabi--Yau manifold $X\to B$ and then take a limit
in which the area of the elliptic fiber approaches zero, yielding
an F-theory compactification.
However, 
adhering to the requirement that $X$ be nonsingular does not
allow for some important features to show up in M-theory, such as
nonabelian gauge symmetry.  An alternate approach, emphasized in
\cite{Grassi:2013kha,Douglas:2014ywa,whatF}
among other places, is to begin with the Weierstrass model
defined by
\begin{equation}
\label{eq:W}
y^2=x^3+fx+g
\end{equation}
and treat the Weierstrass coefficients as directly defining the
F-theory compactification via specification of the (variable) axio-dilaton
in a type IIB compactification on the base $B$.  The total space $\overline{X}$
of a fibration defined in this way is typically singular.

A natural question in either of these approaches is what kind of singularities
to allow.  That is, if we are to compactify M-theory on a space 
$\widetilde{X}$ that is birational to a Weierstrass model $\overline{X}$
(or more generally, whose relative Jacobian is birational to a Weierstrass
model \cite{triples,genus-one}), which singularities should be permitted in $\widetilde{X}$?

One possible answer is to demand that $\widetilde{X}$ admit a resolution 
of singularities $X\to\widetilde{X}$
such that $X$ is a smooth Calabi--Yau manifold.  However, in a number of
recent works including \cite{Morrison-Park,genus-one,mt-sections},
singular spaces $\widetilde{X}$ that do not have
a Calabi--Yau resolution have arisen naturally
in an F-theory or M-theory context.  Such spaces also arise in this
work, with singularities arising over a point in the base of the
elliptic fibration that do not admit a Calabi--Yau
resolution but that do not seem to compromise the physical 
integrity
of the theory.

A perhaps better answer, at least for threefolds\footnote{Note that, as pointed out in
  \cite{nonspherI,c3zn}, the situation is more complicated for
  Calabi--Yau fourfolds.  There are (terminal) singularities such as
  $\mathbb{C}^4/\mathbb{Z}_2$ that seem very well behaved for
  physical purposes, yet which admit neither complex structure
  smoothings \cite{MR0292830} nor Calabi--Yau resolutions
  \cite{MR1016414}.},  is to demand that (after an appropriate partial
resolution)
$\widetilde{X}=\widetilde{X}_0$ have a deformation of complex
structure $\widetilde{X}_t$ that is a Calabi--Yau manifold for each
value of $t > 0$, such that
$\widetilde{X}_0$ is at finite distance from $\widetilde{X}_t$ in the
moduli space metric.  By a result due to Hayakawa \cite{hayakawa} and
Wang \cite{MR1432818}, such $\widetilde{X}$ are characterized by
having {\em Gorenstein canonical singularities}\/ and {\em trivial
  dualizing sheaf}.  {\em Gorenstein}\/ means that there is a natural
anologue of the bundle of holomorphic $n$-forms on $\widetilde{X}$,
known as the {\em dualizing sheaf}, which is a locally free sheaf.
{\em Canonical}\/ means that by blowing up $\widetilde{X}$ we can
never introduce poles into the original $n$-forms.

We call $\widetilde{X}$ a {\em Calabi--Yau variety}\/ if it has
Gorenstein canonical singularities and trivial dualizing sheaf,
and we argue that this is the natural class of spaces on which M-theory
can be compactified.  Among other things, this definition allows
for rational double points in complex codimension two, which correspond
to nonabelian gauge symmetry in the compactified theory.
Allowing such singularities does not seem to compromise the physical
integrity of the theory.

To spell out the meaning of ``canonical singularities''  
in more detail, we can use Hironaka's famous theorem \cite{MR0199184}
to resolve the
singularities on $\widetilde{X}$ by a sequence of 
projective blowups,\footnote{The blowup $\pi:Z\to X$ is
{\em projective}\/ if $Z\subset (X\times \mathbb{P}^N)$ for some $N$,
with the map $\pi$ given by projection to the first factor.  This
condition guarantees that if $X$ is K\"ahler then so is $Z$.
In algebraic geometry, there can be other kinds of blowup, in which
$Z$ is only assumed to be a scheme
(or even just an algebraic space in the sense of Artin \cite{MR0407012}).}
 yielding $\pi:Z\to X$. 
A Gorenstein space $\widetilde{X}$ is {\em canonical}\/ if 
\begin{equation}
\label{eq:canonical}
K_Z = \pi^*(K_{\widetilde{X}}) + \sum a_i D_i
\end{equation}
 with all coefficients $a_i$ non-negative \cite{Reid:c3f}.
The blowup is called {\em crepant}\/ if all $a_i=0$ \cite{pagoda}.
Note that if $\widetilde{X}$ has trivial dualizing sheaf and
$X\to\widetilde{X}$ is a crepant projective resolution of singularities, then
$X$ is a Calabi--Yau manifold.

To formulate the definition of canonical, we need to use non-crepant blowups
in general, and one may wonder about the physical significance of a non-crepant
blowup.  This is best studied by considering a two-dimensional
sigma-model whose target is an arbitrary K\"ahler manifold $Z$
\cite{stabs,c3zn}.  (This is relevant for string compactification
rather than M-theory compactification, but the two are closely
related.)  The behavior of this theory under renormalization is
governed by the canonical divisor $K_Z$.  For an algebraic curve $C$
on $Z$, if $K_Z\cdot C<0$ then the sigma model operator that creates
  $C$ is irrelevant, and under RG flow $C$ contracts to a point.  If
  $K_Z\cdot C>0$ then the operator is relevant and under RG flow, $C$
grows without bound.  Finally, if $K_Z\cdot C=0$ then the operator
that creates $C$ is marginal and under RG flow $C$ remains of finite
volume.

This phenomenon is the essential insight in ``Mori theory''
\cite{MR662120,MR924704}, 
which is a detailed study of the birational geometry of algebraic
threefolds that proceeds by blowing down curves $C$ with $K_Z\cdot C < 0$.  
Moreover, there is an expectation that Mori theory will
be reproduced by ``K\"ahler--Ricci flow'' \cite{MR3185333},
which directly uses Ricci flow
(the one-loop approximation to the sigma model RG flow) to produce new
models, although  the mathematical theory of K\"ahler--Ricci flow
is still under development.

From the physics point of view, one could blow up the original
singular space arbitrarily and then run RG flow on the sigma model.
If $K_Z\cdot C\le0$ for all curves $C$ on $Z$, then we can expect
the RG flow to approach a sigma model on a space in which all the curves
with $K_Z\cdot C<0$ have been shrunk to points.  However, the presence of
any curve $C$ with $K_Z\cdot C>0$ would prevent this from happening.

The endpoint of this process is known from Mori theory, and  the final
space that is reached has ``$\mathbb{Q}$-factorial terminal singularities.''  
{\em Terminal singularities}\/ are
a special class of canonical singularities in which all coefficients $a_i$
in \eqref{eq:canonical}
are strictly greater than zero. 
Such a singularity is {\em $\mathbb{Q}$-factorial}\/ if there are no further
crepant projective blowups that can be made.  
It is also known from Mori theory that for a given variety,
there can be more than one birational
model with $\mathbb{Q}$-factorial terminal singularities.
For Calabi--Yau varieties, any two such models differ by a sequence
of flops \cite{MR924674,[Kol]}.

Putting together the Hironaka blowup followed by the Mori blowdown,
we conclude that any Calabi--Yau variety
has a crepant projective blowup with at worst
$\mathbb{Q}$-factorial
terminal singularities.  
(In fact, using an algorithm of Reid \cite{pagoda}, such
a crepant projective blowup can be made directly, without ``overshooting''
and needing to invoke Mori theory or the RG flow.)
In addition, by a result of Namikawa and Steenbrink \cite{namikawa-steenbrink},
any Calabi--Yau threefold with $\mathbb{Q}$-factorial
terminal singularities can always be deformed to a smooth Calabi--Yau
manifold.
Thus, we do not expect to find such singularities in generic
(maximally Higgsed) 6D F-theory models. Since the corresponding
theorem does not hold for fourfolds, however, such codimension two
singularities in Calabi--Yau fourfolds can arise even in maximally
Higgsed models \cite{Klemm-lry,Taylor-Wang-mc}.

We focus here on {\em maximal crepant projective blowups}, which we
call {\em MCP blowups}\/ for short.  In general, any Calabi--Yau variety
$\widetilde{X}$ of dimension three
has an MCP blowup $X=X_0$ which in turn has a complex structure deformation
to a smooth Calabi--Yau manifold $X_t$.  By Hayakawa and Wang, $X_0$ is
at finite distance from the interior of complex structure moduli space; we then obtain the
original $\widetilde{X}$ from $X_0$ by tuning K\"ahler moduli appropriately.

A crepant projective blowup of a terminal singularity is often called a 
{\em small blowup}\/ (or a {\em small resolution}\/ in case it 
resolves the singularities completely).
There is an important subtlety to be aware of concerning these:
whether or not a given terminal singularity admits
a small projective blowup depends on the existence of certain divisors
that might exist locally (in the complex topology) but not globally.
The divisor must exist globally if the blown up space is to be
a K\"ahler manifold or singular K\"ahler variety.
The endpoint of the Mori program or of K\"ahler-Ricci flow is
 a projective or K\"ahler space, so only global blowups are
permitted.  We will discuss a different perspective on the question of
global blowups, including a physical interpretation, in \S\ref{sec:conifold}.

Returning to F-theory, 
consider a Weierstrass model in the form \eqref{eq:W} with associated
elliptic fibration
$\widetilde{X}\to B$.  Allowing the total space $\widetilde{X}$ to
be a Calabi--Yau variety does not seem to compromise the integrity 
of the F-theory compactification.
Let $X\to \widetilde{X}$ be an MCP blowup.
If the induced map
$X\to B$ has any fiber components of complex dimension two, the
corresponding F-theory model has tensionless strings in its low-energy
spectrum, which indicates that there is a conformal field theory
(CFT) as part of the spectrum \cite{Seiberg-SCFT}.
F-theory models are easier to understand when they are not
coupled to conformal field theories in this way, and fortunately
it is known from
work of  Miranda \cite{MR690264} and Grassi \cite{Grassi91} that
there is a blowup $\widehat{B}\to B$ of the base and an elliptic fibration
$X\to \widehat{B}$ all of whose fibers are one-dimensional.
(Such a fibration is called {\em flat}.)
The corresponding F-theory models have a conventional field-theoretic
description at low energy, which can be understood as an effective description
of the tensor branch of the CFT.
We will use such modified bases and the associated flat fibrations
frequently in this paper.
The necessity for blowing up the base in order to obtain a flat family
can be detected directly from the
Weierstrass model \eqref{eq:W}:  
any point in the base at which $f$ has multiplicity at least
four and $g$ has multiplicity at least six must be blown up.

\section{Fiber products of rational elliptic surfaces} 
\label{sec:products}

We begin in \S\ref{sec:Schoen} by summarizing the construction of
Schoen and the generalization by Kapustka and Kapustka.  In
\S\ref{sec:pairs}, we present the main result of this paper, which is
the set of pairs of singularities that can be consistently combined in
a Calabi--Yau threefold built from a fiber product of rational elliptic
surfaces with section when blowups on the surfaces as well as small resolutions are
allowed.  \S\ref{sec:I-IV} gives an example case worked out in detail
where blowups of the base surface are needed, and \S\ref{sec:blowup-geometry} describes
the geometry on the base of the elliptically fibered threefold in the
remaining cases.  In \S\ref{sec:combinations}, we describe how the
allowed singularity pairs can be combined in different ways using
rational elliptic surfaces from Persson's list to give the complete
set of possible elliptically fibered Calabi--Yau threefolds that can be
constructed in this way.

\subsection{Schoen's construction} 
\label{sec:Schoen} 

A {\it rational elliptic surface with section}\footnote{Such surfaces are sometimes
referred to as ``$dP_9$'' in the physics literature by analogy with
the nomenclature ``$dP_n$'' for a del Pezzo surface obtained as the
blowup of $\mathbb{P}^2$ in $n$ points.}
is a complex surface $A$ that is
constructed by blowing up nine points on $\P^2$, where the first eight
points are generic and the ninth point is the  additional
point that lies on all cubics passing through the first eight points.
A rational elliptic surface is
elliptically fibered over $\P^1$ with section, and the anti-canonical
class $- K_A$ of the surface satisfies $K_A \cdot K_A = 0$.  The class $- K_A$
is effective, and can in fact be used to define the fibration, where
the zero locus of any section of the line bundle $\mathcal{O}(-K_A)$
is a fiber of the fibration.  The fibration consists
of a map 
$\pi_A:A\to \mathbb{P}^1$ whose fibers are elliptic curves.

A rational elliptic surface with section can alternatively be described by
means of a Weierstrass equation
\begin{equation}
 y^2 = x^{3} + f x + g
\label{eq:Weierstrass}
\end{equation}
where $f$ and $g$
are sections of the
line bundles ${\cal  O}(4)$ and  ${\cal O} (6)$, respectively,
on $\P^1$.\footnote{This
can be contrasted with the case of  a Calabi--Yau
variety
that is elliptically fibered over a general
base $B$
(of any dimension), in which 
the coefficients $f, g$ in the Weierstrass equation are
sections of line bundles ${\cal  O} (-4K_B)$, and ${\cal  O}(-6K_B)$.}
For every point  $p\in\mathbb{P}^1$, either $f$ vanishes to order at
most $3$ at $p$, or $g$ vanishes to order at most $5$ at $p$.
The  Weierstrass fibration defines an elliptic surface 
$\pi_{\overline{A}}:\overline{A}\to \mathbb{P}^1$ (after adding the points
at infinity in the fibers) whose total space $\overline{A}$
 is generally singular, with rational double points as 
singularities.  There is a minimal resolution of singularities
$A\to \overline{A}$ (which is unique in the case of surfaces) defining
the smooth\footnote{Some of the fibers of $\pi_A$ may be singular,
but the total space is smooth.}  elliptic surface $\pi_A:A\to \mathbb{P}^1$.

The types of singular fibers that can occur on smooth  elliptic
surfaces with section were classified by Kodaira and are displayed
in Table~\ref{t:Kodaira}.
The criteria distinguishing among the different fiber types 
of a smooth elliptic surface $A$ are expressed in terms of orders of
vanishing of the coefficients $f$ and $g$ of the
associated  Weierstrass equation \eqref{eq:Weierstrass}
as well as the order of vanishing of the discriminant
$\Delta:=4f^3+27g^2$, as shown in the
second column in the Table.
On a smooth elliptic surface $A$, the singular fiber manifests as a
(weighted)
union of intersecting rational curves of self-intersection $-2$,
according to the geometry depicted in column 3 of the Table.

\begin{table}
\begin{center}
  \begin{tabular}{ |c | r | c| c|}
  \bottomrule
  Kodaira type & deg $(f,g,\Delta)$ & geometry & rank \\
  \midrule
  $I_1$ & $(0,0,1)$ & \raisebox{-.3\height}{\includegraphics[width=0.8cm]{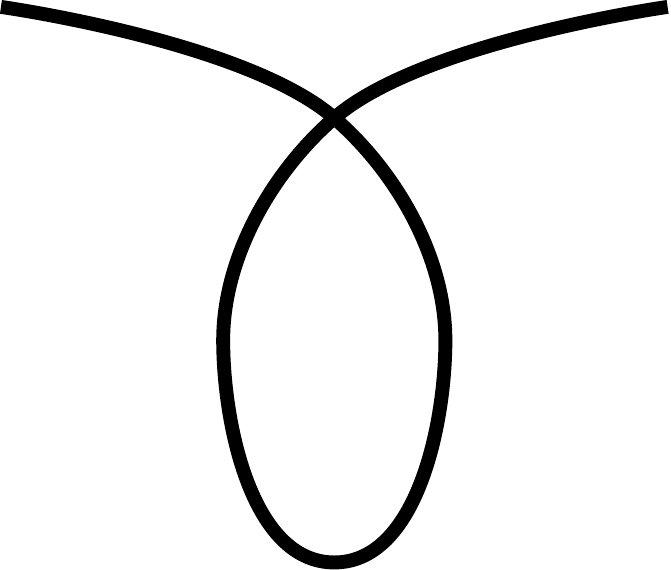}} & 0 \\[10pt]
  $I_n$ & $(0,0,n)$ & \raisebox{-.4\height}{\includegraphics[width=0.8cm]{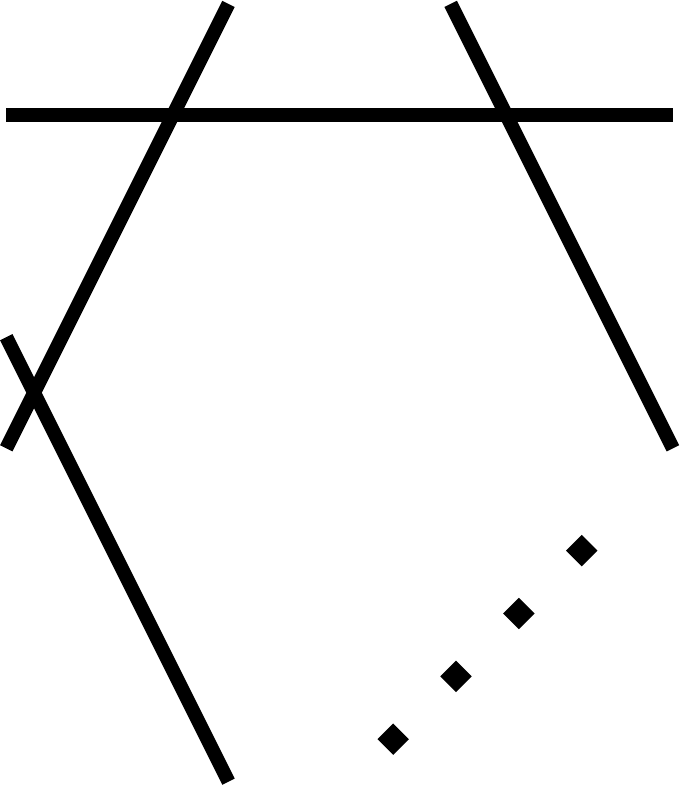}} & $n-1$ \\[12pt]
  $II$ & $(1,1,2)$ & \raisebox{-.3\height}{\includegraphics[width=0.8cm]{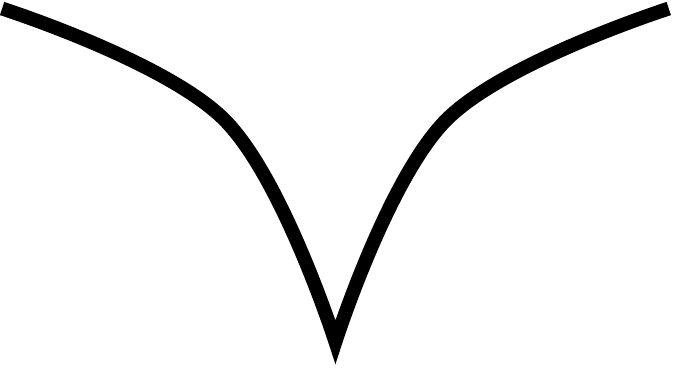}} & $0$ \\[10pt]
  $III$ & $(1,2,3)$ & \raisebox{-.3\height}{\includegraphics[width=0.8cm]{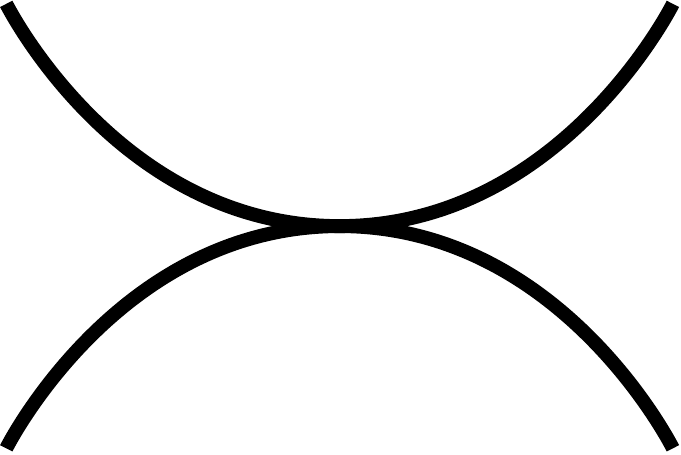}} & $1$ \\[10pt]
  $IV$ & $(2,2,4)$ & \raisebox{-.4\height}{\includegraphics[width=0.8cm]{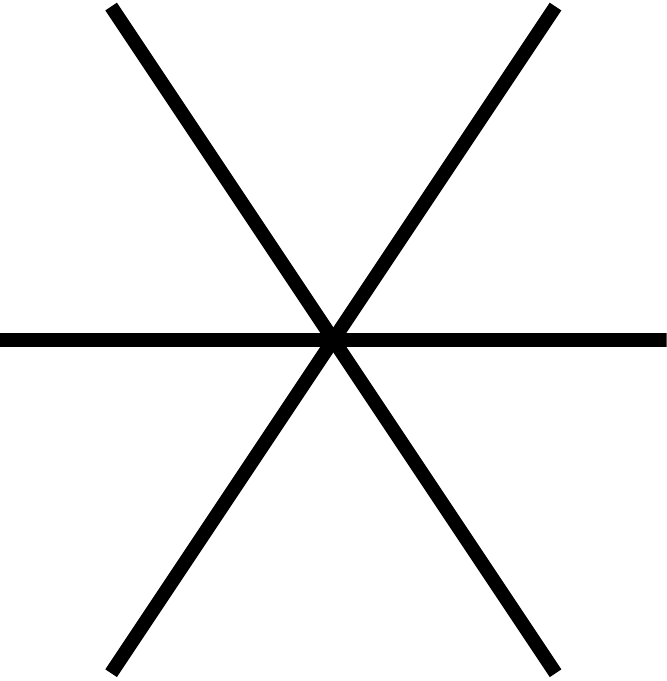}} & $2$ \\[10pt]
  $I_0^*$ & $(2,3,6)$ & \raisebox{-.3\height}{\includegraphics[width=0.8cm]{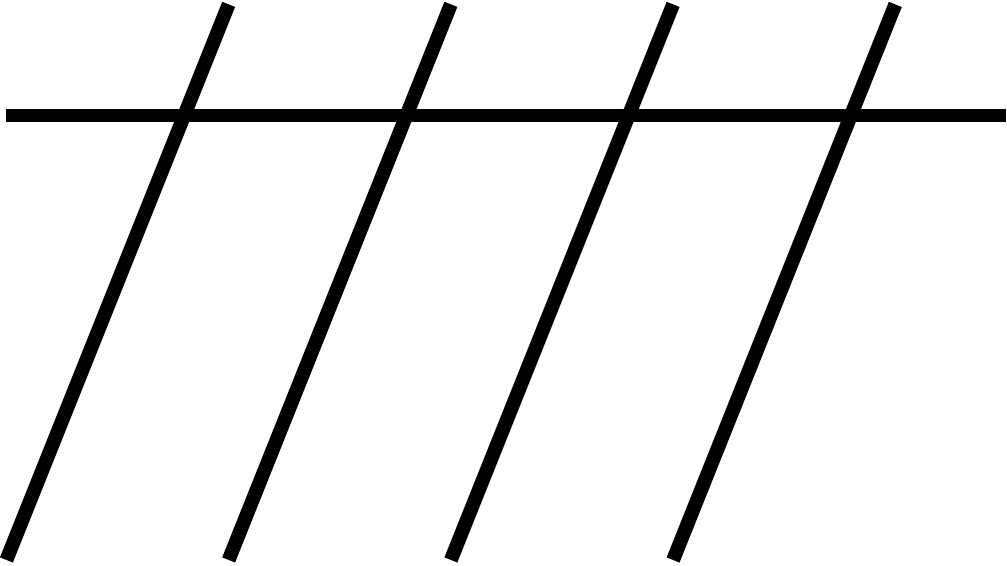}} & $4$ \\[10pt]
  $I_n^*$ & $(2,3,6+n)$ & \raisebox{-.3\height}{\includegraphics[width=2cm]{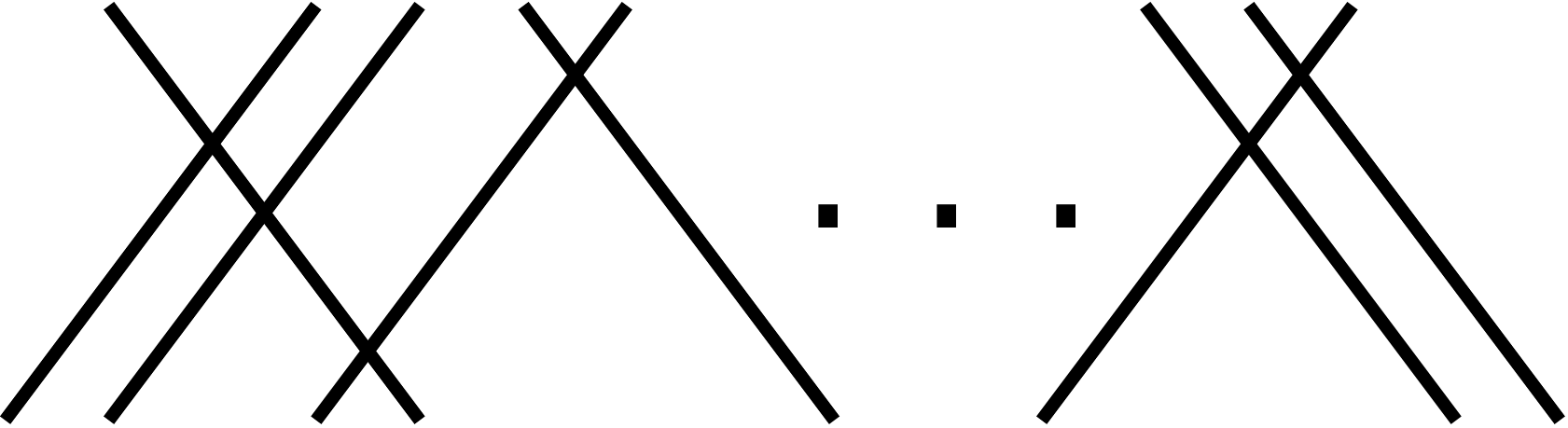}} & $n+4$ \\[10pt]
  $IV^*$ & $(3,4,8)$ & \raisebox{-.3\height}{\includegraphics[width=0.8cm]{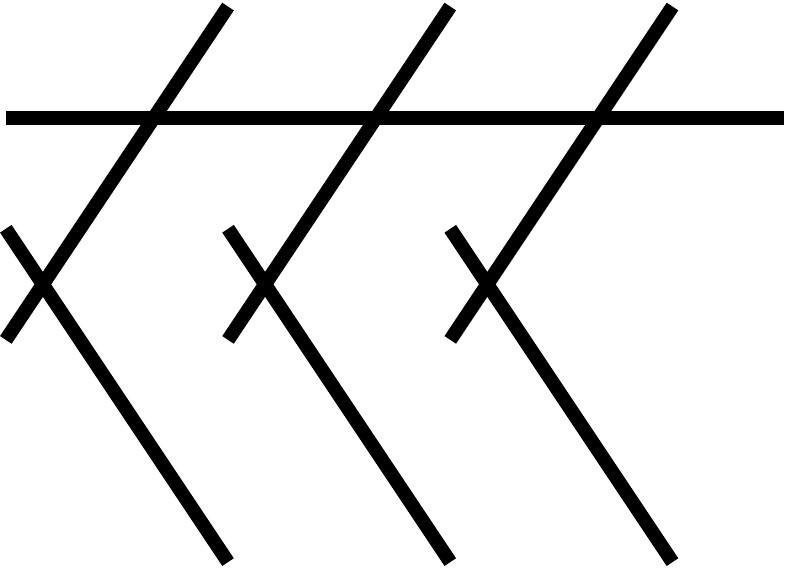}} & $6$ \\[10pt]
  $III^*$ & $(3,5,9)$ & \raisebox{-.4\height}{\includegraphics[width=0.8cm]{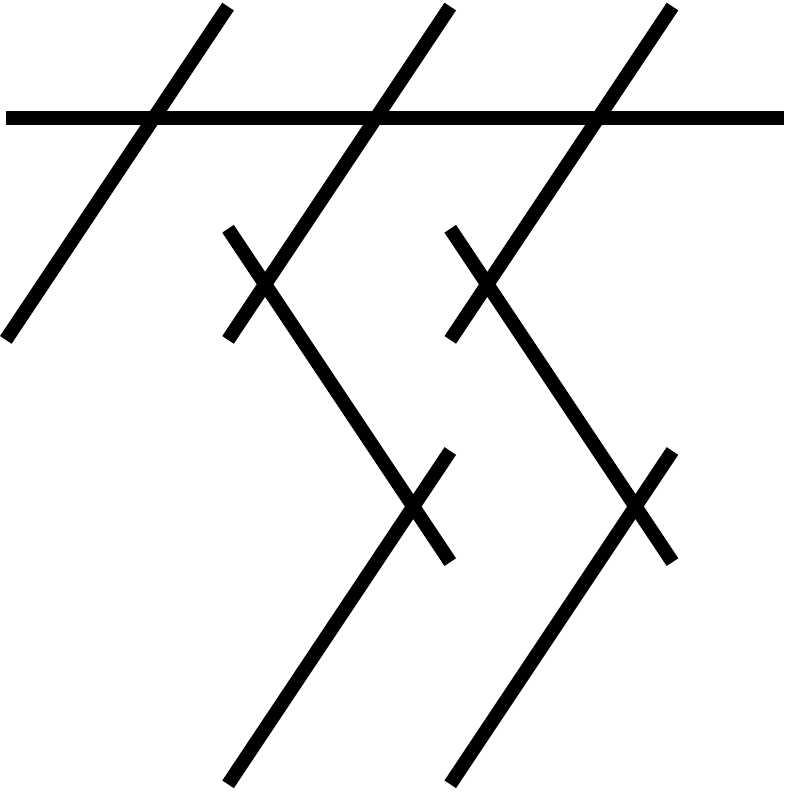}} & $7$ \\[10pt]
  $II^*$ & $(4,5,10)$ & \raisebox{-.4\height}{\includegraphics[width=0.8cm]{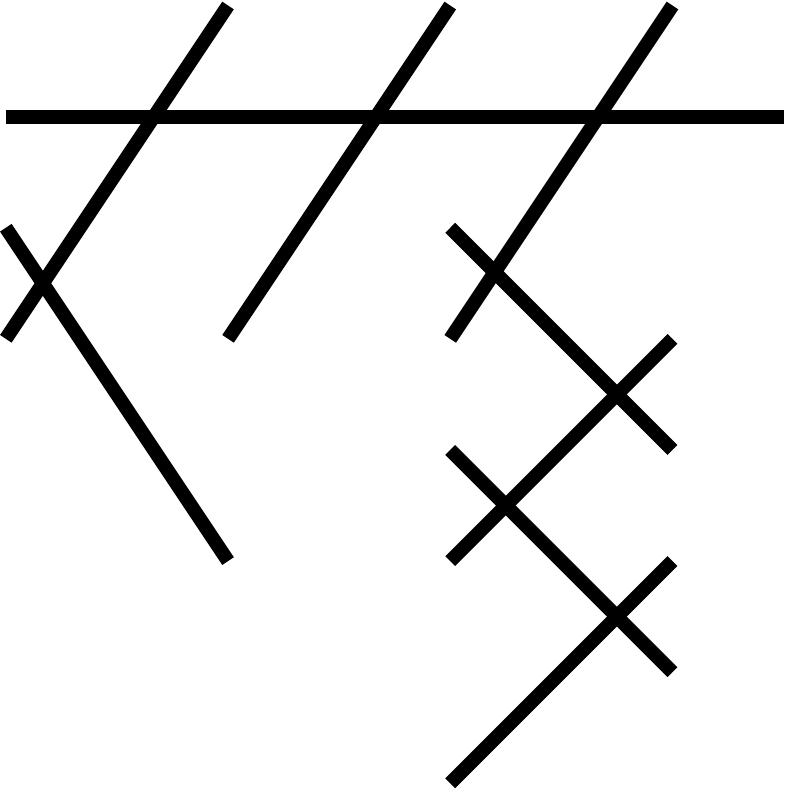}} & $8$ \\[10pt]
  \toprule
  \end{tabular}
\end{center}
\caption[x]{\footnotesize Table of Kodaira types of codimension one
  singularities in elliptic fibrations (with section).   The second column gives degrees
  of vanishing of coefficients $f, g$
in the associated Weierstrass equation (\ref{eq:Weierstrass})
and discriminant $\Delta$ at a point on the base with the given
Kodaira type of singular fiber.  The third column describes the geometry
of the fiber within an elliptically fibered surface as a weighted
combination of intersecting
$-2$ curves.}
\label{t:Kodaira}
\end{table}

A list of the possible singular fiber
configurations on any rational elliptic surface was constructed by
\person\ \cite{Persson}; a nice
alternative description of this list is given in
\cite{Miranda}.

Schoen constructed a class of Calabi--Yau threefolds in terms of fiber
products of rational elliptic surfaces with section.  Given two (smooth)
rational elliptic surfaces with section
\begin{equation}
 \pi_A: A \rightarrow \P^1, \; \; \; \;
\pi_B: B \rightarrow \P^1  \,,
\label{eq:}
\end{equation}
a threefold is constructed as the fiber product space
\begin{equation}
 \widetilde{X} = A \times_{\P^1} B := \{ (u,v)\in A\times B\ |\ \pi_A(u)=\pi_B(v)\} \,.
\label{eq:product}
\end{equation}
This threefold can be viewed as an elliptic fibration over either $A$
or $B$.  When  $A$ and $B$ both have singular fibers over a common
point $p$ in $\P^1$, the
resulting space $\widetilde{X}$ is itself  singular.
Schoen showed in \cite{Schoen} that when the fibers on the two
sides are of type $I_n, I_m$
the threefold $\widetilde{X}$ gives rise to a  smooth elliptically fibered
Calabi--Yau  threefold $X$
after a small projective
resolution of the singularities (unless $n=1$ or $m=1$, in 
which case the small projective resolution exists only if there
is an appropriate ``extra'' divisor on $\widetilde{X}$).\footnote{Note that
in the absence of an appropriate ``extra'' divisor (which in this case
is the graph of an isomorphism between $A$ and $B$),
Schoen showed that a non-K\"ahler small resolution always
exists, but that there is no K\"ahler one.}
Kapustka and Kapustka \cite{Kapustka} generalized this result to include 
the coincident singular fibers
$I_n\times III, I_n\times IV, II \times II,$ and $III \times III$
of $\widetilde{X}\rightarrow \mathbb{P}^1$.
In the following subsection we analyze an even broader class of singularity
combinations for which a Calabi--Yau threefold $ {X}$ can be
constructed as  an elliptic fibration over the base $B$
(or a blowup thereof) by resolving singularities of a fiber product.

\subsection{Allowed singularity pairs} 
\label{sec:pairs} 

\def\smooth{${\mathcal S}$}
\def\nonproj{${\mathcal G}$}
\def\terminal{${\mathcal T}$}
\def\blowup{${\mathcal S}^*$}
\def\afterblowup{${\mathcal G}^*$}

\begin{table}
\begin{center}
\begin{tabular}{| r | c c c c c | c c c c |} 
\hline
B $\backslash$ A &$I_1$ &$I_{n>1}$ & $II$ & $III$ & $IV$ & $I_n^{*}$ & $IV^{*}$ & $III^{*}$ & 
$II^{*}$\\ 
\hline
$I_1$ & \nonproj & \nonproj & \terminal & \nonproj & \nonproj & 
\blowup & \blowup & \blowup & \blowup \\
$I_{m>1}$ & \nonproj & \smooth & \terminal & \smooth & \smooth &
\blowup & \blowup & \blowup & \blowup \\
$II$ & \terminal & \terminal & \nonproj & \terminal & \terminal &
\afterblowup & \blowup & \blowup & \\
$III$ & \nonproj & \smooth & \terminal & \nonproj & \terminal &
\blowup & \blowup & & \\
$IV$ & \nonproj & \smooth & \terminal & \terminal & \blowup &
\blowup & & & \\
\hline
$I_m^{*}$ & \smooth & \smooth & \nonproj & \blowup & \blowup &
&&&\\
$IV^{*}$ & \smooth & \smooth & \smooth & \blowup &&
&&&\\
$III^{*}$ & \smooth & \smooth & \blowup &&&
&&&\\
$II^{*}$ & \smooth & \smooth &&&&
&&&\\
\hline
\end{tabular}
\end{center}
\caption{Singularities of the MCP blowup:
$\mathcal S$mooth, $\mathcal T$erminal $\mathbb{Q}$-factorial, or
$\mathcal G$lobal data needed to decide.  For entries with an asterisk,
a blowup of the base is needed to obtain a flat family.}
\label{t:types}
\end{table}

Given a pair of rational elliptic surfaces $A$ and $B$,
we now study a fiber product $\widetilde{X}=A \times_{\mathbb{P}^1} B$ along
with one of its elliptic fibrations $\widetilde{X}\to B$.  If $A$ and $B$
have any singular fibers whose locations on $\mathbb{P}^1$ coincide,
the total space $\widetilde{X}$ has singularities.
Given that the singular fibers of the $A$ and $B$ fibration at $p$
are of Kodaira type $S_A$ and $S_B$, we denote the coincident singular
fiber of $\widetilde{X}\rightarrow \mathbb{P}^1$
at $p$ by $S_A \times S_B$.
There are three general
possibilities for those singularities:  either they have an MCP blowup
which  still has a flat fibration over the base $B$,
or they have an MCP blowup which admits a flat fibration over
a base $\widehat{B}$ which is a blowup of the original base $B$,
or the singularities are worse than canonical and there is no MCP
blowup.

In fact, the statements in question depend only on a small neighborhood
of the point $p\in\mathbb{P}^1$ over which $A$ and $B$ each have a
singular fiber.  The following proposition explains what happens in each case.

\noindent
{\bf Proposition 1}
{\it Let $\mathbb{D}\subset \mathbb{P}^1$ be a small disk centered
at a point $p$, and let $\widetilde{X}\to B$ be the fiber product of two elliptic surfaces 
$\pi_A:A\to \mathbb{D}$ and $\pi_B:B\to \mathbb{D}$ considered as
an elliptic fibration over $B$.  
Table~\ref{t:types}
indicates the behavior of the singularities of 
$\widetilde{X}$, as a function of the singularity types of the fibers of
$A$ and $B$ at $p$, as follows:
}

{\it
\begin{itemize}
\item
For cases denoted \smooth, there is an MCP blowup  ${X}\to\widetilde{X}$
that is smooth and has
an induced flat elliptic fibration ${X}\to B$,
as shown in \cite{Schoen} and \cite{Kapustka}
for the cases in the upper left box.

\item
For cases denoted \nonproj,  there is an MCP blowup
${X}\to\widetilde{X}$ that has an induced 
flat elliptic fibration ${X}\to B$. The space ${X}$ 
may or may not be smooth, depending on the existence of ``extra''
divisor(s) on $\widetilde{X}$ (which is a global consideration
depending on more information than just the types of the singular
fibers of $A$ and $B$).  Note that either all MCP blowups have
a smooth total space, or none of them do.

\item
For cases denoted \terminal, there is an MCP blowup
 ${X}\to\widetilde{X}$ that has a
flat elliptic fibration ${X}\to B$, but for every MCP blowup,
$X$ is singular,
as shown in \cite{Schoen}, \cite{Kapustka}, and reviewed in
Appendix~\ref{app:terminal}.

\item
For cases denoted \blowup, there is an MCP blowup ${X}\to\widetilde{X}$
that is smooth and has
a flat elliptic fibration  ${X}\to \widehat{B}$ over a base $\widehat{B}$
that is a blowup of the original base $B$.

\item
For the case denoted \afterblowup, 
there is an MCP blowup
 ${X}\to\widetilde{X}$ that
has an induced flat elliptic fibration  ${X}\to \widehat{B}$ over a base $\widehat{B}$
that is a blowup of the original base $B$.
The space ${X}$ 
may or may not be smooth, depending on the existence of ``extra''
divisor(s) on $\widetilde{X}$ (which is a global consideration
depending on more information than just the types of the singular
fibers of $A$ and $B$).  Note that either all MCP blowups have
a smooth total space, or none of them do.

\item
For cases with no entry in Table~\ref{t:types}, the singularities of $\widetilde{X}$ are
worse than canonical and there is no MCP blowup.
These singularities are separated from the bulk of the moduli space by
an infinite distance,  as shown by Hayakawa and Wang.
\end{itemize}
For notational convenience in our later discussion, 
when $B$ requires no blowup we define
$\widehat{B}:=B$ so that in every case we can refer to the flat elliptic fibration
$X\to \widehat{B}$.
}

In all cases other than those without entries, we expect a consistent
F-theory model.

The cases in the upper left box of Table~\ref{t:types} are  treated
in \cite{Schoen} and \cite{Kapustka}.  
We review some of the details in
Appendix~\ref{app:terminal}.
Among other things, ref.~\cite{Schoen}
spells out the ```global considerations'' 
when $A$ and $B$ come from rational elliptic surfaces over $\mathbb{P}^1$
and have singular fibers of types $I_1$ and $I_n$:  the ``extra''
divisor in question is the graph of an isomorphism between $A$ and $B$
(which only exists when they are isomorphic).

There is a simple interpretation in the context of M-theory
of terminal singularities that admit a
small resolution locally, but that do not admit a global K\"ahler resolution
unless there is a global section that can be used to resolve it completely.
A terminal singularity of such type lies above a codimension-two
locus in the base.
There is a subset $\mathcal{S}$ of irreducible curves that come
from the local small resolution of the singularity,
whose elements do not intersect any fibral resolution divisors.
Recall that each fibral resolution divisor represents
an element of the Cartan of the non-abelian gauge algebra.
Since the curves in $\mathcal{S}$ do not intersect any of the fibral
resolution divisors, the only physical scenario in which the singularity can
be completely resolved is when the curves in $\mathcal{S}$
represent matter charged under some abelian gauge factor.
Such a factor can only exist when there are
additional global sections that intersect the curves in $\mathcal{S}$.

It is worth observing that the fiber product $\widetilde{X}$ is a
partial resolution of singularities of the ``Weierstrass fiber product''
\begin{equation}
\overline{X} := 
\overline{A} \times_{\P^1} \overline{B} := \{ (\bar u,\bar v)\in \overline{A}\times \overline{B}\ |\ \pi_{\overline{A}}(\bar u)=\pi_{\overline{B}}(\bar v)\} \,.
\label{eq:Wproduct}
\end{equation}
For analyzing the elliptic fibration $\widetilde{X}\to B$, it is useful
to also introduce 
the intermediate partial resolution ${X'}\to B$ defined by
\begin{equation}
{X'} :=
\overline{A} \times_{\P^1} B := \{ (\bar u,v) \in \overline{A} \times B \ |\
\pi_{\overline{A}}(\bar u) = \pi_B(v) \} \, .
\end{equation}
The fibration ${X'}\to B$ is in Weierstrass form,
and the Weierstrass coefficients $f_{{X'}}$ and $g_{{X'}}$
are completely determined from those of $\overline{A}$ by
\begin{equation}
\begin{aligned}
f_{{X'}} &= \pi_B^*(f_{\overline{A}}) \\
g_{{X'}} &= \pi_B^*(g_{\overline{A}}) \, .
\end{aligned}
\end{equation}
The original fiber product $\widetilde{X}\to B$ is a partial resolution of
singularities of ${X'}\to B$, and the total space ${X'}$
is a partial resolution of singularities of $\overline{X}$.
For most of the paper, we have chosen to refer to the F-theory
compactification on $\widetilde{X}$, as opposed to $X'$, to avoid confusion,
although $X'$ encodes enough data to define an F-theory background.
On the other hand, we have been careful to refer to $X'$ when discussing
various singularities associated to the gauge group and matter content.

We are particularly interested here in the cases
\blowup and \afterblowup, where a blowup of the base is needed to resolve the
singularities.  Such geometries were not considered in
 \cite{Schoen} and \cite{Kapustka}.

\subsection{Example:  $I_0^{*}$ singularities in $B$}
\label{sec:I-IV} 

As an example, we work out in detail the case where $B$ considered as
an elliptic fibration over $\P^1$ contains a
singular fiber of type $I_0^*$ over a point $p \in \P^1$.  In this case, as shown in
Table~\ref{t:Kodaira}, $B$ contains a singular elliptic curve $C$
described by a single $-2$ curve taken with multiplicity two, which
intersects four other $-2$ curves each taken with multiplicity one.
If $A$ also contains a singular fiber at the point $p$, then the fiber
product space $\widetilde{X}$ defined through (\ref{eq:product}) is
singular.
Considering $\widetilde{X}$ as an elliptic fibration over $B$, the type of 
singularity in $A$  over $p$  determines the degrees of vanishing of
$f, g, \Delta$ along the curve $C$
in the Weierstrass model $\widehat{X}\to B$ of $\widetilde{X}$.
For example, if there is a type $III$ singularity in $A$ over $p$,
then the component of $C$ that is a $-2$ curve with multiplicity 2 will
have degrees of vanishing of $f, g, \Delta$ of $(2, 4, 6)$, while the
degree of vanishing over the other $-2$ curves will be $(1, 2, 3)$.  
The multiplicity  of $f, g,  \Delta $
at a point in $B$  where two curves intersect, over each of which the
fibration becomes singular, will generically be the sum of the
degrees of vanishing over the two curves.  Thus, in the case of the type
 $III$ singularity in $A$, the multiplicities of $f, g, \Delta$  at the
intersection points between the $-2$ curves that meet
will be $(3, 6, 9)$.  In
such a case, all the singularities of $\widetilde{X}$
can be resolved by a crepant projective blowup.

\begin{figure}
\begin{center}
\includegraphics[width=3cm]{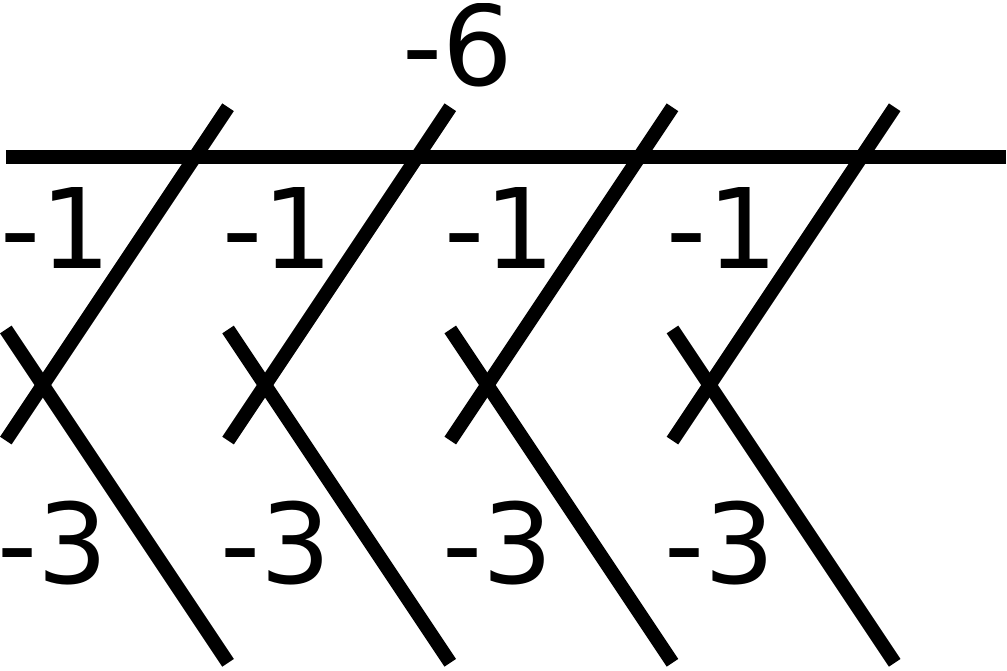}
\end{center}
\caption[x]{\footnotesize The configuration of  intersecting
divisors of negative self intersection on $\widehat{B}$
that replaces the $I_0^*$
configuration of $-2$ curves in the surface $B$ when a fiber product
space $\widetilde{X}$ is formed from $B$ and a rational elliptic surface $A$ with
a type $IV$ singularity over the same point $p$
as the $I_0^*$ singularity in $A$, and $X\to\widetilde{X}$
resolves the singularities, giving
 an elliptic fibration over the blowup $\widehat{B}$.}
\label{f:I-IV}
\end{figure}

If the multiplicities of $f, g, \Delta$ at any point in the surface
$B$ reach or exceed $(4, 6, 12)$, however, then any MCP blowup of the
Calabi--Yau total space creates a new divisor and so the map to $B$ is
not equidimensional.  To avoid this, we blow up the base $B$ before
resolving singularities, giving a new surface $\widehat{B}$.  If $A$
contains a singularity of type $IV$, for example, the degrees of
vanishing on the $-2$ curve in $B$ of multiplicity 2 become $(4, 4,
8)$, while the degrees of vanishing on the other $-2$ curves become
$(2, 2, 4)$, and the multiplicities at the intersection points become
$(6, 6, 12)$.  In this case, each of the intersection points must be
blown up.  This gives four new $-1$ curves in the base (the
exceptional divisors of the blowups), and the self-intersections of
the original $-2$ curves become $-6$ for the curve of multiplicity 2,
and $-3$ for the curves of multiplicity one.  The resulting Calabi--Yau
$X$ is then an elliptic fibration over a new surface $\widehat{B}$
that contains in place of the original union of $-2$ curves of type
$IV$ the configuration of curves shown in Figure~\ref{f:I-IV}.  The
degrees of vanishing of $f, g, \Delta$ over the exceptional curves
arising from the blowup are given by subtracting $(4, 6, 12)$ from the
multiplicities at the original intersection point, so in this case are
given by $(2, 0, 0)$.

Since there are no points on $\widehat{B}$ where $f, g, \Delta $ vanish to
degrees 4, 6, 12, the Weierstrass model ${X'}\to B$ blows up to a Weierstrass elliptic fibration $\widehat{X}\to\widehat{B}$ 
with codimension one Kodaira singularities of types $IV^{*}$ and $IV$
over the $-6$ and $-3$ curves in $B'$ respectively; the standard
resolution of these singularities gives a smooth
Calabi--Yau threefold ${X}$ that is elliptically fibered
over $\widehat{B}$.

If we increase the singularity on $A$ over $p$ further, the fiber
product space no longer has canonical singularities.
For example, if
there is a type $I_0^*$ singularity on $A$, this gives vanishing
degrees of 4, 6, 12 over the $-2$ curve with multiplicity 2 in $B$.
Such a singularity cannot be blown up in a way that gives a Calabi--Yau
total space, regardless of whether points in the base $B$
are blown up.

\subsection{Blowup geometry} 
\label{sec:blowup-geometry} 

\begin{figure}[h!]
\begin{center}
\includegraphics[width=9.9cm]{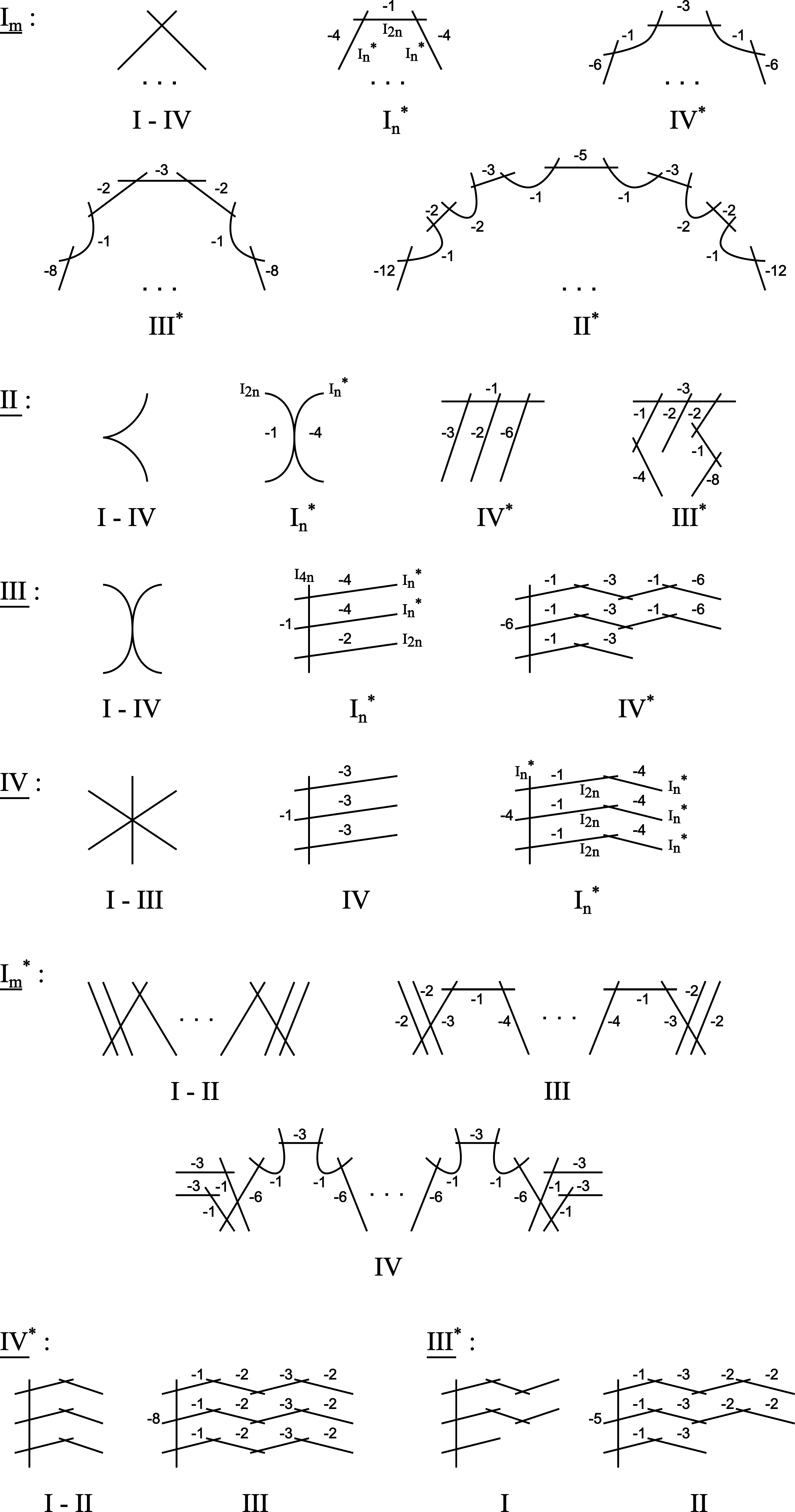}
\end{center}
\caption[x]{\footnotesize Geometries of the blown up
rational curve configurations in $\widehat{B}$
for different B-fiber types. (The B-fiber type is indicated by underlining).}
\label{f:Blowup_Base}
\end{figure}

The other cases described in Proposition 1 can be analyzed in a
parallel fashion.  In each case, thinking of $\widetilde{X}$ as an elliptically
fibered space over the base $B$, a singularity in the elliptic
fibration structure of $A$ corresponds to vanishing degrees of $f, g,
\Delta$ over the corresponding fiber in $B$, according to
Table~\ref{t:Kodaira}.  In each case, this imposes vanishing degrees
on the constituent $-2$ curves in the singular fiber of $B$; 
if the multiplicities
reach 4, 6, 12 at a point than the point in $B$ must be blown up,
while if the
vanishing degrees on a curve reach $4, 6, 12$, then
the space does not have canonical singularities, so no blowup will
produce a Calabi--Yau variety.
Carrying out this algorithm for each of the cases involving a blowup
on $B$, we arrive at the curve geometries illustrated in Figure \ref{f:Blowup_Base}
in the blown up $\widehat{B}$ that supports an elliptically fibered
Calabi--Yau threefold ${X}$ with at worst $\mathbb{Q}$-factorial
terminal singularities.
We note that the un-labeled
curves in the first column of the Figure all have self-intersection $-2$.

There are five basic types of cases shown in Figure~\ref{f:Blowup_Base}.
\begin{enumerate}
\item 
For each of the first four B-fiber types $I$, $II$, $III$, $IV$, 
the cases depicted in the first diagrom of A-fibers require no blowup of the 
base and the analysis is just the one given in \cite{Schoen} and
\cite{Kapustka} and summarized in Proposition 1.  
\item 
For each the last three B-fiber types $I_n^*$, $IV^*$, $III^*$,
the cases with A-fiber of type $I_n$ have gauge groups forming a 
quiver that reproduces
an affine simply laced Dynkin diagram on the base, considered in
\cite{Katz:1997eq}.
\item For all other entries whose A-fiber is not of type $II$ or type $I_n^*$, 
the final blown up configuration consists of non-Higgsable clusters
\cite{clusters}
joined by $-1$ curves (discussed further in \S\ref{sec:combinations}), and as a result,
the resolution of the codimension-one singular locus resolves the
singularities.  The Kodaira fiber types are those dictated by
the non-Higgsable clusters.  Note that in several instances, there are
$-1$ curves of Kodaira type $II$ that do not belong to a non-Higgsable
cluster:  those are $-1$ curves joining a $-5$ curve to a $-3$ curve,
which occur for $II^* \times I_n$ and for $II \times III^*$
(see also \cite{Martini-WT}).  
All other
$-1$ curves are Kodaira type $I_0$ ({\it i.e.}, nonsingular fibers).  
There are some  clusters consisting of only a single $-2$ curve
without a gauge group, as in the $IV^* \times II$ and
$I_0^* \times III$ examples,
and these also have Kodaira type $I_0$.
\item When the A-fiber is of type $II$ and the B-fiber is of type $IV^*$,
the intersections among curves in the base to be concerned with have types $II$-$IV$ and $IV$-$I_0^*$.
These all have crepant projective resolutions \cite{MR690264,alg-geom/9305003}.

On the other hand, when the A-fiber is of type $II$ and the B-fiber
is of type $I_m^*$, the MCP blowup is not smooth.  We consider this
case in Appendix~\ref{app:resolution}.
\item When the A-fiber is of type $I_n^*$, there are two kinds of resolutions
that may be needed.  In most cases,
we have an $I_n^*$ curve transversally intersecting an $I_{2n}$ curve.
(Note that the transversality of the intersection gives
monodromy to the $I_{2n}$ curve so that the corresponding gauge algebra is
$\mathfrak{sp}(n)$.)  The singularities that appear here were explicitly
resolved by Miranda \cite{MR690264}.

On the other hand, for a base with a type $II$ fiber, 
we have an $I_n^*$ curve
that is simply tangent to an $I_{2n}$ curve.
We consider the MCP blowup of this case in
Appendix~\ref{app:resolution}.

\end{enumerate}

\subsection{Combining rational elliptic surfaces} 
\label{sec:combinations}  

Persson's list \cite{Persson, Miranda} gives the 279 possible
combinations of singular fibers that can arise in a rational elliptic
surface.  A wide range of Calabi--Yau threefolds can be constructed by
choosing each of $A$ and $B$ from this list, and then selecting
combinations of coincident singular fibers
that satisfy the constraints of Proposition 1.\footnote{
There is one additional condition that must be considered in general:
whether the locations of the points in the discriminant locus
on each fibration can be tuned to that the desired points coincide between
the two fibrations.
However, in all of our explicit examples, we consider three singular fibers
or fewer, and this is not an issue in those cases.}

As an example, one allowed configuration on Persson's list is a
collection of three type IV singularities.  If we choose both $A$ and
$B$ to be this type of rational elliptic surface, there are four
distinct classes of smooth Calabi--Yau  threefolds that can be constructed,
depending on whether 0, 1, 2, or 3 pairs of type IV singularities on
$A$ and $B$ are chosen to lie over coincident points $p$ in the common
base $\P^1$.

The rank of the Mordell--Weil group of the Calabi--Yau threefold
${X}$ considered as an elliptic fibration over $\widehat{B}$ can be found by
subtracting the rank contribution from each singularity in $A$ as
listed in Table~\ref{t:Kodaira} (in the absence of global sections of $X$
that do not come from pulling back the global sections of $A$). 
Thus, in the preceding example
with three coincident $IV\times IV$ pairs, the
Mordell--Weil group of the resulting ${X}\to \widehat{B}$ is $8-3 \times 2 = 2$.

As another example, we can take $B$ to be a rational elliptic surface
with two $I_0^*$ singular fibers.  We can choose $A$ to have type $IV$
singular fibers at the points corresponding to the two singular fibers
of $B$.  This reduces the rank of the Mordell--Weil group to $4$, for a
generic elliptic fibration over the blown up base $\widehat{B}$, which has
within it two configurations of curves as shown in
Figure~\ref{f:I-IV}.  In addition to the two $IV$ singularities, we
can arrange additional singular fibers in $A$, but the total degree of
vanishing of $\Delta$ of the additional fibers can be at most 4.
We describe this
configuration more explicitly in the following section, as an example
of a case where the base has $\C^{*}$-structure and the generic
elliptic fibration has a Mordell--Weil group with nonzero rank.

In \cite{clusters},  two of the present authors used a physically motivated
approach to identify all possible connected configurations of curves
of negative self intersection  $-2$ or below
that can lie in a surface $B$ that supports an
elliptically fibered Calabi--Yau threefold with section.  Such
configurations consist of  single curves of negative self intersection
$-3, \ldots, -8, -12$, linear chains of curves of self intersection
$(-3, -2), (-3, -2, -2), (-2, -3, -2)$, and general configuration of
$-2$ curves.  Furthermore, there are strong restrictions on which
combinations of these ``non-Higgsable clusters'' can be connected by $-1$ curves, as
detailed in \cite{clusters}.  As mentioned in the previous section,
all of the configurations of curves in
blown up bases $\widehat{B}$ as described above consist of various combinations
of these clusters, connected in allowed ways by $-1$ curves.
We note that while these configurations of curves are associated
to non-Higgsable gauge groups in the low-energy theory \cite{clusters},
the manifolds obtained by resolving
the coincident singular fibers may have enhanced gauge symmetry
that can be Higgsed further down. For example, any of the base
singularities coincident with an $I_n$ or $I_n^*$
singularity with large enough $n$ along the
$A$-direction has enhanced gauge symmetry.
The theories whose gauge/matter sector consist of only
non-Higgsable clusters can be obtained from these manifolds
by further setting the complex structure to a generic point, and
thereby completely Higgsing the gauge groups that could be
gotten rid of this way. Only after this Higgsing do we arrive at
a generic elliptic fibration over the same base $\widehat{B}$ obtained by the
resolution procedure.

\subsection{Conifold transitions, the Higgs mechanism,  and massless hypermultiplets}
\label{sec:conifold}

One of the most subtle aspects of Mori theory is the issue of
small resolutions (or small partial resolutions) of terminal singularities.
As explained 
in \S\ref{sec: canonical}, 
when there is no
crepant projective blowup possible, the singularity is called 
{\em $\mathbb{Q}$-factorial}, and this is a property we expect thanks
to Mori theory.  Whether there exists a crepant projective blowup
 is a question that cannot be decided by simply
analyzing a small neighborhood of the terminal singular point in question:
from the point of view of algebraic geometry, the crucial question is
whether or not a certain global divisor exists on the singular space.

This issue is
familiar from its appearance in the first example of a conifold transition
studied in the physics literature 
\cite{Candelas:1988di,Candelas:1989ug,Greene:1995hu}.
In that example, a limiting quintic threefold contains $16$ conifold points,
all located in a single $\mathbb{P}^2$ contained within the limiting
quintic.  When the $\mathbb{P}^2$ is blown up, all $16$ conifold points
are resolved to $\mathbb{P}^1$s, but they only contribute a single K\"ahler class to the
blown up threefold.  In other words, these small blowups are not independent:  all $16$
must be done at once, or none at all can be done.
In particular, all of the $\mathbb{P}^1$s lie in a common homology class.

Furthermore, as emphasized in
\cite{Greene:1996dh},  each conifold point
has an $S^3$ ``vanishing cycle'' in any nearby nonsingular quintic threefold.
But these $S^3$'s are also not independent in homology:  there is a linear
relation among them which implies the existence of a $4$-chain whose
boundary is that linear relation.  In the conifold limit, the $4$-chain
becomes a $4$-cycle, namely, the divisor which is to be blown up.
Thus, an alternate description of the global issue is the question of
whether the vanishing cycles of the individual conifold points are linearly
independent in homology or not.

This transition is interpreted in \cite{Greene:1995hu,Greene:1996dh}
as an instance of the Higgs mechanism for a $\mathfrak{u}(1)$ gauge symmetry.\footnote{The
discussion in \cite{Greene:1995hu,Greene:1996dh} is in terms of type II
string theory, but we give an M-theory version here \cite{Aspinwall:2000kf}.}
The blown up space has a $\mathfrak{u}(1)$ gauge symmetry and $16$ massive hypermultiplets
of charge $1$, represented by M2-branes wrapping $\mathbb{P}^1$s.  
When we blow down, the $16$ hypermultiplets become massless.
As discussed in \cite{Greene:1995hu,Greene:1996dh}, the space of
flat directions which enable the transition can be identified with the
space of homology relations 
among the $\mathbb{P}^1$s.  If there were only a single $\mathbb{P}^1$
charged under the $\mathfrak{u}(1)$ (and hence no homology relation), there would
be no flat direction and no transition.\footnote{This follows from
a mathematical result of Friedman \cite{MR848512}, which states
that   given a collection of disjoint contractible
rational curves $C_1$, \dots $C_k$ on a Calabi--Yau threefold $X$, the
singular space 
$\overline{X}$  obtained by contracting these
curves  can be smoothed by a complex structure
deformation if and only if there is a relation
$
\sum_{j=1}^k a_j [C_j] = 0
$
among the homology classes $[C_j]$ such that all coefficients $a_j$ are
nonzero. This
does not contradict the result of Namikawa and Steenbrink
\cite{namikawa-steenbrink} because
$\overline{X}$ is not $\mathbb{Q}$-factorial.}
Note that on the Higgs branch, the massless hypermultiplets which arose
from flat directions no longer have an interpretation in terms of wrapped
branes.

More generally,\footnote{The mathematics of the general situation
was analyzed in \cite{clmnone,clmntwo,MR848512} and the physics
was discussed in \cite{looking}.}
we can consider a transition in which the complex
structure of a Calabi--Yau manifold is deformed until it aquires
$\delta$ conifold points whose vanishing cycles admit $\rho$ homology
relations.  At the transition point, the $\rho$ 4-chains will all become
4-cycles realized by divisors, which both enhance the gauge symmetry by
$\mathfrak{u}(1)^{\oplus \rho}$ and provide the geometric ingredients
needed for 
performing the small blowups.
The intersections of
the $\mathbb{P}^1$s with the new divisors after blowing up determine
the charges and the homology classes, and we see that there must be
$\delta-\rho$ homology relations among the $\mathbb{P}^1$s.

If we approach the transition from the Coulomb branch, {\it i.e.}, approach
the singular space by blowing down a collection of $\delta$ $\mathbb{P}^1$s,
we get one massless hypermultiplet for each of the $\mathbb{P}^1$s,
each charged under the $\mathfrak{u}(1)$.  There are $\delta-\rho$ flat directions, 
corresponding to the 
$\delta-\rho$ homology relations among the $\mathbb{P}^1$s, leaving only
$\delta-\rho$ ``new'' massless hypermultiplets on the Higgs branch.
On the other hand, if we approach the 
transition from the Higgs branch by ``unHiggsing,'' {\it i.e.}, approach
the singular space by complex structure deformation, we again
associate one massless hypermultiplet to each of the $\delta$ conifold points.
Out on the Higgs branch they are uncharged; however, the charges at
the transition point are associated to the coefficients in the corresponding
homology
relation among the vanishing cycles (since those coefficients go over to 
intersection numbers once the 4-chain becomes a 4-cycle
\cite{Greene:1996dh}).

Note that the extreme cases $\rho=0$ and $\rho=\delta$ are both allowed. 
If $\rho=0$ then no blowup is possible and we have only a Higgs branch.
On the other hand, if $\rho=\delta$ then the neutral massless
hypermultiplets are identical
on Higgs and Coulomb branches. In other words, the Coulomb branch contains
everything and there is no Higgs branch {\em per se}.

For more general terminal singularities, we expect a similar pattern:
in the limit of a complex structure deformation, (1) each vanishing
cycle should be associated to a massless hypermultiplet
\cite{Grassi-Weigand}, (2) each independent homology relation among
vanishing cycles should determine a $\mathfrak{u}(1)$ in the limiting
theory, and (3) the charges of each massless hypermultiplet should be
determined by the coefficient of its vanishing cycle in the
appropriate homology relations.

To illustrate this,
let us consider the simplest case where a $\mathbb{Q}$-factorial terminal 
singularity arises in the classes of manifolds of interest. Let us take a 
family $X_t$ of manifolds where the $A$ and $B$ fibrations are generic with 
twelve $I_1$ singularities. Let us assume that at $t=0$, there is a 
coincident $I_1 \times I_1$ fiber over a single point on the 
base $\mathbb{P}^1$. Upon moving to this point by taking $t \rightarrow 0$, 
a single 3-cycle vanishes and shrinks into a conifold singularity. 
The discriminant locus of $X_0 \rightarrow B$ has a single nodal 
singularity above which the conifold singularity is located. There is, 
however, no global divisor that can be utilized to resolve this singularity, 
and thus there is no enhanced gauge symmetry at this point. Following the 
discussion presented above, we find that a single neutral massless 
hypermultipet is localized at this conifold singularity.

These types of singularities can also arise at points containing  matter 
charged under nonabelian groups, particularly in cases where the Kodaira 
singularity type associated with the nonabelian algebra is non-minimal.
In Appendix B, we describe another class of examples where extra states 
arise at the intersection of type $IV$--$IV$ tunings of the gauge 
algebra $\gsu_2\oplus \gsu_2$.
Another example of this was encountered in \cite{Grassi:2014zxa}, 
where an additional uncharged scalar at a $III$--$IV$ intersection fills 
out the desired matter content of the standard model
gauge algebra $\gsu_2\oplus \gsu_3$.

\section{Elliptic fibrations over bases with a $\C^{*}$-structure} 
\label{sec:application}

An interesting set of examples of the threefolds described in this
paper  were recently encountered from a somewhat different direction.
Using the analysis of \cite{clusters} describing allowed
configurations of curves of negative self intersection in any given
base, a complete list of toric bases for elliptically fibered
Calabi--Yau threefolds was given in \cite{toric}.  This work was
extended in \cite{Martini-WT} to include all  bases that admit a
(single) $\C^{*}$ action.\footnote{Such surfaces have been studied
for decades (see {\it e.g.}\ \cite{MR0284435,MR0292832}) and have recently
been recognized as a special case of  ``T-varieties'' \cite{MR2975658}.}  
Of the more
than 100,000 $\C^{*}$ bases in this class, a small number (thirteen) have the
property that the generic elliptic fibration over these bases has a
nonzero Mordell--Weil rank.  These bases are tabulated in 
Appendix~\ref{sec:appendix-abelian}.  
We will refer to these bases as $\widehat{B}_1$, \dots, $\widehat{B}_{13}$
where the data describing $\widehat{B}_k$ is given in equation
(\ref*{sec:appendix-abelian}.k).

Each of these bases can be described as a blowup of a Hirzebruch
surface $\F_m$ with two disjoint sections $\Sigma_0, \Sigma_\infty$
and some number $N$ of fibers along which points either on one of the
sections or at an intersection of exceptional divisors are blown up,
which maintains the $\C^{*}$ structure.  Thus, each such base can be
characterized by the self intersections $n_0, n_\infty$ of the two
sections, and $N$ chains of divisors of given negative self
intersections that connect the two sections.
Each of these bases and the corresponding elliptically
fibered Calabi--Yau threefolds correspond to one of the fiber product
constructions described  in the previous section.

\subsection{Examples} 
\label{sec:c-example} 

As an example, consider the base \baselink{13}.
This base contains two $I_0^*$ configurations of $-2$ curves,
connected by four  $-1$ curves.  This is therefore a rational elliptic
surface with two $I_0^*$ singular fibers.  The Hodge numbers of the
Calabi--Yau threefold that is a generic elliptic fibration over a
rational elliptic surface are $h^{1, 1} = h^{2, 1} = 19$.  The
Mordell--Weil rank of this threefold  is $r = 8$.
An explicit determination of the moduli in the Weierstrass model for
elliptically fibered Calabi--Yau threefolds over this base using the
methods developed in \cite{Martini-WT} confirms that the Weierstrass
coefficients can be described as
\begin{eqnarray}
f & = & f_0z^4+ f_1z^3w + \cdots + f_4w^4\label{eq:w-46}\\
g & = & g_0z^6+ g_1z^5w + \cdots + g_6w^6\,,  \nonumber
\end{eqnarray}
where $z, w$ are homogeneous coordinates on the base $\P^1$
over which \baselink{13} is itself an elliptic fibration.  These
can be tuned, for example, to produce a type $II^{*}$ singularity at
a point in $\P^1$ by arranging the coefficients so that
$f = a (z - \alpha w)^4, g = b (z - \alpha w)^{5}(z - \beta w)$.  Such
a tuning corresponds to having an additional type $II^*$ singular
fiber in $A$  at a point on $\P^1$ away from the singular fibers of
\baselink{13}.  Since the rank reduction from the
$II^*$ singularity is 8, the resulting Calabi--Yau manifolds have
Mordell--Weil rank 0; in physics terminology, the 8 abelian gauge
fields associated with a reduction of F-theory on \baselink{13} can all be
simultaneously ``unHiggsed'' to form an $E_8$  gauge group.

Now, consider the base \baselink{12}.  This base is identical to
\baselink{13} except that one of the $I_0^*$ configurations of $-2$
curves has been replaced by the configuration of divisors shown in
Figure~\ref{f:I-IV}, corresponding to the tuning of a type $IV$ (2, 2,
4) vanishing on one of the singular fibers of the rational elliptic
surface \baselink{13}.  This corresponds to a fiber product
construction where the base $B$ has two $I_0^*$ singular fibers and
the base $A$ has a single type $IV$ fiber at the same point as one of
the singular fibers on $B$.  The Calabi--Yau threefold $\widehat{X}$ given
by resolving a generic elliptic fibration over the base \baselink{12}
generically has Mordell--Weil rank 6, where two of the sections have
been lost from the type $IV$ tuning.  The Weierstrass moduli for
elliptic fibrations over \baselink{12} are those in (\ref{eq:w-46}),
with the first (or last) two coefficients in both $f$ and $g$ dropped.
In this situation again the abelian gauge fields can be all
simultaneously unHiggsed.  This can be done by tuning an additional
$IV^*$ at a further point on the $\P^1$, corresponding to the
combination $IV, IV^*$ that is available on the Persson list; in the
physical theory, this combines the remaining 6 abelian gauge fields
into an $E_6$ gauge group.

Finally, consider the base \baselink{8}.  In this base,
both $I_0^*$ fibers have been given type $IV$ singularities, so each
cluster of $-2$ curves has been replaced by the configuration of
divisors in Figure~\ref{f:I-IV}.  The resulting Calabi--Yau threefold
has Mordell--Weil rank 4, and the Weierstrass coefficients are
parameterized as
\begin{eqnarray}
f & = & f_2z^2 w^2\\
g & = &  g_2z^4w^2 + g_3z^3w^3+ g_4z^2 w^4 \,.
\end{eqnarray}
We can now ask how much of the remaining Mordell--Weil rank can be
removed by tuning an additional singularity over a point in $\P^1$.
The discriminant is essentially a quartic, so the most we could hope
for would be to tune a type $I_4$ singularity of rank 3.  Even this
cannot be done, however, since to give  $\Delta$ four equal roots
would require setting $f = 0, g = a z^2 w^2 (z - \alpha w)^2$, which would be
a (2, 2, 4) singularity having rank only 2.  Thus, in this case, in
the physics language there are two abelian gauge factors that cannot
be removed by unHiggsing.
The fact that the type $I_4$ singularity cannot be tuned over this
base matches with the fact that the singularity combination $IV, IV,
I_4$ does not appear in Persson's list.

We have therefore seen that the three bases \baselink{13}, \baselink{12},
and \baselink{8}
are a sequence of bases corresponding to progressive tunings of
higher singularities in a rational elliptic surface $A$ that we fiber
product with the original base $B$ given by \baselink{13}.

\subsection{The other bases}

A similar analysis shows that the other 10 bases listed in
Appendix~\ref{sec:appendix-abelian} have similar descriptions through
sequences of blowups of a rational elliptic surface $B$ associated
with fiber products with different rational elliptic surfaces $A$.
The difference in the other cases is that the rational elliptic
surface $B$ itself does not have a description as a $\C^{*}$ base.  We
consider each family in turn.
\vspace*{0.05in}

\noindent
{\bf \baselink{11} $\rightarrow$ \baselink{10}, \baselink{9} $\rightarrow$ \baselink{5}} \\
The base \baselink{11} is constructed from a type $IV$ fiber with a $IV$
tuning  on the left and a type $IV^*$ fiber on the right, with generic
Mordell--Weil rank 6.  \baselink{10} comes from a type $III$ tuning on the
right (reducing the Mordell--Weil rank by 1), and \baselink{9} comes from an
increased tuning to $I_0^{*}$ on the left (reducing the Mordell--Weil
rank by another 2).  The base \baselink{5} comes from combining these
tunings, with a final Mordell--Weil rank of 3.
The resulting $f, g$ each have only two terms, so the highest
nonabelian rank
that can be reached by further tuning at an additional point in $\P^1$
is rank 1 from a $(1, 2, 3)$ tuning.  Thus, again,
in this case the original base
associated with \baselink{5} has two abelian gauge fields that cannot be
removed by unHiggsing. 
Again, this matches with the Persson list, which does not allow any
combination of singularities with total rank more than 1 beyond the
combination $III, I_0^*$.  
\vspace*{0.05in}

\noindent
{\bf \baselink{6} $\rightarrow$ \baselink{7}, \baselink{3} $\rightarrow$ \baselink{4}} \\
The base \baselink{6} is built from a type $III$ fiber on the left, and a
$III^{*}$ fiber on the right, with an $I_0^*$ tuning on the left,
giving Mordell--Weil rank 4.  \baselink{7} comes from a $II$ tuning on the
right, \baselink{3} comes from a $IV^{*}$ tuning on the left, and \baselink{4}
follows from both tunings, with a Mordell--Weil rank of 2.  The
resulting $f, g$ have respectively one and two terms.  In this case,
an  $I_2$ singularity can be tuned by choosing $f$ precisely to flip
the sign of the middle term  in the discriminant $4f^3+27g^2$.  This
corresponds to the allowed combination $II, IV^{*}, I_2$ in Persson's
list, giving in this case a single abelian field corresponding to the
base \baselink{4} that cannot be unHiggsed.
\vspace*{0.05in}

\noindent
{\bf \baselink{2} $\rightarrow$ \baselink{1}} \\
Finally, the base \baselink{2} is built from a type $II$ fiber with a
$IV^{*}$ tuning on the left and a type $III^{*}$ fiber on the right.
The Mordell--Weil rank of 2 is reduced to 1 on base \baselink{1}, where the
tuning on the left is enhanced to $III^{*}$.  The resulting $f, g$
have 2 terms each, so it is possible to tune an additional $III$ or
$I_2$ singularity at another point on the base $\P^1$, reducing the
Mordell--Weil rank to 0. Again, this matches with the fact that the
$III^{*}, III$ and $III^{*}, I_2$ singularity combinations are both
allowed on Persson's list.
\vspace*{0.1in}

Thus, all of the bases identified in \cite{Martini-WT} that have
$\C^{*}$ structure and nonzero Mordell--Weil rank everywhere in the
Calabi--Yau moduli space fit into the classification of models
described in this paper.  Matching with the results of
\cite{mt-sections}, in the case where the Mordell--Weil rank over a
given base is generically 1, there is a tuning of the Weierstrass
degrees of freedom that transforms the extra section into a vertical
divisor associated with an additional Kodaira type singularity.  In
cases where the Mordell--Weil rank is generically greater
than one, however, we have found several situations where the entire
Mordell--Weil rank cannot be removed by tuning moduli to convert the
sections to vertical divisors.  In all the cases we have studied,
there is a perfect matching between the possible tunings that remove
Mordell--Weil rank and Persson's list of allowed singularity types for
the rational elliptic surface $A$ with which we can take the fiber product.

\section{Aspects of abelian gauge symmetry}
\label{sec:abelian}

The generalization of Schoen manifolds that we study
generically have non-trivial Mordell--Weil groups.
This implies that many of the F-theory compactifications
on these manifolds have abelian gauge symmetry.
In this section, we study aspects of the abelian gauge
symmetry of these vacua.%
\footnote{Throughout this section, we refer to the gauge
algebra, rather than the gauge group of the theory, to avoid
subtleties involving the global structure of the gauge group.
We also only concern ourselves with the ``free part'' of the
Mordell--Weil group of an elliptic fibration $X$, which is obtained
by quotienting the group by its torsion subgroup.
We denote this group $\MW(X)$, as opposed to the Mordell--Weil
group $MW(X)$.}

For the purposes of studying the abelian gauge symmetry
of an F-theory compactification on the MCP blowup $X$ of
$\widetilde{X} = A \times_{\mathbb{P}^1} B$, it is useful to
start with the family of varieties
$X_t = \widetilde{X}_t = \overline{X}_t =
A_t \times_{\mathbb{P}^1} B$,
where $A_t$ is a generic elliptic fibration over $\P^1$ for $t > 0$,
none of whose
singular fibers (which are all $I_1$) are coincident with those
of $B$. We can arrive at $\widetilde{X}=\widetilde{X}_0$ by tuning the
complex structure coefficients of $X_t$, which either corresponds
to unHiggsing matter charged under a gauge group, turning
off a Higgs branch operator in a strongly coupled SCFT
\cite{Seiberg-SCFT,SCFT-1,SCFT-2}, or both.
We view the F-theory compactification on the MCP blowup
$X \rightarrow \widetilde{X}$ as a theory obtained from
$X_t$ by first moving to a special point $t=0$ in the Higgs branch
moduli space which has enhanced gauge symmetry and/or
a strongly coupled sector, and then moving on the tensor branch
of that strongly coupled sector, if applicable, corresponding to
blowing up one or more points on $B$.

For a generic fibration $A_t$ the Mordell--Weil group
$MW(X_t)$ is free and has rank 8. All the generators
of this group can be obtained by pulling back the
generators of $MW(A_t)$.
$\overline{X}_t$ is a smooth Calabi--Yau manifold with
no codimension-two singular fibers, and thus
nothing is charged under these $\mathfrak{u}(1)$s---%
the 8 $\fu(1)$s are non-Higgsable.
Upon going to the locus $\widetilde{X}$, two things can
happen that can affect the abelian gauge group of the theory:
\begin{enumerate}
\item $A$ develops singular fibers beyond $I_1$ and $II$
(which can either coincide with singular fibers of $B$ or not),
contributing negatively to the MW rank.
\item A global divisor, which can be used to blow up a collection
of codimension-two singular fibers, develops, contributing positively
to the MW rank.
\end{enumerate}

In event 1, a subset of the $\mathfrak{u}(1)$ components
of the gauge algebra enhances into a non-abelian gauge component
$\widetilde{\mathfrak{g}}$, whose gauge algebra can be read
off of the singular fibers of $A$. 
The subset of the $\mathfrak{u}(1)$ components that
are enhanced can be identified with
the Cartan subalgebra of $\widetilde{\mathfrak{g}}$.
Recall that the divisors of a rational surface form the
lattice with the inner product $\widetilde{U} \oplus (-E_8)$ with
\be
\widetilde{U}= 
\begin{pmatrix}
0&1\\1&-1
\end{pmatrix} \,,
\ee
which is given by the intersection pairing of the surface.
The sublattice $\widetilde{U}$ is spanned by the fiber and the
zero-section of the rational surface, while the $E_8$ lattice
is spanned by the rest of the divisors. Denoting the set of
irreducible components of all the singular fibers to be $T_A$, the
Cartan matrix of $\widetilde{\mathfrak{g}}$ can be identified
by the intersection matrix of $T_A$. 
Meanwhile, the orthogonal complement $T_{A,\perp}$ of $T_A$ with respect to the $(-E_8)$ lattice is called the essential sublattice of the N\'eron-Severi group of $A$, and can be identified with the ``free part'' $\widehat{MW}(A)$ of the Mordell--Weil group of $A$ \cite{Shioda}.
These elements pulled back to the MCP blowup of
$X \rightarrow \widetilde{X}$ span the subgroup $\MW_A(X)$, 
whose rank we denote $r_A$, of $\MW(X)$.
These are to be identified with
the $\mathfrak{u}(1)$s in $X_t$ that did not enhance to anything
upon moving to the locus $\widetilde{X}$.

Event 2 introduces additional $\mathfrak{u}(1)$
factors into the fray. A simple example is when the complex
structure of $A$ is tuned to be isomorphic to that of $B$.
$\widetilde{X}$ then has 12 $I_1 \times I_1$ singularities,
with codimension-two singularities localized at the 12 nodal
points of the $I_1$ fibers in the base. In this case, as originally
explained in \cite{Schoen}, the singularities
can be resolved by blowing up along the graph of
the isomorphism, which is just the image $\Gamma$ of the points
$ (x,x) \in A \times B$ under the map $A \times B
\rightarrow A \times_{\mathbb{P}^1} B$. Thus  $X$
has MW rank 9, and the gauge algebra of the theory is given by
$\fu(1)^{\oplus 9}$. Note that $\Gamma$ passes through
all the 12 singularities, thus implying that there are 12
hypermultiplets carrying unit charge under the new
$\fu(1)$.

While $\widetilde{X}$ is a singular space, we note that the
complex structure of the $A$ and $B$ fibrations
and the K\"ahler modulus of the base manifold $B$
suffice to define a sensible physical theory, as long as all the
coincident singular fibers of $\widetilde{X}$ are one of the allowed types
in Table \ref{t:types}. One may have concerns about the existence of
codimension-two singularities on $B$ that require blowing
up the base, but these singularities merely indicate that the effective
theory can be understood as a gauge theory with hypermultiplets and
SCFTs coupled to gravity, with subgroups of the global symmetry
groups of the SCFTs possibly gauged.%
\footnote{Some basic checks on the consistency of these
effective theories have been carried out in \cite{SCFT-2}.}

With this picture in mind, let us now focus our attention to
certain aspects of the abelian gauge group of the variety $X$.
In \S\ref{sec:height}, we study the abelian
anomaly coefficients of F-theory compactified on $X$. This corresponds
to the height-pairing matrix of the rational sections of the threefolds
\cite{intersection}.  In
\S\ref{sec:non-enhanceable}, we study the problem of
enhanceability of the abelian gauge group, {\it i.e.}, whether
generators of the Mordell--Weil group can be localized to Kodaira
singularities by tuning the complex structure.  In particular, we show
that there are many manifolds for which there exist factors of the
abelian gauge group that cannot be ``unHiggsed,'' or equivalently,
``enhanced'' into a non-abelian gauge group without introducing
problematic singularities.

\subsection{Abelian anomaly coefficients}
\label{sec:height}

In this section, we mainly study the anomaly coefficients
of the $\fu(1)$ components of the gauge algebra
that correspond to the generators of the subgroup
$\MW_A(X)$ of the free part $\MW(X)$ of $MW(X)$. As noted
previously, these $\fu(1)$s are the $\fu(1)$s in the compactification
on $X_t$ that did not enhance to a non-abelian gauge symmetry
at the $\widetilde{X}$ locus. At the end of the section, we present
a computation of the anomaly
coefficients of the MW rank 9 example introduced above.

Before going on further, let us recall some basic facts about
abelian anomaly coefficients in F-theory compactifications
on Calabi--Yau threefolds. The anomaly coefficients
are bilinear forms on the abelian gauge algebra that are
valued in the second homology of the base manifold.
More concretely, labeling the components of the
abelian gauge algebra by $M, N, P, Q \cdots$, the anomaly coefficients
take the form $b_{MN}$, where each $b_{MN}$
is a divisor class in the base of the elliptic fibration \cite{Park-Taylor}.
Given the K\"ahler class $j$ of the base, the kinetic terms
for the $\fu(1)$ gauge fields are given by
\be
j \cdot b_{MN}\, F^M \wedge * F^N \,,
\ee
where $F^M$ is the field-strength two-form for the $M$th
$\fu(1)$. Furthermore, the abelian anomaly equations are given by
\be\label{u1 anom}
-6K \cdot b_{MN}
= \sum_I q^I_M q^I_N \,,
\qquad
b_{MN} \cdot b_{PQ}
+b_{MP} \cdot b_{NQ}
+b_{MQ} \cdot b_{NP}
= \sum_I q^I_M q^I_N q^I_P q^I_Q \,,
\ee
where $\cdot$ here denotes the intersection pairing in the base.
The variable $I$ summed over on the right-hand side of the
equations indexes all the individual hypermultiplet
components in the theory, and $q^I_M$ is the charge
of the $I$th hypermultiplet under the $M$th $\fu(1)$.
Writing the rational section corresponding to the
$M$th abelian gauge component as $S_M$,
the anomaly coefficient can be computed geometrically by
the formula \cite{intersection}
\be
b_{MN} = - \pi_B(\sigma(S_M) \cdot \sigma(S_N))
\ee
where $\sigma$ denotes the Shioda map \cite{Shioda,
intersection, Morrison-Park} and the dot on the right-hand side of this
equation is the intersection pairing within the Calabi--Yau
manifold. $\pi_B$ is the projection map to the base.

Let us now label the (free) basis of $\MW(A) \cong
\MW_A(X) \subset \MW(X)$
by the indices $m, n, p, q, \cdots$ and proceed to compute $b_{mn}$,
which is the projection of the bilinear form $b_{MN}$
on $\MW(X)$ (viewed as a vector space) to the subspace
$\MW_A(X)$. Recall from the discussion at the beginning of
the section that for sections $s_m$ of $A$, the Shioda pairing
is given by
\be
\sigma(s_m) \cdot \sigma(s_n) = (T_{A,\perp})_{mn} \,,
\ee
where $T_{A,\perp}$ is the orthogonal complement of
the lattice spanned by singular fibers of $A$ with respect
to $(-E_8)$. Now let us denote the pull-back of $s_m$ to $X$
by $S_m$. Then the main claim of this section is that
\be\label{ab anom}
b_{mn} = - \pi_B(\sigma(S_m) \cdot \sigma(S_n)) =
-(T_{A,\perp})_{mn} F_{\widehat{B}} \,,
\ee
where $F_{\widehat{B}}$ is the fiber class of the
base $\widehat{B}$. It is convenient
to express this equation in the notation
\be\label{b result}
b_A = -T_{A,\perp} \otimes F_{\widehat{B}} \,.
\ee
The self-intersection of $F_{\widehat{B}}$ being zero,
the anomaly equations
\eq{u1 anom} imply that none of the $r_A$ $\fu(1)$s
have charged matter, and are thus non-Higgsable.

The derivation of equation \eq{ab anom} is quite simple
from the point of view described at the beginning of the section.
As explained, the theory compactified on $X_t$ has
gauge algebra $\fu(1)^{\oplus8}$. The cohomology class
of the sections of $A_t$ are merely products of the
sections of $A_t$ with the class of the fiber in the base.
Thus the gauge anomaly coefficient for this theory is given by
\be\label{X0}
b_{t>0} = E_8 \otimes F_B \,,
\ee
where $F_B$ is the class of the base $B$.
There are no codimension-two singularities in $X_t$,
thus no charged matter under the $\fu(1)$s---all 8 $\fu(1)$s
are thus non-Higgsable.
This is consistent with the anomaly equations, since
$F_B^2 = -K \cdot F_B =0$. 
The simplest way to confirm equation \eq{X0} is to tune
the complex structure of $A_t$ such that there is an $E_8$
singularity that does not coincide with any of the $B$ singularities.
The theory becomes an $E_8$ theory with a single adjoint
hypermultiplet. The $\fu(1)$s of $X_t$ can be identified with
the Cartan subalgebra of the $\mathfrak{e}_8$ algebra,
and the $\mathfrak{e}_8$ brane is located along the fiber class $F_B$.
By Higgsing the adjoint hypermultiplet of the $\mathfrak{e}_8$ theory,
we arrive at the abelian theory with anomaly coefficients \eq{X0}.

Now starting from $X_t$, we tune the complex structure of the
$A$-fibration $t \rightarrow 0$ to arrive at $\widetilde{X}$. Then a subset of
the abelian components of $A_0$ become enhanced to a
non-abelian gauge group. These components can be identified
with the subspace $T_{A}$ of $(-E_8)$. Denoting the sections
of $A_t$ that span the subset $T_{A}$ as $\widetilde{s}_i$,
the sections $\widetilde{S}_i$ obtained by pulling back these
sections to $X_t$ become vertical when we arrive at
$\widetilde{X}$. Meanwhile, the $\fu(1)$ components
of $A_t$ orthogonal to the $\fu(1)$ components
that enhance at the $\widetilde{X}$ locus remain decoupled
from the unHiggsing process, and are unaffected.
Thus these abelian components
correspond to the sections obtained by pulling back the sections
$T_{A,\perp}$ to $\widetilde{X}$, {\it i.e.}, elements of
$\MW_A(\widetilde{X})$, and their anomaly coefficients
are given by
\be\label{pre-blow-up}
\widetilde{b}_A = -T_{A,\perp} \otimes F_B \,.
\ee
When the MCP blowup $X \rightarrow \widetilde{X}$ does not
require blowups of the base, equation \eq{pre-blow-up}
implies equation \eq{b result}. When the blowup
$\widehat{B} \rightarrow B$ is required, the effective theory
of the F-theory compactification on $\widetilde{X}$ has
a strongly coupled SCFT, and the blowup of the base
corresponds to moving on the tensor branch of the SCFT.
Now moving on this tensor branch does not affect the
abelian gauge fields corresponding to elements of
$\MW_A(\widetilde{X})$ at all, since none of the operators
of the SCFT are charged under them. Thus
$\MW_A(\widetilde{X}) \cong \MW_A({X})$,
and the anomaly
coefficients of these gauge fields are given by equation
\eq{b result} as claimed:
\be\nonumber
b_A = -T_{A,\perp} \otimes F_{\widehat{B}} \,,
\ee
$F_{\widehat{B}}$ being the pull-back of the class
$F_B$ to $\widehat{B}$, which is orthogonal to the
resolution divisors introduced in the blow up. Recall that the
volumes of the resolution divisors correspond to the
vacuum expectation values of the tensor branch operators.

A crucial point in this picture is the fact that a $\fu(1)$ of
${X}_t$, at any point in the complex structure moduli space,
cannot have anything charged under itself unless it enhances
to a non-abelian gauge symmetry. Thus none of the charged
matter appearing at the $\widetilde{X}$ locus is charged
under the $\fu(1)$s corresponding to the sections $\MW_A(X)$,
since none of them have enhanced to a non-abelian gauge
symmetry. This also indicates that, although additional rational
sections can appear at the $\widetilde{X}$ locus, they cannot
obstruct the result \eq{b result}.

Let us conclude this section by discussing $X$ obtained from
resolving $\widetilde{X}$ with isomorphic $A$ and $B$
fibrations that
are generic. Let us denote the eight sections obtained by
pulling back sections $s_m$ of $A$ to $X$ by $S_m$,
and the additional section, the graph of the isomorphism, by $S_0$.
As derived previously, the anomaly coefficients only involving the
sections $S_m$ are given by
\be
b_{mn} = (E_8)_{mn} \, F_B \,,
\ee
where we take $m$ to label the roots of $E_8$.
We can utilize the result \cite{Morrison-Park}
\be
b_{MN} = - \pi(S_M \cdot S_N) -K
+ S_M \cdot Z + S_N \cdot Z
\ee
where the formula becomes simpler in the absence of any
non-abelian gauge symmetry. $S_M \cdot Z$ denotes the class of
the divisor along which the section $S_M$ intersects the zero
section $Z$.

For the rational surface $K = -F_B$. Now
$- \pi(S_0 \cdot S_0) = K = -F_B$, since $S_0$ itself is a
rational surface, and $S_0$ being the graph of the isomorphism
yields $S_0 \cdot Z = Z_B$, where $Z_B$ is the class of
the base of the rational surface. We then arrive at
\be
b_{00} = 2F_B + 2Z_B \,.
\ee
Meanwhile,
\be
S_m \cdot Z= (s_m \cdot z) F_B  = 0
\ee
for the base $z$ of the rational surface $A$, since
the sections $s_m$ of $A$ do not intersect $z$.
Also, the fact that $S_0$ is the graph of the isomorphism
implies that
\be
\pi(S_m \cdot S_0) = [s_m] \,.
\ee
We then arrive at
\be
b_{0m} = F_B + Z_B -[s_m] = -[\sigma_B(s_m)]
\ee
where $\sigma_B$ here is the Shioda map on the rational
surface $B$ \cite{Shioda}. Recall that
\be\label{sigma(s)}
[\sigma_B(s_m)] \cdot F_B  =[\sigma_B(s_m)] \cdot Z_B = 0 \,,
\quad  
[\sigma_B(s_m)] \cdot [\sigma_B(s_n)] = - (E_8)_{mn} \,.
\ee
To summarize, we find that
\be\label{rank 9 case}
b_{00} = 2F_B + 2Z_B \,, \quad
b_{0m} =  -[\sigma_B(s_m)]\,, \quad
b_{mn} = (E_8)_{mn} \, F_B \,.
\ee

Let us check the anomaly equations using the anomaly
coefficients \eq{rank 9 case}. Recall that there are 12
charged hypermultiplets in the spectrum with charge 1 under
$\fu(1)_0$ corresponding to the section $S_0$. This is because
the graph $S_0$ passes through all 12 codimension-two
singularities sitting on top of the nodal points of the $I_1$
singularities in the base. The anomaly equations only involving
indices $m > 0$ are trivially satisfied, so let us only check those
involving the index $0$. Since all of the 12 hypermultiplets are
solely charged under $\fu(1)_0$, we get
\bea
-K \cdot b_{00} = F_B \cdot b_{00} &= 2 \,, &
-K \cdot b_{m0} = F_B \cdot b_{m0} &= 0 \,, \\
b_{00}^2 & = 4 \,, &
3 b_{00} \cdot b_{0n} &= 0 \,, \\
b_{00} \cdot b_{mn} + 2 b_{0n} \cdot b_{0m} &=0 \,,&
b_{0m} \cdot b_{np} + b_{0n} \cdot b_{mp} + 
b_{0p} \cdot b_{mn}  &= 0\,. 
\eea
These equations can be confirmed by utilizing the relations
\eq{sigma(s)} and the intersection numbers
\be
F_B \cdot F_B = 0 \,, \quad
Z_B \cdot Z_B = -1 \,, \quad
F_B \cdot Z_B = 1 \,.
\ee

\subsection{Non-enhanceable $\fu(1)$ components}
\label{sec:non-enhanceable}

We begin this section by studying an
example of an F-theory compactification
with an abelian gauge symmetry,
that cannot be unHiggsed, or enhanced,
to a ``sensible" theory with
non-abelian gauge symmetry.
After examining a particular
example in detail and explaining the obstructions
to enhancement, we conclude by commenting on
the proliferation of manifolds with such sections
among the
elliptically fibered Calabi--Yau manifolds
constructed as an MCP blowup of fiber
product spaces of elliptic rational surfaces.

Before moving on further, let us define what we mean by a 
``non-enhanceable'' abelian gauge component.
We say that a $\mathfrak{u}(1)$ component of the gauge algebra of an
F-theory background $\widehat{X} \rightarrow \widehat{B}$
is {\it non-enhanceable} when
it cannot be enhanced to a non-abelian gauge symmetry
by tuning moduli of the theory, parameterized by
neutral hypermultiplets, to a point within finite distance
from the interior of  moduli space. In geometric terms, a $\mathfrak{u}(1)$
is non-enhanceable when the
corresponding rational section of the $\mathfrak{u}(1)$
cannot be converted to a vertical divisor by tuning moduli without moving
an infinite distance within the complex structure moduli space of 
the MCP blowup $X \rightarrow \widehat{X}$.
While it may be possible that for certain $X$, the $\mathfrak{u}(1)$
does not enhance at any point in the closure of the moduli
space \cite{Morrison-Park-2},
in all the examples we consider, the non-enhanceability comes from
the fact that the point of enhancement lies at infinite distance.

Now given an F-theory background with gauge algebra
$\mathfrak{g} \oplus \mathfrak{u}(1)$, it might be the case that
the $\mathfrak{u}(1)$ component is non-enhanceable by our definition,
but becomes enhanceable upon breaking $\mathfrak{g}$ into a subgroup
by giving supersymmetric expectation values to charged hypermultiplets:
\be
\mathfrak{g} \oplus \mathfrak{u}(1) 
\quad \rightarrow \quad
\mathfrak{g}' \oplus \mathfrak{h}' \oplus \mathfrak{u}(1) 
\quad \rightarrow \quad
\mathfrak{g}' \oplus \mathfrak{h} \,,
\ee
where the first arrow denotes the breaking of $\mathfrak{g}$
to $\mathfrak{g}' \oplus \mathfrak{h}'$ and the second arrow denotes the enhancement
of $\mathfrak{h}' \oplus \mathfrak{u}(1)$ to $\mathfrak{h}$. Here,
$\mathfrak{g}' \oplus \mathfrak{h}' \subsetneq \mathfrak{g}$, $\mathfrak{h}' \oplus \mathfrak{u}(1)  \subsetneq \mathfrak{h}$, where $\mathfrak{h}$
is a non-abelian algebra. This process, however, cannot happen
when the initial theory is non-Higgsable. Geometrically, this is when
the initial background is a generic fibration over a given base.
Note that in this case the $\mathfrak{u}(1)$ component
in question is always non-Higgsable.

The model whose abelian gauge symmetry
we examine in detail is the generic elliptic fibration 
\eq{eq:A8} over \baselink{8},
which is a non-Higgsable model.
Let us understand how
certain abelian gauge symmetries of the model
could be enhanced, but others can not.
Recall from section \eq{sec:c-example},
the Weierstrass model of the $A$-fibration of
this theory can be expressed as
\be
y^2 = x^3 + f x +g
\ee
with
\bea
f &= f_2 w^2 z^2 \\
g &= g_2 w^4 z^2 + g_3 w^3 z^3 + g_4 w^2 z^4
\label{Wcoeff}
\eea
for projective coordinates $z$ and $w$ on the
$\mathbb{P}^1$ base. As before, we denote the
fiber product of $A$ and $B$ as $\widetilde{X}$.
The Mordell--Weil group of
the elliptic fibration is generated by four rational
sections, and hence the gauge algebra of the F-theory
model compactified on this manifold has  four
abelian components in it. We show how this abelian component
can be enhanced to a non-abelian gauge component in
various ways, and show that in fact
two $\fu(1)$ components are non-enhanceable, {\it i.e.},
cannot be un-Higgsed to a non-abelian gauge component.

The basis of the group of sections can be chosen
from the six sections
\bea
s_i~:~x &= \alpha_i w z, \quad y = \sqrt{g_2} w^2 z + \sqrt{g_4} wz^2  \\
t_i~:~x &= \beta_i w z, \quad y = \sqrt{g_2} w^2 z - \sqrt{g_4} wz^2  \,,
\label{sections}
\eea
where $\alpha_i$ and $\beta_i$ denote the
three roots of the cubic polynomials
\be
\alpha^3 + f_2 \alpha + (g_3-2 \sqrt{g_2 g_4})=0
\label{alpha}
\ee
and
\be
\beta^3 + f_2 \beta + (g_3+2 \sqrt{g_2 g_4})=0 \,,
\label{beta}
\ee
which are distinct for generic values of the
coefficients. We note that the six sections obey
the relations
\be
s_1+s_2+s_3=0, \qquad
t_1+t_2+t_3=0 \,.
\ee
We therefore may choose, for example,
$s_1,s_2, t_1$ and $t_2$ as the generators of
the Mordell--Weil group.

The discriminant locus of this manifold
is given by
\bea
&4f^3 + 27g^2\\
&= 27 w^{4}z^4
\left[ g_2 w^2 + (g_3+{2i \ov 3\sqrt{3}}f_2^{3/2}) wz + g_4 z^2 \right]
\left[ g_2 w^2 + (g_3-{2i \ov 3\sqrt{3}}f_2^{3/2}) wz + g_4 z^2 \right] \,.
\label{disc}
\eea
Note that there are two $IV$ fibers of the
$A$-fibration located at the points $z=0$ and
$w=0$. These coincide with the loci of the
$I_0^*$ singularities of the $B$-fibration.
Upon inspection of equation \eq{disc},
we see some manifest ways of enhancing the
abelian gauge symmetry to a non-abelian gauge
symmetry, which we organize in the following way:
\begin{itemize}
\item $\mathfrak{su}(2)$ : $4f_2^3 + 27 (g_3 - 2\sqrt{g_2 g_4})^2 =0$.
\item $\mathfrak{su}(2)$ : $4f_2^3 + 27 (g_3 + 2\sqrt{g_2 g_4})^2 =0$.
\item $\mathfrak{su}(2)^{\oplus 2}$ : $g_3 =0,~f_2^3 = -27g_2 g_4$.
\item $\mathfrak{su}(3)$ (type $IV$) : $f_2=0,~g_3^2 = 4g_2 g_4$.
\end{itemize}
Note that the maximal rank enhancement
in this list is of rank 2. Further enhancement
is not possible---they lead to singularities that lie at
infinite distance from the interior of moduli space, as we show shortly.

It is instructive to work out how each of the abelian
gauge symmetries enhances to $\mathfrak{su}(2)$. Let us take
two roots $\alpha_1$ and $\alpha_2$ of \eq{alpha}
and examine the limit where the $\fu(1)$ associated
to the section $(s_1 - s_2)$ is un-Higgsed.
The complex numbers $f_2$ and $g_i$ parametrizing
the Weierstrass coefficients can then be traded for
$\alpha_1$, $\alpha_2$, $\sqrt{g_2}$ and $\sqrt{g_4}$
in the following way:
\bea
f &= -(\alpha_1^2+\alpha_1 \alpha_2+\alpha_2^2) w^2z^2 \\
g&= g_2 w^4z^2 +
(2\sqrt{g_2g_4}-\alpha_1\alpha_2(\alpha_1+\alpha_2)) w^3z^3
+ g_4 w^2z^4 \,.
\eea
The rational section $(s_1 -s_2)$, according to standard rules
of adding sections \cite{Silverman}, is given by
\be
[X,Y, Z] =\left[
\, c_3^2 -{2 \ov 3} b^2 c_2\,,\,\, -c_3^3 + b^2 c_2 c_3\,,\,\, b \,
\right]
\ee
in projective coordinates of the $A$-fiber.
Here, the coefficients $c_i$ and $b$ are given by
\be
c_2 ={3 \ov 2} (\alpha_1 + \alpha_2) w z, \quad
c_3 = 2\sqrt{g_2} w + 2\sqrt{g_4} z, \quad
b = (\alpha_1-\alpha_2) \,.
\ee
The Weierstrass model can then be written in the form
\be
y^2 = x^3 + (c_1 c_3-b^2 c_0 - {c_2^2 \ov 3}) x
+ \left(
c_0c_3^2 -{1 \ov 3} c_1c_2c_3 + {2 \ov 27} c_2^3
-{2 \ov 3} b^2 c_0 c_2 + {1 \ov 4} b^2 c_1^2
\right)
\label{pre-enhancement}
\ee
for
\be
c_0 = {1 \ov 4} w^2z^2, \quad
c_1 =0 \,.
\ee
This fits nicely into the form of the elliptic fibration
with Mordell--Weil rank-one in \cite{Morrison-Park}.%
\footnote{It has recently been shown  \cite{Morrison-Park-2}
that elliptic fibrations with Mordell--Weil rank-one do
not necessarily have to be a Jacobian of a
$\mathbb{P}^{112}$ model.}
This form, in particular, is convenient for enhancing
the $\fu(1)$ gauge algebra into an $\mathfrak{su}(2)$---we can simply
take $b= (\alpha_1 -\alpha_2)$ to be zero. By tuning
$b$ to zero, the Mordell--Weil section becomes trivial,
and an $I_2$ singularity appears at the locus
\be
c_3 =2\sqrt{g_2} w + 2\sqrt{g_4} z=0 \,.
\label{su21}
\ee
This enhancement corresponds to tuning the parameters
$f_2$ and $g_i$ so that
\be
4f_2^3 + 27 (g_3 - 2\sqrt{g_2 g_4})^2 =0 \,.
\ee
We can further enhance the gauge algebra to $\mathfrak{su}(3)$
by taking $\alpha_2-\alpha_3$ to zero. In order for
this to happen, $f_2$ must additionally be set
to zero. The gauge divisor is, as before,
given by the elliptic fiber of the base lying
above the locus \eq{su21}.

Similarly, the $\fu(1)$ associated to the section
$(t_1-t_2)$ is enhanced by taking $(\beta_1-\beta_2)$
to zero. The enhancements corresponds to the tuning 
\be
4f_2^3 + 27 (g_3 + 2\sqrt{g_2 g_4})^2 =0 \,.
\ee
This leads to an $I_2$ singularity to appear at
\be
\sqrt{g_2} w - \sqrt{g_4} z=0
\label{su22}
\ee
By tuning $f_2$ to zero, the Mordell--Weil
rank reduces further by one, and the singularity
of the $A$-fibration enhances to a $IV$ singularity
at this point.
On the other hand, we can choose to enhance
the $\fu(1)$ associated to section $(s_1-s_2)$
rather than $(t_2-t_3)$. In this case, the $A$-fibration
has two $I_2$ singularities, each at the loci
\eq{su21} and \eq{su22}.

We may attempt to un-Higgs abelian
gauge symmetries further. For example, we
can make
the Mordell--Weil group trivial by setting
either
\be
f_2 =
g_2 =
g_3 =0 \,,
\ee
or
\be
f_2 =
g_3 = 
g_4 =0 \,.
\ee
Note that in the former/latter case,
the $A$-fibration develops a $IV^*$
singularity at the locus $w=0$/$z=0$,
respectively. This means that there exists
a coincident singular fiber of $\widetilde{X}$
given by the product $IV^* \times I_0^*$
at the corresponding point, which does not have
an MCP blowup.
Another option is to set
\be
g_2 =g_4 =0 \,.
\ee
In this case, the $A$-fibration develops
two $I_0^*$ singularities, each at $w=0$
and $z=0$. The coincident singular fibers
at these points, which are both
given by the product $I_0^* \times I^*_0$,
also lie at infinite distance
from the interior of the moduli space.

In fact, certain $\fu(1)$s
cannot be un-Higgsed without giving rise to
singularities that lie at infinite distance to begin with.
The sections $(s_i - t_j)$, and $(s_i+t_j)$
are of this sort.
For definiteness, let us try to un-Higgs
the section $(s_1 - t_1)$. To do so,
we can write the Weierstrass equation
of the $A$-fibration in the form
\eq{pre-enhancement} with
\bea
c_0={1 \ov 4} w^2 z^2, \quad
c_1=
\left( {\alpha_1^2+\alpha_1\beta_1 +\beta_1^2
+f_2 \ov 2 \sqrt{g_2}} \right) wz^2,\\
c_2 = {3 \ov 2}(\alpha_1+\beta_1) wz,\quad
c_3 = 2\sqrt{g_2} w,\quad
b=(\beta_1 - \alpha_1) \,.
\eea
To enhance this $\fu(1)$, we must tune $b$
to zero. It is simple to see that for this
tuning, there is a coincident singular fiber
$I_0^* \times I_0^*$
at $w=0$, which is {not allowed}.
Similarly, attempting to un-Higgs the
section $(s_1+t_1)$ gives rise to a
coincident singular fiber
$I_0^* \times I_0^*$ at
$z=0$.
We therefore see that two of the four
$\fu(1)$ components of this theory are
non-enhanceable.

As can be seen from the example we
have studied,
a simple way of discerning whether
a model has a non-enhanceable abelian
component is to attempt to fully enhance the
abelian gauge symmetry and look for
obstructions. This is in fact the essence
of the brief argument presented in
\S\ref{sec:c-example}.
Given a manifold constructed
by performing an MCP blowup on a fiber
product space of
surfaces $A$ and $B$, the abelian gauge
symmetry can be enhanced by decreasing the
Mordell--Weil rank of the $A$ fibration
by making the singularities of $A$ worse.
There are only a limited number of ways
of doing so.
In particular, let us assume that there are $k$
coincident singular fibers
\be
S_{1} \times \Sigma_1, \cdots, S_k \times \Sigma_k \,,
\ee
such that the resolution of each results in at least
one rigid curve, {\it i.e.}, negative intersection curve, in the
base over which there is a non-abelian gauge symmetry.
This is true, for example, when the MCP blowup of the
coincident singular fiber requires blowing up the base.
In this case, the singularities $S_i$ are stuck, and cannot
be moved by tuning complex structure parameters.
Furthermore, they are allowed to only get worse upon
enhancement. If enhancing
a certain abelian component
worsens the singularity $S_i$
to $S_i'$, and if $S_i' \times \Sigma_i$ is a singularity
that does not admit an MCP blowup,
that component can be identified
as a non-enhanceable $\fu(1)$. The ``vacuum''
obtained by enhancing the abelian component
does not have an effective description as a
six-dimensional supergravity theory coupled
to gauge fields, hypermultiplets or SCFTs.

Since there are only 16 $A$-fibrations
in Persson's list with Mordell--Weil rank zero,
it is particularly easy to detect whether there
exists at least one abelian component that
is non-enhanceable. This can be done by
first identifying which of the 16 fibrations
the current $A$-fibration can
enhance into, and see if the coincident
singular fibers of the enhanced model
admit an MCP blowup.
For example, for the generic elliptic fibration
over \baselink{8}, or model \eq{eq:A8},
there are two coincident singular fibers
given by the product $IV \times I_0^*$.
Of the 16 fibrations of Mordell--Weil rank
zero, the only two $A$-fibrations the
model \eq{eq:A8}
can possibly enhance into are the ones
with either the singularities $IV^*$
and $IV$ or two
$I_0^*$ singularities,
which we have already
observed in the detailed analysis.
In both cases, there
exists a coincident singular fiber that
cannot be resolved without moving an
infinite distance within moduli space.

We thus see that there can
exist a plethora of models with
non-enhanceable components of the
abelian gauge algebra. For example, let
us consider the class of
Calabi--Yau varieties $X$
obtained through an MCP blowup of
a fiber product space
where the $A$-fibration has the singular fiber list:
\be
A ~:~ \big[ I_1^* ,\, IV ,\, I_1 \big] \,.
\ee
We also assume that the $B$ fibration
has a $IV$ fiber coincident
with the $I_1^*$ fiber of $A$, and either an
$I_n^*$ or a $IV$ fiber coincident with the
$IV$ fiber of $A$. Thus there are two coincident
singular fibers:
\bea
(I_1^* \times IV,~ IV \times IV )
\quad \text{or} \quad
(I_1^* \times IV,~ IV \times I_n^* ) \,.
\eea
We assume that there are not any additional
coincident singular fibers. There are 8 entries in
Persson's list that can be used as the base $B$
that satisfies this criterion.

$X$ has MW rank 1,
whose corresponding section comes from
pulling back the unique free generator of the
MW group of the $A$ fibration. The corresponding
$\fu(1)$ gauge symmetry is always non-enhanceable.
The way to see this is to first acknowledge that the
resolution $X \rightarrow \widetilde{X}$ requires
the blowup of the base for the two coincident
singular fibers, thus making the $I_1^*$ and $IV$
singularities of the $A$-fibration stuck.
Now of the 16 fibrations of the rational surface that has
MW rank zero, the only one that the 
given $A$-fibration can enhance into has the singular
fiber list
\be
\big[ IV^* ,\, IV \big] \,.
\ee
Thus the $I_1^* \times IV$ singularity of $X$
becomes a $IV^* \times IV$ singularity, which does
not yield a resolution with a sensible six-dimensional
effective description.

We note that while model \eq{eq:A8} is a non-Higgsable model,
the previous example is not. Let us end the section with
another non-Higgsable model with a non-enhanceable
$\mathfrak{u}(1)$. Let us consider when the $A$ fibration
has the singular fibers
\be
\big[I_0^*,\, IV,\, I_1,\, I_1 \big]
\ee
as well as for the $B$ fibration. Let the coincident singular fibers
be given by
\be
(I_0^* \times IV , \,
IV \times I_0^*) \,. 
\ee
The Mordell--Weil rank of the resolution of this fiber product
space is given by $2$. Resolution of the coincident singular
fibers requires
blowing up the base, and thus the singularities $I_0^*$
and $IV$ are stuck and can only get worse.
The only Mordell--Weil rank-zero
fibrations that the $A$ fibration can enhance to have the singular
fiber list
\be
\big[I_0^*,\, I_0^*\big] \quad \text{or} \quad
\big[IV^*,\, IV \big] \,.
\ee
Thus there is either a $I_0^* \times I_0^*$ fiber or
$IV^* \times IV$ fiber, either of which lies
at infinite distance from the interior of moduli
space. Thus this model must have a non-enhanceable
$\mathfrak{u}(1)$.

\section{$A$-$B$ duality}
\label{sec:AB}

In this section, we examine the duality of
manifolds under the exchange of $A$
and $B$ fibers of the Calabi--Yau threefolds
obtained by the $A$-$B$ construction
studied so far.
Given  a manifold that admits
multiple elliptic fibrations, the physics of an
F-theory model associated to this manifold depends on the choice
of the fiber. For a fiber product space
\begin{equation}
 \widetilde{X} = A \times_{\P^1} B \,,
\end{equation}
we choose the elliptic fiber of the $A$ direction
to be the F-theory fiber.
We could have performed an MCP blowup
of the coincident
singular fibers of $\widetilde{X}$ in multiple
ways. We have, however, chosen
to perform the blowup these
singularities in a way consistent
with the elliptic fibration structure,
so that the MCP blowup is still a flat
elliptic fibration with the fiber aligned
along the $A$ direction.
Let us denote the manifold obtained
by such a resolution by $X$.
The base $\widehat{B}$ of $X$ is given by
blowing up the base $B$ as explained in
detail in \S\ref{sec:products}.
Meanwhile, we can consider a dual
variety $\widetilde{X}^D$ such that its elliptic fibration
structure is given by
\begin{equation}
 \widetilde{X}^D = A^D \times_{\P^1} B^D \equiv B \times_{\P^1} A \,,
\end{equation}
whose MCP blowup is given by ${X}^D$
such that the fiber is aligned in the $A^D=B$
direction.
The base $\widehat{B^D}$ of this manifold is given by
a blow up of $B^D = A$.

While $\widetilde{X}$ and $\widetilde{X}^D$
are the same singular variety, their MCP blowups
$X$ and $X^D$ are in general
different since each one is compatible with a different elliptic fibration.
Upon compactifying F-theory on $X\to \widehat{B}$
and $X^D\to \widehat{B^D}$, we arrive at different
six-dimensional theories. The tensor, gauge
and matter content in general will be different.
Upon compactifying the theory further
on a circle,
however, the five-dimensional effective theories
become dual to each other. This is because the
two distinct varieties come from performing two distinct
MCP blowups on a single singular variety.
Put differently, $X$ and $X^D$ sit at
different regions of the movable cone 
\cite{MR924674,beyond}
({\it i.e.}, the extended K\"ahler moduli space)
 of a single variety. As we will see
shortly, they are in fact related by flops.\footnote{This also follows
from general results in Mori theory \cite{MR924674,[Kol]} .}

\begin{figure}[!t]
\centering\includegraphics[width=15cm]{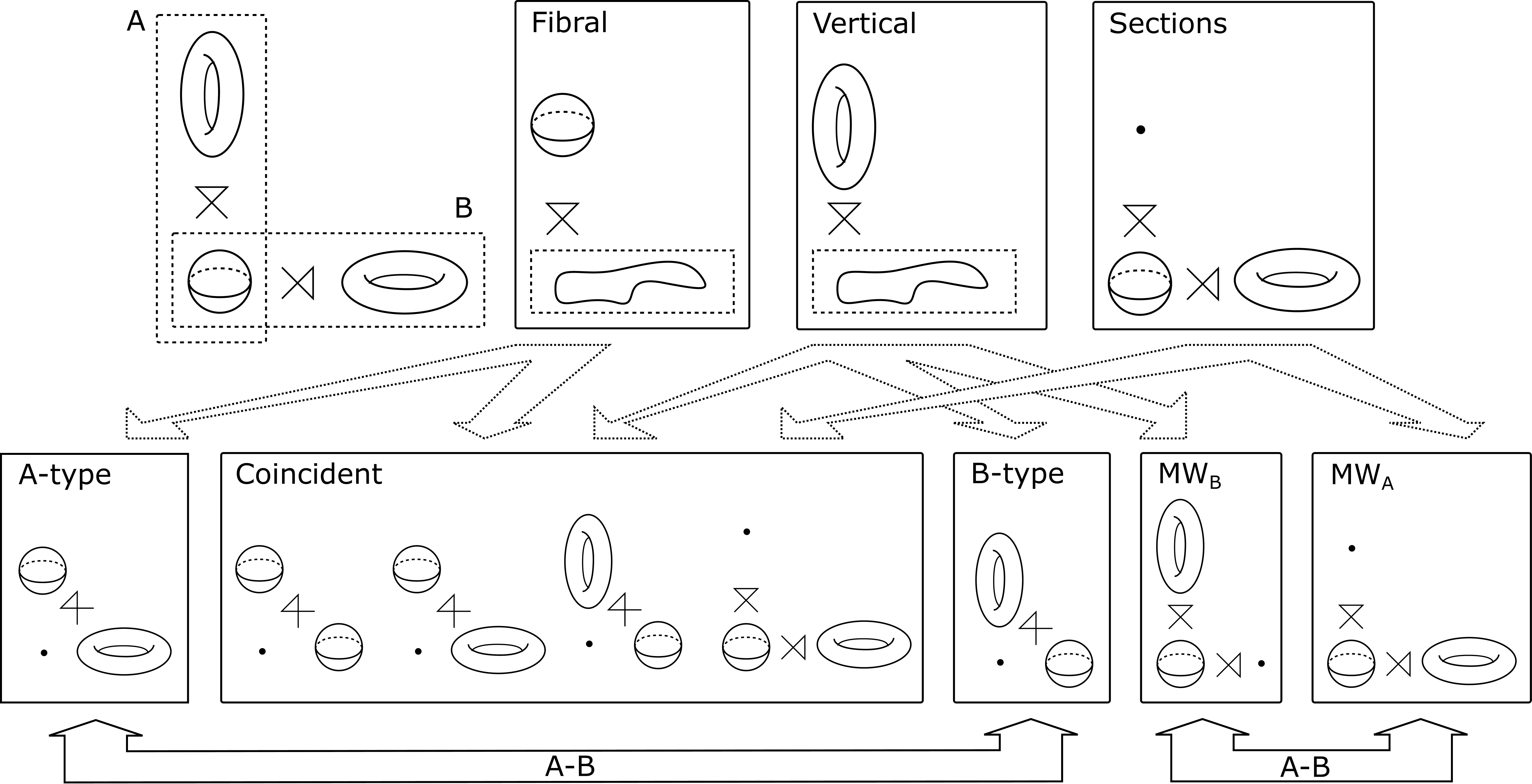}
\caption{\small A classification of divisors
in the resolution of an elliptic fibraiton over
a rational surface, and their exchange
properties under $A$-$B$ duality. The divisors
are fibrations, whose bases and generic fibers
we have depicted. The
fibral divisors are either resolution
divisors of $A$-type or coincident
singular fibers, while the vertical divisors
can be of $MW_B$-type or resolution
divisors of either coincident singular fibers
or B-type singularities. The sections either are
$MW_A$-type divisors, or a global divisor,
the blow up along which resolves multiple
codimension-two singularities at the
coincident loci.
The solid arrows with the
label ``$A$-$B$'' denote the exchange
under $A$-$B$ duality.}
\label{f:divisors}
\end{figure}

Let us understand how various
elements of the second cohomology of
$X$ and $X^D$ map into each
other under these flops, which we
denote ``$A$-$B$ duality.''%
\footnote{For technical reasons, we restrict
our discussion from this point on to $\widetilde{X}$
that have honest resolutions to a smooth CY
manifold. We expect these results to generalize
to MCP blowups.}
A useful way of classifying the second
cohomology, or equivalently, 
the four-cycles of the manifold, is to divide them
into five groups:
resolution divisors of $A$ singularities,
$B$ singularities, and coincident
singular fibers and divisors
of type $MW_A$ and $MW_B$.
The names of the first two types
of divisors are self-explanatory.
The resolution divisors of the coincident
singular fibers can be localized at the
locus of the singular fiber, or be a global
section that (partially) resolves multiple
codimension-two singularities.
The $MW_{A}$($MW_B$)-type divisors
are four-cycles that come from
pulling back a section (including the
zero section) of the $A$($B$)-fibration
with respect to the projections to each
rational surface.
It is straightforward to understand how the
different classes of divisors are mapped
into each other under $A$-$B$ duality.
The resolution divisors of $A$-singularities
are exchanged with the resolution
divisors of $B$-singularities, while
the resolution divisors of the coincident
singular fibers are exchanged among 
themselves. Meanwhile, the divisors
of types $MW_A$ and $MW_B$ are
also exchanged.

Another useful way of classifying the
divisors is to divide them into three groups,
according to the Shioda-Tate-Wazir
theorem \cite{MR0429918,stw}:
fibral divisors, vertical divisors, and
sections. The fibral divisors are
divisors that have the structure of a
rational curve fibered along a
divisor in the base. Vertical divisors
are those obtained by fibering the
elliptic fiber over divisors in the base.
The sections are four-cycles that
are parallel to the base manifold,
and consist of the zero section and
the generators of the Mordell--Weil
group. A schematic depiction of each
type of divisor is given in Figure
\ref{f:divisors}.
Then, from the point of view of $X$
we find the following:
\ben
\item
The fibral divisors consist of
resolution divisors of coincident
singular fibers and $A$-singularities.
\item
The vertical divisors consist of
resolution divisors of coincident
singular fibers and $B$-singularities,
and divisors of type $MW_B$.
\item
The sections of the either divisors of 
type $MW_A$, or come from global
resolution divisors that resolve multiple
codimension-two singularities localized
at coincident singular fibers.
\een
The exchange of the various divisors
under $A$-$B$ duality is depicted in
Figure \ref{f:divisors}.

As mentioned earlier, we can engineer
little string theories \cite{LST} by taking
a certain scaling limit of the F-theory
compactification on these manifolds.
The $A$-$B$ duality
then becomes a $T$-duality of the
little string theory.

In what follows, we examine
some examples to see how $A$-$B$
duality can be understood in terms of
flops. The first example is the case when
$A$ and $B$ have a single coincident
singular fiber given by the product
$I_0^* \times I_n$, that is,
we assume all other singularities
are either of type $I_1 \times \cdot$ or
$\cdot \times I_1$.  Next,
we examine the case that $A$ and $B$
have a single coincident singular fiber
$I_0^* \times IV$. We end the section by
relating the $A$-$B$ dualities with
$T$-dualities of little string theories.

\subsection{Example: Model with coincident singular fiber $I_0^* \times I_n$}

Let us consider a Calabi--Yau
manifold $\widetilde{X}$
with one coincident singular fiber
$I_0^* \times I_n$
over the point $p$ of the base
$\mathbb{P}^1$. Let us denote the
elliptically fibered
manifold obtained by resolving
$\widetilde{X}$ so that its fiber
is aligned in the $A$-direction as $X$.
The $B$-fibration, which is to be considered
as the base of $X$, has a reducible fiber
that decomposes into $n$ $(-2)$-curves
at locus $p$. On each of the $n$
components, there lies a $I_0^*$
singularity. This implies that the
intersection of the $(-2)$ curves
must be further resolved. The singularities
can be fully resolved by $n$ blowups
by $(-1)$-curves at each intersection.
The base is then given by $\widehat{B}$, which
is a rational surface
blown up at $n$ points.
There are no singular fibers sitting above
a generic point of these $(-1)$-curves.
On the other hand,
the $n$ now-$(-4)$ curves carry
$\mathfrak{so}(8)$ gauge-symmetries. Upon
resolving each of these $I_0^*$
singularities, we obtain $4n$
fibral divisors. Meanwhile,
there are $(10+n)$ fibral divisors,
$(10-n)$ of which are of type $MW_A$.
$2n$ of the divisors come from resolving
the coincident singular fiber.
Finally, there are $(9-4)=5$ sections
of the theory. Hence the number of
divisors are given by
\bea
\text{Fibral} ~:~ 4n,\quad
\text{Vertical} ~:~10+n,\quad
\text{Sections} ~:~5 \,.
\\
\text{Coincident} ~:~4n+2n=6n,\quad
MW_B~:~10-n, \quad
MW_A~:~5 \,.
\eea
The gauge algebra of the F-theory
compactification on $X$ is given by
\be
\gg_X = \mathfrak{so}(8)^{\oplus n}
\oplus \fu(1)^{\oplus 4} \,.
\ee
There are $T_X = (9+n)$ tensor multiplets
in the theory, as the rational surface $B$
has been blown up $n$ times to reach $\widehat{B}$.
This number is always given by one less than the
vertical divisor count.

\begin{figure}[!t]
\centering\includegraphics[width=15cm]{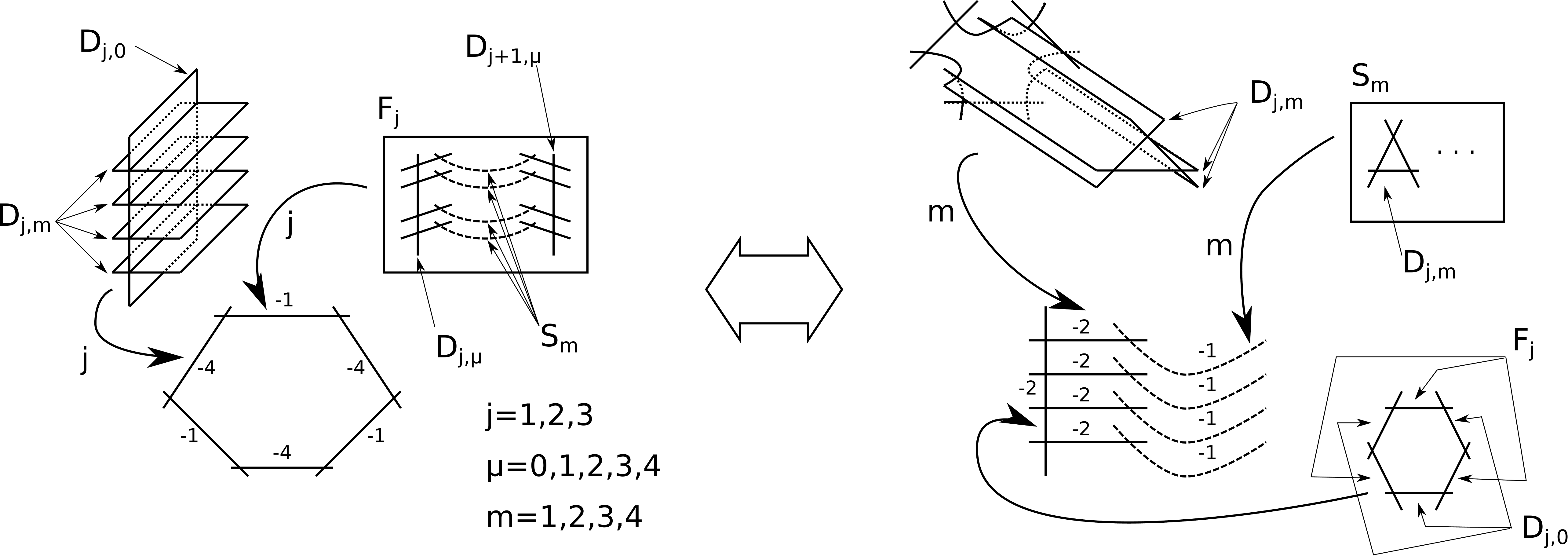}
\caption{\small The behavior of divisors of $X$ (left) and
$X^D$ (right) under flops for distinct resolutions of
a coincident singular fiber given by $I_0^* \times I_3$.}
\label{f:I0star-In-flops-a}
\end{figure}

Now let us consider the dual resolution
$X^D$ of this manifold, where $B^D=A$
is taken to be the base. The $I_0^*$ singularity
of the $A^D=B$-fibration implies that there is a
reducible fiber, which, upon resolution
can be shown to consist of five $(-2)$-curves.
On four of the curves, there
is an $I_n$ singularity, while on one, there is
an $I_{2n}$ singularity. The resolution
of these singularities yields $4(n-1)+(2n-1)$
fibral divisors that come from resolving the
coincident singular fiber.
Meanwhile, there are 10 vertical divisors.
Five of these vertical divisors, which lie above
the aforementioned $(-2)$-curves,
arise in the process of resolving the
coincident singular fiber. The remaining five
are of type $MW_B^D$. Meanwhile, there
are $(10-n)$ $MW_A^D$-divisors.
To summarize, the divisor count is given as the
following:
\bea
\text{Fibral} ~:~ 6n-5,\quad
\text{Vertical} ~:~10,\quad
\text{Sections} ~:~(10-n) \,.
\\
\text{Coincident} ~:~4(n-1)+(2n-1)+5=6n,\quad
MW_B^D~:~5, \quad
MW_A^D~:~(10-n) \,.
\eea
Note that the divisor count is exchanged
as described previously. The gauge algebra
of the $X^D$ compactification is given by
\be
\gg_{X^D} = \mathfrak{su}(2n) \oplus
\mathfrak{su}(n)^{\oplus 4} \oplus
\fu(1)^{\oplus (10-n)} \,,
\ee
while the number of tensor multiplets $T_{X^D}$
is $9$.

Let us now examine how the various divisors
in the two manifolds are related to each other
by flops. We denote the divisors that are
involved in the flops by $D_{j,\mu}$, $F_j$
and $S_m$ where $j$ runs from $1$ to $n$
and $\mu$ runs from $0$ to $4$. We label
the non-zero indices of $\mu$ by $m$.
The geometry of the divisors are depicted
in Figure \ref{f:I0star-In-flops-a} in the case
that $n=3$.

In $X$, the divisors $D_{i,\mu}$ are
obtained by fibering rational curves that
compose the $I_0^*$ fiber,
labeled by $\mu$,
over the $(-4)$ curves,
labeled by $j$, in the base.
The central component of the $I_0^*$ fiber
is given the index $\mu=0$ while the
other rational curves are labeled by $m$.
$F_j$ are rational surfaces obtained by
fibering the elliptic fiber over the $(-1)$
curves in the base. This rational surface
has two $I_0^*$ fibers at the loci where the
$(-4)$ curve meets the $(-1)$ curves.
The components of these $I_0^*$ fibers
within $F_j$ can be understood as the
intersection curves between $F_j$ and the
divisors $D_{j,\mu}$ and $D_{j+1,\mu}$.
$S_m$ are rational sections of $X$
that intersect $F_j$ along its sections,
which are represented as a dotted lines in
Figure \ref{f:I0star-In-flops-a}.
This section intersects the curves
$D_{j,m} \cap F_j$ and $D_{j+1,m} \cap F_j$
within $F_j$.

\begin{figure}[!t]
\centering\includegraphics[width=15cm]{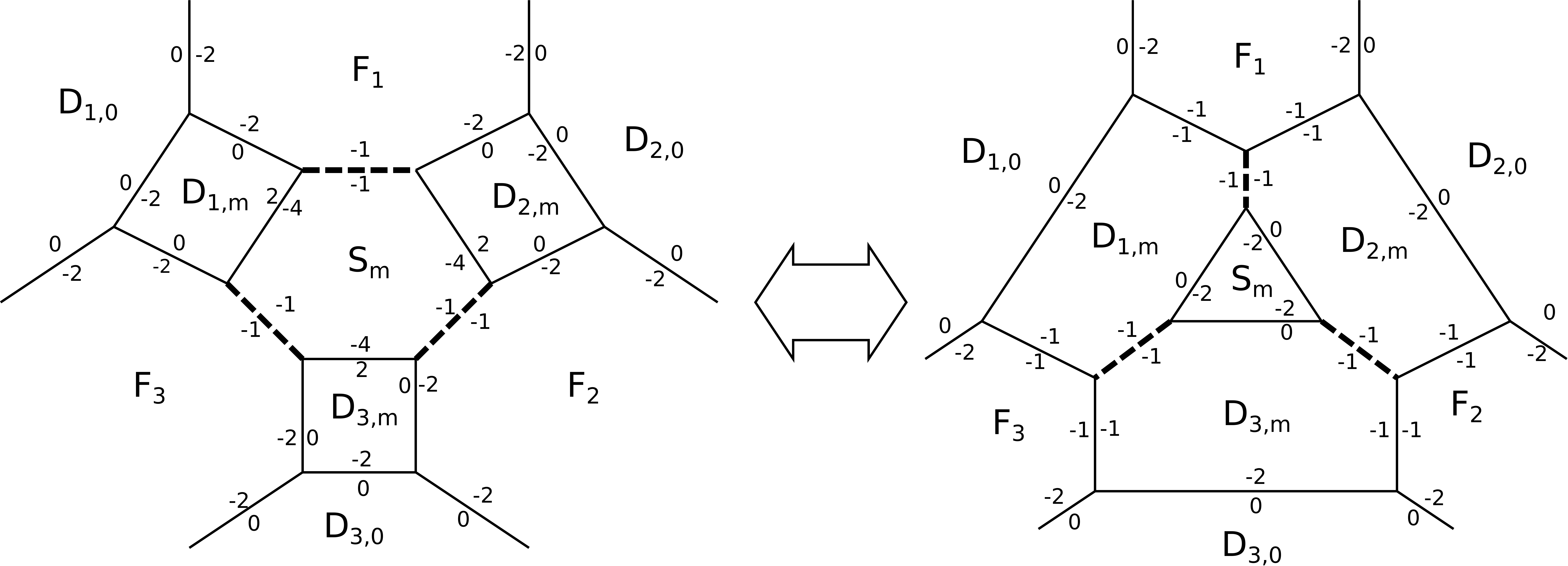}
\caption{\small Flops relating $X$ (left) and
$X^D$ (right).}
\label{f:I0star-In-flops-b}
\end{figure}

The manifold $X^D$ can be obtained
from $X$ by flopping along $4n$ curves.
The flops are illustrated in Figure
\ref{f:I0star-In-flops-b} for $n=3$.
Before describing the flops
in detail, let us explain the information this
Figure contains.
The faces of the diagrams
correspond to the divisors, while the edges
along which two faces meet denote the
curves along which the two corresponding
divisors intersect. The numbers on each side
of an edge denote the self-intersection of
the curve within the corresponding divisor.
For example, the curve at the intersection of
$D_{j,m}$ and $S_m$ in $X$
have self-intersection number $2$ within $D_{j,m}$,
and is a $(-4)$-curve from the point of view of
$S_m$. The two numbers labeling a curve
adds up to $(-2)$ via the adjunction formula,
since the divisors and curves live inside a
Calabi--Yau total space. Two edges meet when
the corresponding curves have intersection
number-one in the divisor both of the curves
lie within. The self-intersection number of
curves change after a flop, as curves within
the divisors are blown up or down during
a flop transition. The intersection numbers
are modified accordingly with the blowups
and blow-downs---if a point on the curve
is being blown up, the self intersection number
decreases by one, while when an adjacent
curve gets blown down, it increases by one.

$X^D$ is obtained from $X$ by flopping
the curves that are represented by dotted
lines in Figure \ref{f:I0star-In-flops-b}.
Such flops are carried out for each
$m = 1,\cdots,4$, resulting in
$4n$ curves being flopped.
This corresponds to blowing down four
curves in each $F_j$, $n$ curves in each
$S_m$ and blowing up two curves in
each of the $D_{j,m}$. $X^D$ is then
an elliptic fibration over $\widehat{B^D}=A$, which
is a rational surface with a resolved $I_0^*$
fiber.

In $X^D$, the divisors $D_{j,m}$ are
obtained by fibering the components of
the $I_n$ fiber over the non-central components
of the $I_0^*$ fiber in the
base $A$. The divisors $D_{j,0}$, together
with $F_j$ consist the $2n$ divisors obtained
by fibering the rational curves of the
$I_{2n}$ fiber over the central component of
the $I_0^*$ fiber in the base.
The $S_m$ now are vertical divisors. They are
rational curves with a resolved $I_n$ fiber.
The the components of the $I_n$ fiber within
$S_m$ lie at the intersection between $S_m$
and $D_{j,m}$. This structure is
summarized in the right-hand side of
Figure \ref{f:I0star-In-flops-a}.

Let us examine the topology change
of the divisors. $S_m$ in $X$ was a copy
of the base $\widehat{B}$ of $X$, which is a rational
surface blown up at $n$ points. During the flop,
$S_m$ undergoes $n$ blow-downs and
becomes a rational surface with an
$I_n$ fiber. Meanwhile, the $D_{j,m}$, which
were minimal ruled surfaces in $X$, are
blown up at two points. These two blowups,
which sits above the intersection locus of
the $m$-th non-central component of the $I_0^*$
fiber with the central component, is responsible
for the bifundamental matter between the
$I_{2n}$ and $I_n$ gauge groups. The
$3n$ curves responsible for generating the
bifundamental matter can be observed at the
boundary of the outer-ring of $2n$ divisors
consisting of $D_{j,0}$, $F_j$ $(j=1,\cdots,n)$
and the inner-ring of $n$ divisors 
$D_{j,m}$ in Figure
\ref{f:I0star-In-flops-b}.

\subsection{Example: Model with coincident singular fiber $I_0^* \times IV$}

\begin{figure}[!t]
\centering\includegraphics[width=15cm]{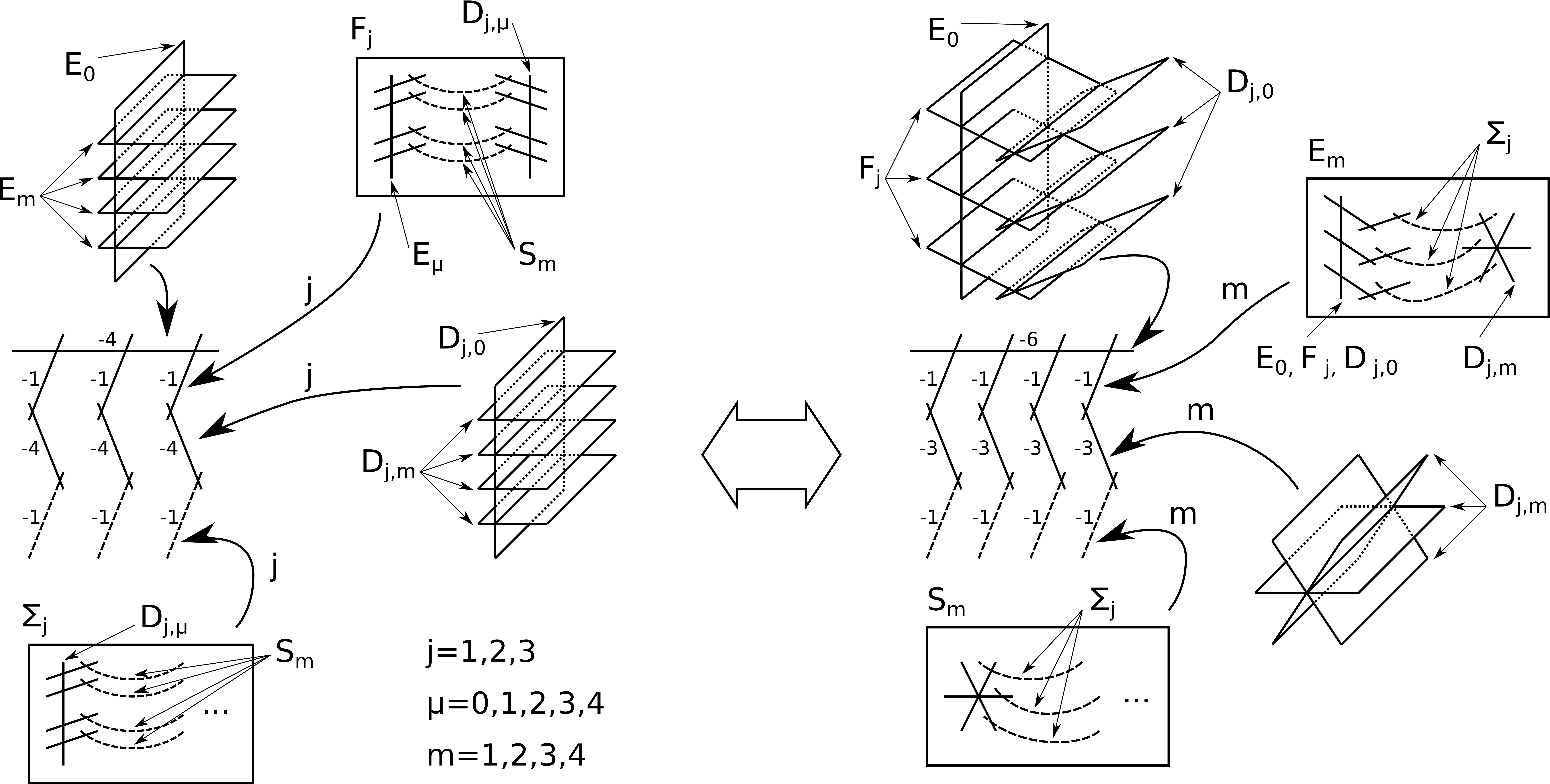}
\caption{\small The behavior of divisors of $X$ (left) and
$X^D$ (right) under flops for distinct resolutions of
a coincident singular fiber $I_0^* \times IV$.}
\label{f:I0star-IV-flops-a}
\end{figure}

Let us now consider a Calabi--Yau
manifold $\widetilde{X}$
with a single coincident singular fiber
$I_0^* \times IV$
over a point $p$ in the base
$\mathbb{P}^1$. As before, we
denote the manifold obtained by 
resolving $\widetilde{X}$ so that
the elliptic fiber is aligned in the
$A$-direction to be $X$.
The base $\widehat{B}$ is obtained by
blowing $B$ up at four points, where
$B$ is a rational surface with a resolved
$IV$ fiber.
This leads to a configuration of
a central $(-4)$ curve and three branches
that consist of a $(-1)$ and $(-4)$ curve,
labeled by $j$.
For each $(-4)$ curve, there is a $I_0^*$
fiber fibered over it.
The configuration of these curves in the
base are depicted on the left-hand side of
Figure \ref{f:I0star-IV-flops-a}.

We denote the divisors involved in the flops
by $E_\mu$, $D_{j,\mu}$, $S_m$ and $\Sigma_j$.
These divisors
are also depicted on the left-hand side of Figure
\ref{f:I0star-IV-flops-a}.
The index $j$ runs from $1$ to $3$,
while $\mu$ runs from $0$ to $4$. As before,
the non-zero indices of $\mu$ are denoted by
$m$. $E_\mu$ are minimal ruled
surfaces obtained by fibering
the components of the $I_0^*$ fibers along
the central $(-4)$ curve in the base,
while $D_{j,\mu}$ come from fibering the
components of the $I_0^*$ fibers along the
other $(-4)$ curves. The surfaces obtained
by fibering the central component of the $I_0^*$
fiber carry the index $\mu=0$.
Meanwhile, $F_j$ are rational surfaces sitting
above the $(-1)$ curves connecting the central
$(-4)$ curves to the $(-4)$ curves in the branch.
They contain two $I_0^*$ fibers whose components
lie at the intersection of $F_j$ and $E_\mu$
or $D_{j,\mu}$. For each $j$, there also exists
a vertical divisor $\Sigma_j$ that lies above a
$(-1)$ divisor in the base that intersects
the $(-4)$ curves at a point. $\Sigma_j$ are
also rational surfaces with a $I_0^*$ fiber
whose components are intersections of
$\Sigma_j$ with $D_{j,\mu}$.
The sections $S_m$ of $X$
intersect $F_j$ and $\Sigma_j$ along its
sections, represented as a dotted lines in
Figure \ref{f:I0star-IV-flops-a}.
This section intersects the curves
$E_{m} \cap F_j$ and $D_{j,m} \cap F_j$
within $F_j$ and
$D_{j,m} \cap \Sigma_j$
within $\Sigma_j$.
The divisor count of this manifold is
given by the following:
\bea
\text{Fibral} ~:~ 16,\quad
\text{Vertical} ~:~14,\quad
\text{Sections} ~:~5 \,.
\\
\text{Coincident} ~:~16+7=23,\quad
MW_B~:~7, \quad
MW_A~:~5 \,.
\eea
The gauge algebra of F-theory compactified
on $X$ is given by
\be
\gg_{X} =
\mathfrak{so}(8)^{\oplus 4} \oplus
\fu(1)^{\oplus 4} \,,
\ee
while $T_X = 13$.

\begin{figure}[!t]
\centering\includegraphics[width=15cm]{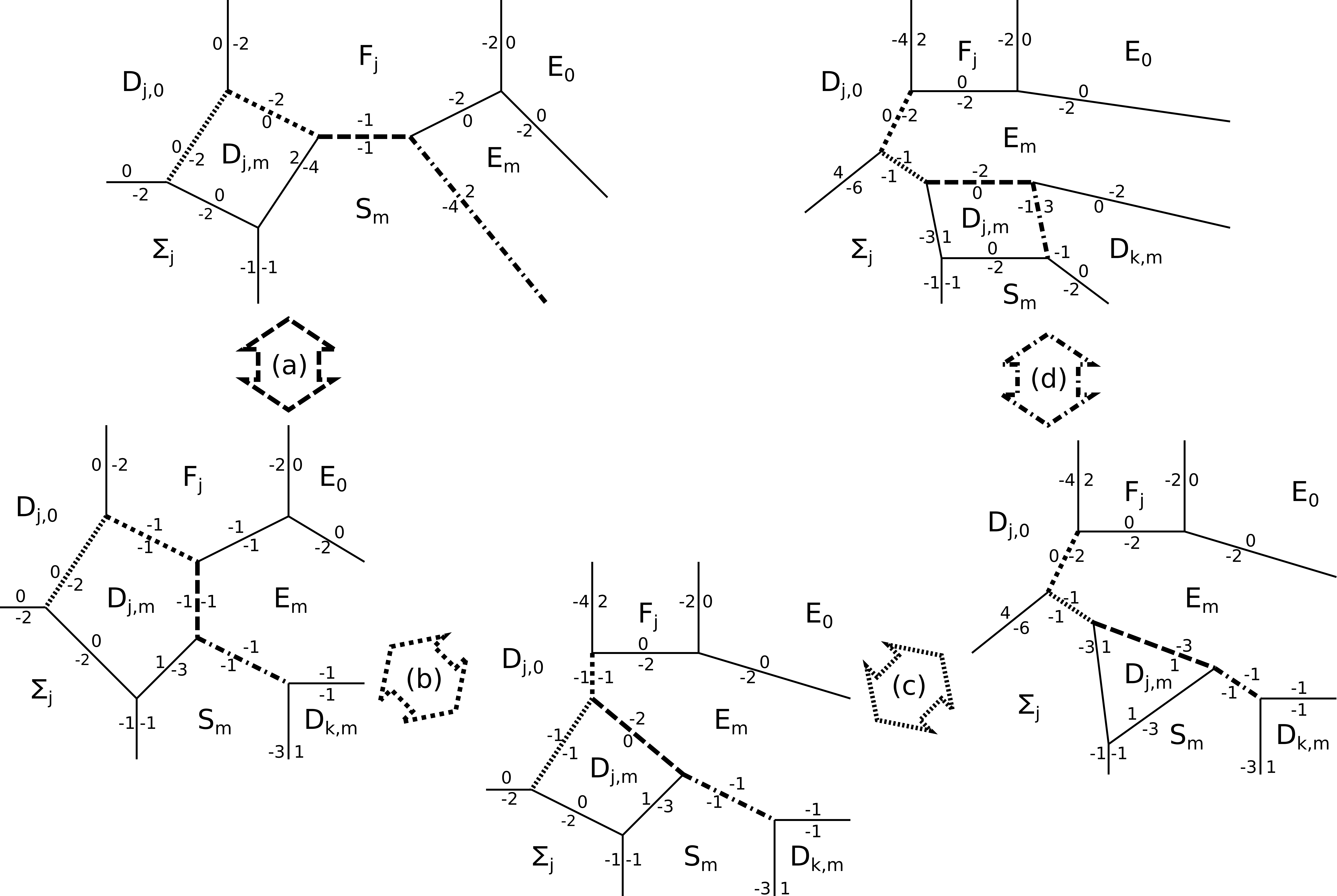}
\caption{\small Flops relating $X$ (upper-left) and
$X^D$ (upper-right). The dash-style of the arrows
indicate which curve is being flopped.}
\label{f:I0star-IV-flops-b}
\end{figure}

In the dual resolution $X^D$ of this manifold, 
the base $B^D=A$ has a single singular $I_0^*$
fiber, which can be resolved into a Kodaira
fiber consisting of five $(-2)$-curves.
There is an $IV^*$ singularity
along the central component of the Kodaira fiber,
while there are $IV$ singularities along the rest
of the four components, labeled by $m$.
The base manifold must
be blown up at the four points the four $(-2)$
components intersect the central component
to resolve the codimension-two singularities.
The configurations of these curves within the
base can be seen on the right-hand side of
Figure \ref{f:I0star-IV-flops-a}.

Let us now describe the 40 flop transitions
that take $X$ to $X^D$. These are depicted
in Figure \ref{f:I0star-IV-flops-b}.
In (a), the 12 curves at the intersection of
$F_j$ and $S_m$ are flopped. These curves
are the dotted curves on $F_j$ on the
left-hand side of Figure \ref{f:I0star-IV-flops-a}.
Note that the intersection numbers of
the curve $E_m \cap S_m$ jumps by three,
as this curve intersects three of the blown
down curves, namely $F_j \cap S_m$
for $j=1,2,3$. Then the 12 curves
at the intersection of $F_j$ and $D_{j,m}$
are flopped in (b). The self-intersection
number of the curve
$F_j \cap D_{j,0}$ jumps by four, as
it intersects four of these curves---%
$F_j \cap D_{j,m}$ for $m=1,\cdots, 4$.
In (c), the 12 curves that lie at the intersection
of $D_{j,0}$ and $D_{j,m}$ are flopped.
The self-intersection number of
$\Sigma_j \cap D_{j,0}$ now jump by
four, as it intersects four of the curves
being flopped.
Finally, in (d), the 4 curves at the intersection
of $S_m$ and $E_m$ are flopped.
An interesting point to see is that
by blowing down $S_m \cap E_m$
the three divisors $D_{j,m}$ for given
$m$ meet at a point, forming a singular
type-$IV$ Kodaira fiber, as depicted in
the right-hand side of Figure \ref{f:I0star-IV-flops-a}.

After the flops, $F_j$, $E_0$ and $D_{j,0}$
become minimal ruled surfaces that constitute
a type-$IV^*$ Kodaira fiber fibered over the $(-6)$
curve in the base manifold of $X^D$.
The minimal curves
$D_{1,m}$, $D_{2,m}$ and $D_{3,m}$
are obtained by fibering the components
of the type-$IV$ Kodaira fibers over
the four $(-4)$ curves labeled by $m$.
The sections $S_m$ of $X$, each of
which undergoes three blow-downs during
the flops, become vertical
divisors of $X^D$, which are rational surfaces.
These surfaces have a single type-$IV$ fiber
whose components lie at the intersection
of $S_m$ and $D_{j,m}$.
$\Sigma_j$, which were vertical divisors of
$X$, now become sections of $X^D$,
which are given by copies of the base
$\widehat{B^D}$, which is a rational surface
blown up at four points.
Note that each $\Sigma_j$ is blown
up four times during the flop transitions.
$E_m$, which were rational surfaces in
$X$, are blown up eight times and become
rational surfaces in $X^D$. These rational
surfaces are elliptic fibrations
over the $(-1)$ curves in the base that connect
the central $(-6)$ curve to the $(-3)$ curves.
Each $E_m$ has two singular fibers,
each of type $IV^*$
and $IV$. The components of
the $IV^*$ fiber are given by the intersection
curves of $E_m$ with $E_0$, $E_j$ and $D_{j,0}$,
while the components of the $IV$ fiber can
be identified with the curves at the intersection
loci of $E_m$ with the divisors $D_{j,m}$.
The number of each type of divisor
is given by the following:
\bea
\text{Fibral} ~:~ 14,\quad
\text{Vertical} ~:~14,\quad
\text{Sections} ~:~7 \,.
\\
\text{Coincident} ~:~14+9=23,\quad
MW_B^D~:~5, \quad
MW_A^D~:~7 \,.
\eea
The divisor count is exchanged
with respect to that of $X$ as
expected. Compactifying F-theory on
$X^D$, we get a theory with $T_{X^D} = 13$
and gauge algebra
\be
\gg_{X^D} = \mathfrak{e}_6 \oplus
\mathfrak{su}(3)^{\oplus 4} \oplus
\fu(1)^{\oplus 6} \,.
\ee

\subsection{Little string theories}

We review from \cite{LST} the construction of little string theories from
F-theory.  The starting point is an F-theory model with base $\widehat{B}$ 
that has a map $\varphi:\widehat{B}\to\mathbb{D}$
to a small disk.
The metric on $\widehat{B}$ is scaled so that the induced area 
of $\mathbb{D}$ goes to infinity, while the area of a fiber 
$\varphi^{-1}(t)$ remains finite.  That finite area is proportional
to the scale of the little string theory.  D3-branes wrapping
the components of the central fiber $\varphi^{-1}(0)$ provide
the tensionless strings in the theory.  Little string theories
are expected to have T-duality upon compactification on
a circle \cite{Seiberg:1997zk,Intriligator:1997dh,Intriligator:1999cn}.

The general fiber of $\varphi$ can either have genus zero or genus one.
The key construction produces F-theory models over such a base
$\widehat{B}$ with a map $\varphi$ for which the general fiber has genus one.  
If the fibration $\varphi$ has a section\footnote{Note that
there are some cases, explored in \cite{LST}, in which $\varphi$ has
a multi-section rather than a section.} (as is the case in this
paper), then the 
dualizing sheaf $\mathcal{O}_{\widehat{B}}(K_{\widehat{B}})$ of
$\widehat{B}$ is the pullback of a line bundle on the disk $\mathbb{D}$.
This implies that the Weierstrass coefficients $f_{\widehat{B}}$ and 
$g_{\widehat{B}}$ are determined
by corresponding Weierstrass coefficients $f_{\mathbb{D}}$ and
$g_{\mathbb{D}}$ on the disk.  Those Weierstrass coefficients
determine an elliptic fibration $A\to\mathbb{D}$, and the F-theory model
for the little string theory is a fiber product.

Our $A$-$B$ duality now immediately yields the expected T-duality of these
models, which applies to the circle-compactified theories.  The T-dual
of the little string theory built as above, is the little string theory
constructed from the elliptic fibration over $A$ (possibly after blowing
up $A$, as we have discussed).

\section{Conclusions}
\label{sec:conclusions}

In this paper we have constructed a general class of elliptic
Calabi--Yau threefolds using fiber products of rational elliptic
surfaces.  These threefolds have the feature that the elliptic
fibration has multiple independent sections at all points in the
complex structure moduli space, corresponding in F-theory to
non-Higgsable abelian gauge components in the associated low-energy 6D
supergravity theory.  From the mathematical point of view these
manifolds are interesting as a generalization of the constructions of
Schoen \cite{Schoen} and Kapustka and Kapustka \cite{Kapustka}.  From
the physics point of view, these are interesting as they give
low-energy theories with non-Higgsable abelian components.

One significant feature of the models we have studied here is that
they contain a range of singularity types, some of which include
terminal singularities with no crepant projective resolution.  These
singularities, which occur at codimension two in the base, are not known
to cause problems with the associated physics theory, so that elliptic
Calabi--Yau threefolds with these types of singularities 
fit
naturally in the class of physically sensible geometries that can be
used in F-theory.  Other codimension two singularities without a
Calabi--Yau resolution have been encountered in 6D F-theory models,
such as in the context of discrete
gauge symmetries \cite{Braun-Morrison, mt-sections, Anderson-ggk,
  Mayrhofer-ptw, Klevers-discrete, Klevers-discrete-2, fglmm}.  A related
observation has been made for 4D F-theory models, which is that many
elliptic Calabi--Yau fourfolds have singularities that
either can be tuned or are present everywhere in complex structure
moduli space, which do not admit crepant projective resolutions, but which seem
innocuous from a physics point of view; this can happen either
in codimension two on the base \cite{Klemm-lry,Taylor-Wang-mc} or in codimension
three  \cite{MR0292830, MR1016414}.
A further understanding of how 
singularities without a Calabi--Yau resolution fit into the landscape
of F-theory compactifications and associated geometry should shed
light on both the physics and mathematics of these
configurations.\footnote{As this work was being completed we learned
  of recent work making progress in this direction
\cite{Grassi-Weigand}.}

In terms of physics applications, one potentially interesting direction in
which to extend this work is to 4D theories, where combining a
non-Higgsable $\mathfrak{u}(1)$ factor with a non-Higgsable 
$\mathfrak{su}(3) \oplus \mathfrak{su}(2)$
\cite{Grassi:2014zxa, Taylor-Wang-mc} would give a family of vacua
where the standard model gauge group would arise at a generic point in
the moduli space, without requiring tuning.

The constructions developed here provide a large class of examples of
theories in which there are one or more $\fu(1)$ components that exist
generically in the moduli space over a given base, and where tuning
moduli to enhance to a nonabelian gauge symmetry following
\cite{Morrison-Park, mt-sections} gives rise to additional
singularities without a Calabi--Yau resolution.  In some cases these
singularities can be removed by blowing up points on the base but in
other cases they cannot, and lie at infinite distance in moduli
space. One way in which it has been suggested such an
unHiggsing structure may be relevant is in classifying models with
abelian components by first classifying nonabelian models and then
Higgsing  ({\it e.g.} \cite{Cvetic:2015uq}), although it has also been
suggested that there can exist certain models with abelian gauge
symmetry that are unattainable this way \cite{Morrison-Park-2}.
The examples given here
show explicitly that for the approach of obtaining abelian models by Higgsing
to be complete, it would be necessary to
include in the classification of nonabelian models certain kinds of
singularities without Calabi--Yau resolutions, corresponding to
unphysical 6D models at infinite distance from the interior of 
moduli space. While in principle this may be possible it
would significantly complicate this approach to the classification of
F-theory models with abelian symmetries.

Another particularly interesting feature of these constructions is the
presence of two distinct elliptic fibrations for each of the resulting
Calabi--Yau threefolds.  This leads to an interesting T-duality type
structure for the associated little string theories, which would be
interesting to explore further.
\vspace*{0.2in}

{\bf Acknowledgements}: We would like to thank
Lara Anderson,
Antonella Grassi,
Gabriella Martini, Tony Pantev and Yinan Wang for helpful
discussions.  This research was supported by the DOE (USA) under contracts
\#DE- SC00012567,
\#DOE-SC0010008 and
\#DE-FG02-92ER-40697,
 by the National
Science Foundation (USA) under Grant Nos. PHY-1066293 and PHY-1307513,
and by the Centre National de la Recherche
Scientifique (France).
All of us would like to thank the Aspen Center for Physics for
hospitality during part of this work, and DRM would like to thank
the Simons Center for Geometry and Physics and the Institut Henri
Poincar\'e for hospitality during
other parts of the work.

\appendix

\section{Terminal singularities and small resolutions}
\label{app:terminal}

As we saw in \S\ref{sec:conifold}, the global questions about the
existence of crepant projective blowups can either be phrased in
terms of algebraic geometry (does a certain divisor exist?) or 
in terms of topology (is there a homology relation among vanishing
cycles?).  One advantage of the algebraic geometry approach is
that for certain terminal singularities the question can be resolved
locally without needing global data.

In fact, some
terminal singularities  are known to have no crepant 
blowup (projective or not) 
 based solely
on their local equation.  A class of these was analyzed by Reid \cite{pagoda}:
singularities of the form
\begin{equation}
x^2+y^2+u^2+v^k=0
\end{equation}
admit no crepant blowup
when $k$ is odd, and may or may not admit a 
crepant projective
blowup (depending on global conditions about divisors) when $k$ is
even.
Other terminal singularities that admit no projective crepant blowup include
\begin{equation}
x^2+y^2+u^3+v^4=0,\ 
x^2+y^3+u^3+v^3=0,\ \text{ and }
x^2+y^3+u^3+v^4=0,\
\end{equation}
while some that may or may not admit a crepant projective blowup depending on
global conditions are
\begin{equation}
x^2+y^2+u^3+v^3=0,\ \text{ and }
x^2+y^2+u^4+v^4=0\,.
\end{equation}

The global conditions enter as follows:  higher order terms in the local
equations can obstruct a factorization which is visible at lower order.
For example, a node on a plane curve
whose leading order terms are $x^2+y^2$ may have a
complete equation of the form $x^2+y^2+x^3+y^3$, which is irreducible.  
The latter describes
a node whose two local branches are connected globally, as one would
see in a Kodaira fiber of type $I_1$.  In fact, this phenomenon is the
cause of the appearance of global conditions in the $I_1$ case but not
in the $I_n$ cases for $n>1$:  in the latter cases, the nodes appearing
in the singular fiber each represent the intersection of two distinct
curves (so factorization is possible) while in the former case, the
node is part of an irreducible curve and factorization is impossible.

The reason for an unknown ``global condition'' even in these cases, is
that there may be some other divisor passing through the singular point,
the blowup of which would resolve the singularity.  All we can say for
certain is that the obvious divisors related to factorization of the
given local equation do not exist globally.

The singularity 
$x^3 + y^3 + u^3 + v^3$
is somewhat different: it is associated to 
a blowup of the base because there is a crepant projective blowup 
of the singularity that introduces a divisor.

We now review the singularities occuring in the fiber product contructions
described in the upper left box of Table~\ref{t:types}; these
singularities were originally described in \cite{Schoen} and \cite{Kapustka}
(using slightly different notation).  Each of the
two elliptic surfaces is nonsingular, and we must consider the possible
singular points on each fiber in question.  The surface in question
will have a description of the form $\varphi(x,y)=t$, and the second
surface will have a description of the form $\psi(u,v)=t$, with the
two parameters $t$ being identified.  The singularity on the fiber product
then takes the form 
\begin{equation}
\varphi(x,y)-\psi(u,v)=0.
\end{equation}
The corresponding local equation $\varphi(x,y)$ takes the form $x^2+y^2$
for a node (normal crossing points in $I_m$); it takes the form $x^2+y^3$ for
a cusp (as in type $II$); it takes for form $x^2+y^4$ for a tacnode
(as in type $III$); and it takes the form $x^3+y^3$ for an ordinary
 triple point (as in type $IV$).

\begin{table}
\begin{center}
\begin{tabular}{| r | c c c c c |} 
\hline
B \!$\backslash$\! A \!\!&$I_1$ &$I_{n>1}$ & $II$ & $III$ & $IV$ \\ 
\hline
$I_1$ & $x^2{+}y^2{-}u^2{-}v^2$ & $x^2{+}y^2{-}u^2{-}v^{2}$ 
& $x^2{+}y^2{-}u^2{-}v^3$ & $x^2{+}y^2{-}u^2{-}v^4$ & $x^2{+}y^2{-}u^3{-}v^3$ \\
\!\!$I_{m>1}$\!\! & \!$x^2{+}y^{2}{-}u^2{-}v^2$\!
& \!\!$x^2{+}y^{2}{-}u^2{-}v^{2}$\!\!
& \!\!$x^2{+}y^{2}{-}u^2{-}v^3$\!\!
& \!\!$x^2{+}y^{2}{-}u^2{-}v^4$\!\!
& \!\!$x^2{+}y^{2}{-}u^3{-}v^3$\!\! \\
$II$ & $x^2{+}y^3{-}u^2{-}v^2$ & $x^2{+}y^3{-}u^2{-}v^{2}$ 
& $x^2{+}y^3{-}u^2{-}v^3$ & $x^2{+}y^3{-}u^2{-}v^4$ & $x^2{+}y^3{-}u^3{-}v^3$ \\
$III$ & $x^2{+}y^4{-}u^2{-}v^2$ & $x^2{+}y^4{-}u^2{-}v^{2}$ 
& $x^2{+}y^4{-}u^2{-}v^3$ & $x^2{+}y^4{-}u^2{-}v^4$ & $x^2{+}y^4{-}u^3{-}v^3$ \\
$IV$ & $x^3{+}y^3{-}u^2{-}v^2$ & $x^3{+}y^3{-}u^2{-}v^{2}$ 
& $x^3{+}y^3{-}u^2{-}v^3$ & $x^3{+}y^3{-}u^2{-}v^4$ & $x^3{+}y^3{-}u^3{-}v^3$ \\
\hline
\end{tabular}
\end{center}
\caption{Local equations of singularities in fiber products of 
nonsingular surfaces.}
\label{t:new}
\end{table}

We reproduce the upper left of Table~\ref{t:types} in
Table~\ref{t:new}, 
this time displaying the local equation.
The singularities that occur are all of the types analyzed in this
appendix, yielding most of the results displayed in
Table~\ref{t:types}.
(Note that since all variables are complex, the signs are not
relevant.)
There are some cases in Table~\ref{t:types} where an appropriate global 
divisor is
known to exist, showing that a crepant projective
resolution exists.  The divisors
in question are provided by divisors that are fiber components for
one of the elliptic fibrations.  Such divisors exist globally, and can be used
to perform the crepant resolutions (unless the fiber
is irreducible {\it i.e.}, cases $I_1$ and $II$).

\section{Two explicit MCP blowups}
\label{app:resolution}

A fiber product in which one fiber has type II and the other fiber has type
$I_n^*$ presents computational challenges for finding an MCP blowup
no matter which of the two 
elliptic fibrations is being studied.
In this appendix, we give some of the details in each of these cases.

We begin with a base containing a type $I_m^*$ fiber ({\it i.e.}, a configuration
of rational curves whose dual graph is an affine $\widehat{D}_{m+4}$
diagram) whose elliptic fibration has a fiber of type $II$ along the
multiplicity one components of the $\widehat{D}_{m+4}$ graph.  Let
$\psi$ be a function on the base which vanishes along the
$\widehat{D}_{m+4}$ graph with the appropriate multiplicities:
\begin{equation}
\psi = \prod \psi_j^{m_j}
\end{equation}
where $\{\psi_j=0\}$ is the curve corresponding to  
the $j^{\text{th}}$ node in the graph and
$m_j$ is its Dynkin index.  In the Weierstrass equation, $\psi$ divides
$f$ and $\psi$ divides $g$.  Thus, when $m_j=1$ we have a Kodaira
fiber of type $II$, but when $m_j=2$ we have a Kodaira fiber of type $IV$.
Our challenge is to find a maximal crepant projective
blowup of this Weierstrass model.

As pointed out in \cite{alg-geom/9305003}, the ``collision'' of two fibers
of type II, or of two fibers of type IV, has a leftover terminal singularity
after the codimension two (in the total space) singularities have been
resolved.  We will carry this out explicitly in our example.

The Weierstrass equation takes the form
\begin{equation}
y^2=x^3+c_1\psi x + c_2 \psi
\end{equation}
for some constants $c_1$, $c_2$ with $c_2\ne0$.  Near a point of intersection
of two of the curves over which the fibration has type $IV$, we may write
$\psi = s^2t^2\widehat{\psi}(s,t)$, where $\{s=0\}$ and $\{t=0\}$ are
 the two curves.  (That is, $s=\psi_j$, $t=\psi_{j+1}$ for some $j$.)
Note that the function $\widehat{\psi}(s,t)$
cannot be a square, since $\psi$ is not a square.

We blowup the locus $x=y=s=0$ and consider the chart with
coordinates $\tilde x = x/s$, $\tilde y = y/s$, $s$, and $t$.  The
proper transform of the equation is 
\begin{equation}
\tilde y^2 = s \tilde x^3  + c_1 st^2 \widehat{\psi} \tilde x + c_2 t^2 \widehat{\psi}.
\end{equation}
We then blowup the locus $\tilde x = \tilde y = t = 0$ and consider
the chart with cooordinates
$\bar y = \tilde y/\tilde x$, $\bar t = t/\tilde x$, $\tilde x$, and $s$.
This time, the proper transform of the equation is
\begin{equation}
\bar y^2 = s \tilde x (1  + c_1 t^2 \widehat{\psi}) + c_2 t^2 \widehat{\psi},
\end{equation}
which has a conifold singularity at the origin.  Notice that the exceptional
divisor for each of the blowups is irreducible since 
$\widehat{\psi}$ is not a square; this implies that the corresponding
gauge algebra on each type $IV$ locus is $\mathfrak{su}(2)$.
Also, since $\widehat{\psi}$ is not a square, the terms
\begin{equation}
\bar y^2 - c_2 t^2 \widehat{\psi}
\end{equation}
cannot be factored.  Indeed, without an additional (global) divisor which
is not visible in the construction, this conifold point cannot be
resolved.  Thus, we have produced a crepant projective blowup which
is not smooth, and which may or may not be maximal depending on the
existence of a global divisor.

\begin{figure}
\begin{center}
\includegraphics[width=6cm]{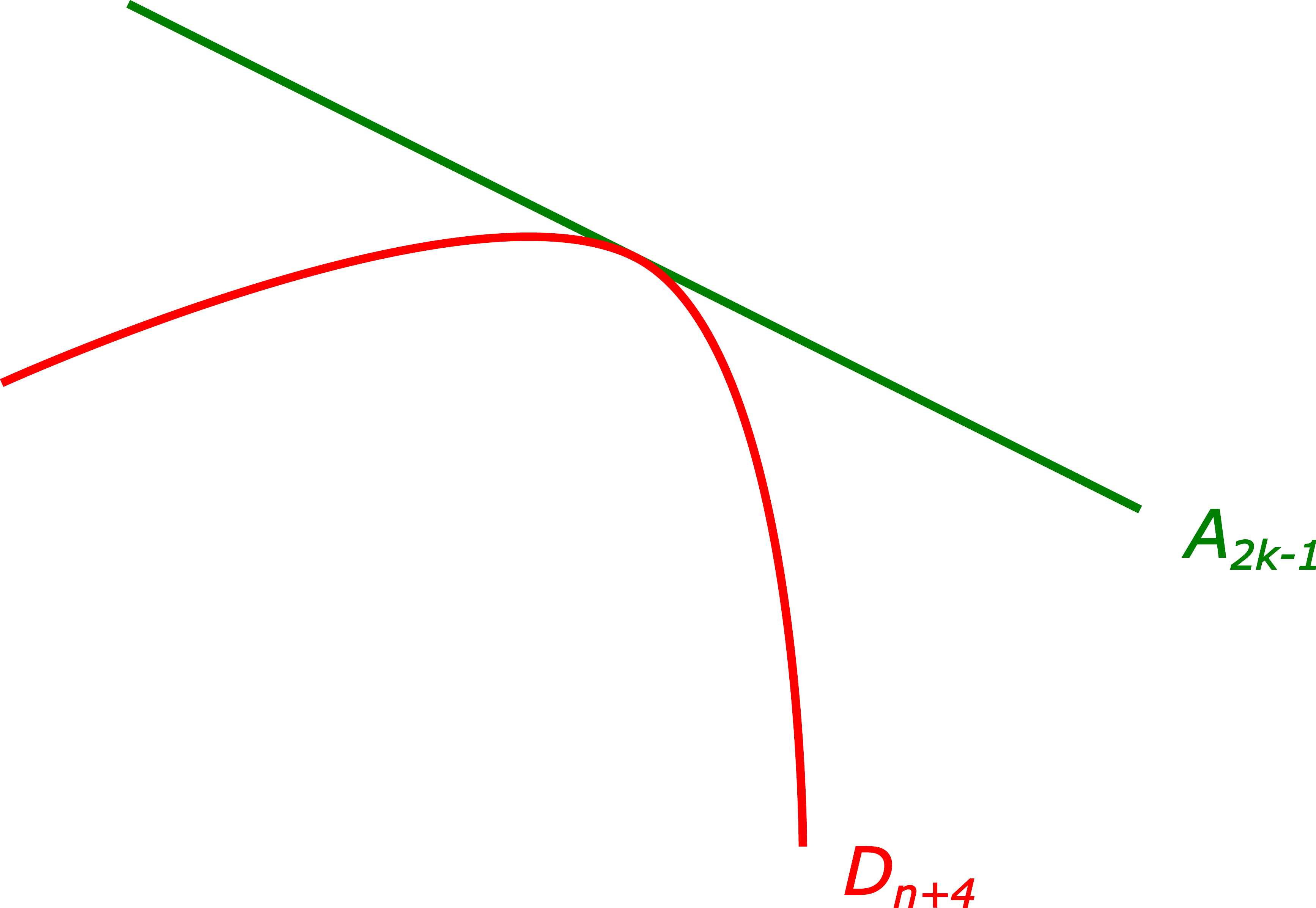}
\end{center}
\caption[x]{\footnotesize The original singular locus.}
\label{f:singular-locus}
\end{figure}

We now turn to a base containing a type II fiber ({\it i.e.}, a cuspidal
curve) whose elliptic fibration
has a fiber of type $I_n^*$ along the cuspidal curve.  At generic points
of the curve, $f$ and $g$ have multiplicities $2$ and $3$ but at the
singular point of the cuspidal curve, the multiplicities are $4$
and $6$.  Thus, the singular point must be blown up.

The blown up base will have two rational curves which are tangent to
each other.  Over the proper transform of the original cuspidal curve
the elliptic fibration has a fiber of type $I_n^*$; but over the other
curve the fiber has type $I_{2n}$.  Since the curves are tangent, we
cannot simply quote known results to understand the crepant 
(projective) blowups.

The singularity that appears at the intersection point
is among those considered by
 Katz and Vafa \cite{Katz:1996xe}, who found a local equation describing
the singularity in the form
\begin{equation}
x^2+y^2z -z^{n+3}(z+t^2)^{2k}  \,,
\end{equation}
with a $D_{n+4}$ singularity (Kodaira type $I_n^*$) along $x=y=z=0$,
and an $A_{2k-1}$ singularity (Kodaira type $I_{2k}$) along $x=y=z+t^2=0$.
We assume that $n\geq1$ (since the case $n=0$ is trivial), and $k\geq1$.
We will analyze this singularity by induction on $k$; the application to
our specific problem is the case $k=n$.  The singularity is illustrated
in Figure~\ref{f:singular-locus}.

Note that the case of transverse intersection (the one already treated 
by Miranda \cite{MR690264}) is obtained from this by substituting $s=t^2$.
We will comment on the differences as we go along.\footnote{Note that
Miranda's 1983 analysis produces a fiber in codimension two that is 
not of the expected Kodaira type, anticipating the phenomenon rediscovered in
\cite{matter1} and \cite{Esole:2011sm}.}

Our strategy is to resolve the $A_{2k-1}$ singular locus, step by step.
To this end, we blow
up $x=y=z+t^2=0$, and consider two coordinate charts.  The key coordinate 
chart for understanding the inductive structure has coordinates 
$\tilde x=x/(z+t^2)$, $\tilde y=y/(z+t^2)$, $z$, and $t$ with equation
\begin{equation}
\label{eq:blowup}
\tilde x^2+\tilde y^2z -z^{n+3}(z+t^2)^{2k-2} ,
\end{equation}
which is a singularity of the same form but with a lower value of $k$.
Note that if $k=1$, then \eqref{eq:blowup} takes the simpler
form
\begin{equation}
\tilde x^2+\tilde y^2z -z^{n+3},
\end{equation}
which has a curve of $D_{n+4}$ singularities at $\tilde x = \tilde y = z = 0$ 
(and arbitrary $t$).

\begin{figure}
\begin{center}
\includegraphics[width=6cm]{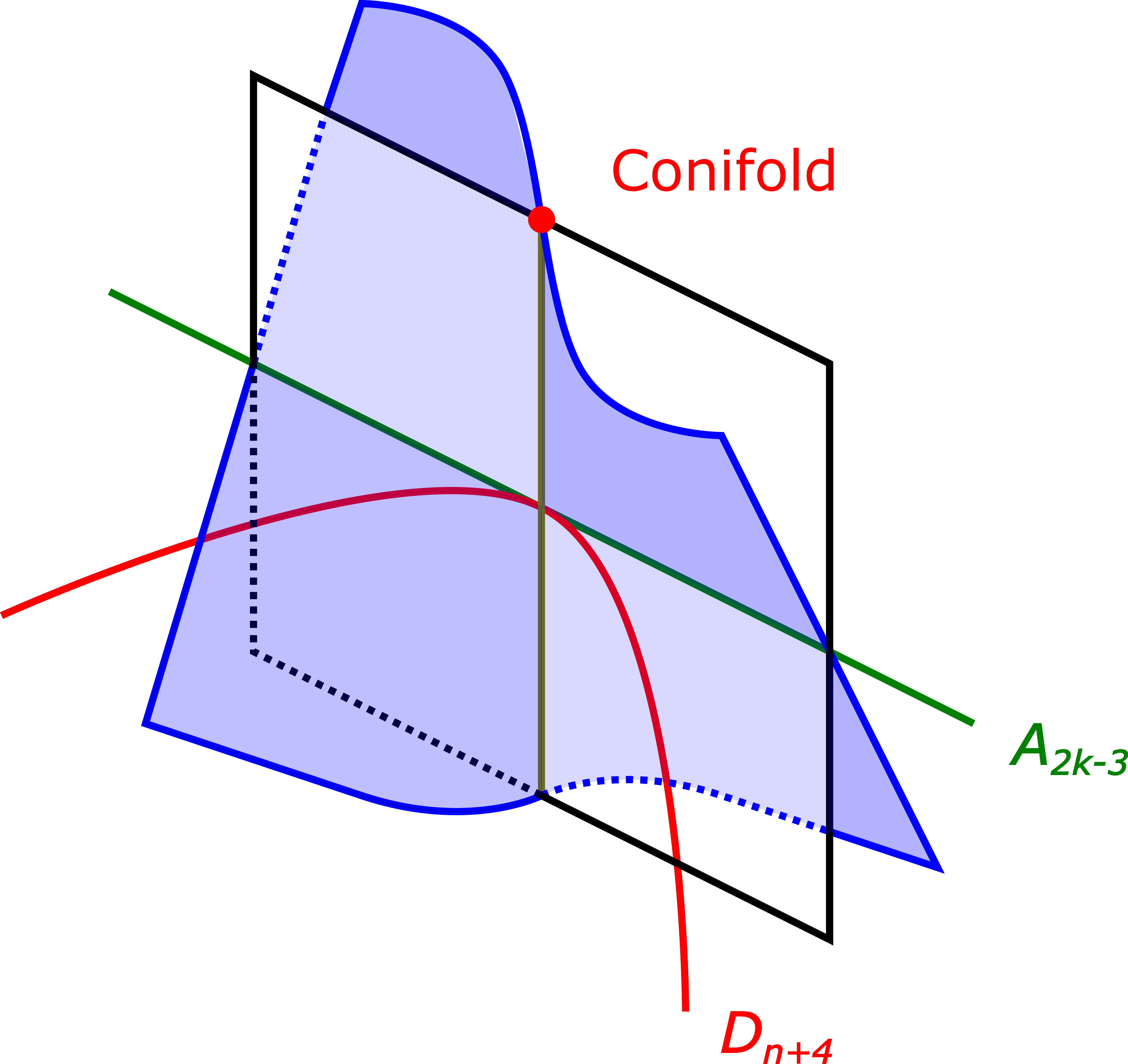}
\end{center}
\caption[x]{\footnotesize The first blowup.}
\label{f:first-blowup}
\end{figure}

The exceptional divisor in this chart is described by $z+t^2=0$.  Substituting
that into the equation, if $k>1$ we find two components
\begin{equation}
z+t^2=\tilde x\pm t\tilde y=0.
\end{equation}
(Note that in the transverse intersection case, with $s=t^2$, there is
only a single irreducible component.)  The two components 
meet at $z=\tilde x=t=0$.
This is illustrated in Figure~\ref{f:first-blowup}.

\begin{figure}
\begin{center}
\includegraphics[width=5cm]{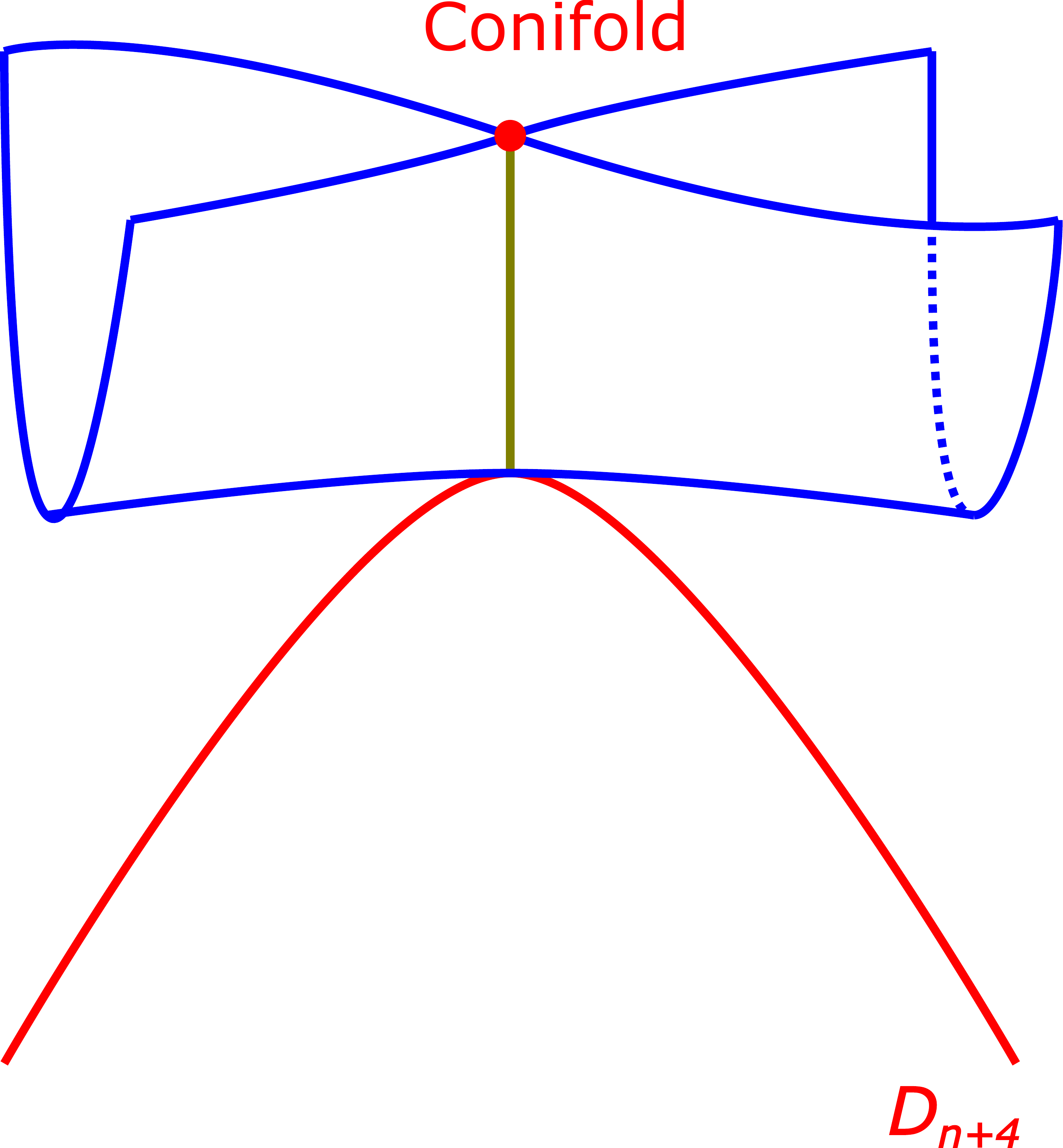}
\end{center}
\caption[x]{\footnotesize The final $A_\ell$ blowup.}
\label{f:final-blowup}
\end{figure}

On the other hand, if $k=1$, the exceptional divisor takes the form
\begin{equation}
z+t^2=\tilde x^2- \tilde y^2 t^2 +(-1)^nt^{2n+6}=0 \,,
\end{equation}
which is an irreducible surface, singular along $z=\tilde x=t=0$.
(When we use $s$ rather than $t^2$, the surface has an isolated
singularity at $\tilde x = \tilde y = z = s = 0$.)
This is illustrated in Figure~\ref{f:final-blowup}.

The other important coordinate chart\footnote{There is a third coordinate
chart, but no new singularities appear there.} has coordinates
$\bar x = x/y$, $\bar z = (z+t^2)/y$, $y$, and $t$ with equation
\begin{equation}
\label{eq:otherchart}
\bar x^2 + y \bar z - t^2 - (y\bar z-t^2)^{n+3}y^{2k-2}\bar z^{2k} .
\end{equation}
This time, the exceptional divisor is described by $y=0$ and we again
see two components when $k>1$
\begin{equation}
y=\bar x\pm t=0,
\end{equation}
meeting at $\bar x = y = t = 0$.
(There is only a single component when we use $s$ rather than $t^2$.)
However, when $k=1$ the exceptional divisor 
\begin{equation}
y=\bar x^2 -t^2+(-1)^nt^{2n+6}\bar z^2
\end{equation}
is irreducible, and is singular along $\bar x = y = t = 0$.  (It is 
nonsingular when we use $s$ rather than $t^2$.)

For any value of $k$, the threefold in this chart has a conifold singularity
at the origin.  If for $k\ne1$ we rewrite \eqref{eq:otherchart} in factored form
\begin{equation}
(\bar x+t)(\bar x-t) + y\left( \bar z   - 
(y \bar z-t^2)^{n+3}y^{2k-3}\bar z^{2k}\right),
\end{equation}
we see that a small resolution can be obtained by blowing up
$x+t=y=0$, which is one of the components of the exceptional
divisior, {\it i.e.}, it exists globally as a divisor.

When $k=1$, however, there is no obvious divisor to use for the global
small resolution of the conifold point.

Note that if we use $s$ rather than $t^2$, \eqref{eq:otherchart}
is nonsingular and a small resolution is not needed.

Thus, by induction on $k$ we can resolve the $I_{2k}$ locus almost
completely,
leaving only a single conifold point as well as 
the $I_n^*$ locus.  The singular $I_n^*$ locus is just a product of a
$D_{n+4}$ singularity with the parameter curve, so its resolution
proceeds just like the resolution of $D_{n+4}$.  This leaves only
the final conifold point unresolved, and depending on (unknown) global
data it may not be possible to resolve it.

\section{$\C^{*}$ bases with generic nonzero Mordell--Weil rank}
\label{sec:appendix-abelian}

This appendix contains the list of  13 $\C^{*}$ bases identified in 
\cite{Martini-WT} over which a generic
elliptic fibration has Mordell--Weil group of nonzero rank $r$.  
In each example, $N$ chains of curves of negative self intersection
connect curves $\Sigma_0$ and $\Sigma_\infty$ with self-intersection
$n_0, n_\infty$ respectively.  In each case, the Hodge numbers $h^{1,
  1}, h^{2, 1}$ of the generic elliptically fibered Calabi--Yau
threefold over that base are given.

\begin{eqnarray}
\label{eq:A1}
r = 1: &  & N = 3, 
n_0 = -2, n_\infty = -3; \; \;
\ho = 34,\htt = 10
\\
\small
{\rm chain}\ 1: &  &  (-2, -1, -2) \nonumber\\
{\rm chain}\ 2: &  & (-2, -2, -1, -4, -1) \nonumber\\
{\rm chain}\ 3: &  & (-2, -2, -2, -2, -2 -1, -8, -1, -2)  \nonumber
\normalsize
\end{eqnarray}

\begin{eqnarray}
\label{eq:A2}
r = 2: &  & N = 3, 
n_0 = -1, n_\infty = -2; \; \;
\ho = 24,\htt = 12
\\
\small
{\rm chain}\ 1: &  &  (-2, -1, -2) \nonumber\\
{\rm chain}\ 2: &  & (-3, -1, -2, -2) \nonumber\\
{\rm chain}\ 3: &  & (-6, -1, -2, -2, -2, -2, -2)  \nonumber
\normalsize
\end{eqnarray}

\begin{eqnarray}
\label{eq:A3}
r = 2: &  & N = 3, 
n_0 = -2, n_\infty = -6; \; \;
\ho = 46,\htt = 10
\\
\small
{\rm chain}\ 1: &  &  (-2, -1, -3, -1) \nonumber\\
{\rm chain}\ 2: &  & (-2, -2, -2, -1, -6, -1, -3, -1) \nonumber\\
{\rm chain}\ 3: &  & (-2, -2, -2, -1, -6, -1, -3, -1)\nonumber
\normalsize
\end{eqnarray}

\begin{eqnarray}
\label{eq:A4}
r = 2: &  & N = 3, 
n_0 = -5, n_\infty = -6; \; \;
\ho = 61,\htt = 1
\\
\small
{\rm chain}\ 1: &  &  (-1, -3, -1, -3, -1) \nonumber\\
{\rm chain}\ 2: &  & (-1, -3, -2, -2, -1, -6, -1, -3, -1) \nonumber\\
{\rm chain}\ 3: &  & (-1, -3, -2, -2, -1, -6, -1, -3, -1)\nonumber
\normalsize
\end{eqnarray}

\begin{eqnarray}
\label{eq:A5}
r = 3: &  & N = 3, 
n_0 = -4, n_\infty = -8; \; \;
\ho = 62,\htt = 2
\\
\small
{\rm chain}\ 1: &  &  (-1, -4, -1, -2, -3, -2, -1) \nonumber\\
{\rm chain}\ 2: &  &  (-1, -4, -1, -2, -3, -2, -1) \nonumber\\
{\rm chain}\ 3: &  &  (-1, -4, -1, -2, -3, -2, -1) \nonumber
\normalsize
\end{eqnarray}

\begin{eqnarray}
\label{eq:A6}
r = 4: &  & N = 3, 
n_0 = -1, n_\infty = -2; \; \;
\ho =  25,\htt = 13
\\
\small
{\rm chain}\ 1: &  &  (-2, -1, -2) \nonumber\\
{\rm chain}\ 2: &  & (-4, -1, -2, -2, -2) \nonumber\\
{\rm chain}\ 3: &  & (-4, -1, -2, -2, -2)  \nonumber
\normalsize
\end{eqnarray}

\begin{eqnarray}
\label{eq:A7}
r = 4: &  & N = 3, 
n_0 = -1, n_\infty = -5 \; \;
\ho = 40,\htt = 4
\\
\small
{\rm chain}\ 1: &  &  (-2, -1, -3, -1) \nonumber\\
{\rm chain}\ 2: &  &  (-4, -1, -2, -2, -3, -1) \nonumber\\
{\rm chain}\ 3: &  &  (-4, -1, -2, -2, -3, -1)  \nonumber
\normalsize
\end{eqnarray}

\begin{eqnarray}
\label{eq:A8}
r = 4: &  & N = 4, 
n_0 = -6, n_\infty = -6; \; \;
\ho = 51,\htt = 3
\\
\small
{\rm chain}\ 1: &  &  (-1, -3, -1, -3, -1) \nonumber\\
{\rm chain}\ 2: &  & (-1, -3, -1, -3, -1) \nonumber\\
{\rm chain}\ 3: &  &  (-1, -3, -1, -3, -1) \nonumber\\
{\rm chain}\ 4: &  & (-1, -3, -1, -3, -1)  \nonumber
\normalsize
\end{eqnarray}

\begin{eqnarray}
\label{eq:A9}
r = 4: &  & N = 3, 
n_0 = -2, n_\infty = -4; \; \;
\ho = 35,\htt =  11
\\
\small
{\rm chain}\ 1: &  &  (-2, -2, -1, -4, -1) \nonumber\\
{\rm chain}\ 2: &  & (-2, -2, -1, -4, -1) \nonumber\\
{\rm chain}\ 3: &  & (-2, -2, -1, -4, -1)\nonumber
\normalsize
\end{eqnarray}

\begin{eqnarray}
\label{eq:A10}
r = 5: &  & N = 3, 
n_0 = -1, n_\infty = -8; \; \;
\ho = 51,\htt = 3
\\
\small
{\rm chain}\ 1: &  &  (-3, -1, -2, -3, -2, -1) \nonumber\\
{\rm chain}\ 2: &  &  (-3, -1, -2, -3, -2, -1) \nonumber\\
{\rm chain}\ 3: &  &  (-3, -1, -2, -3, -2, -1)  \nonumber
\normalsize
\end{eqnarray}

\begin{eqnarray}
\label{eq:A11}
r = 6: &  & N = 3, 
n_0 = -1, n_\infty = -2; \; \;
\ho = 24,\htt = 12
\\
\small
{\rm chain}\ 1: &  &  (-3, -1, -2, -2) \nonumber\\
{\rm chain}\ 2: &  & (-3, -1, -2, -2) \nonumber\\
{\rm chain}\ 3: &  & (-3, -1, -2, -2)  \nonumber
\normalsize
\end{eqnarray}

\begin{eqnarray}
\label{eq:A12}
r = 6: &  & N = 4, 
n_0 = -2, n_\infty = -6; \; \;
\ho = 35,\htt = 11
\\
\small
{\rm chain}\ 1: &  &  (-2, -1, -3, -1) \nonumber\\
{\rm chain}\ 2: &  & (-2, -1, -3, -1) \nonumber\\
{\rm chain}\ 3: &  &  (-2, -1, -3, -1) \nonumber\\
{\rm chain}\ 4: &  & (-2, -1, -3, -1)  \nonumber
\normalsize
\end{eqnarray}

\begin{eqnarray}
\label{eq:A13}
r = 8: &  & N = 4, 
n_0 = -2, n_\infty = -2; \; \;
\ho = 19,\htt = 19
\\
\small
{\rm chain}\ 1: &  &  (-2, -1, -2) \nonumber\\
{\rm chain}\ 2: &  & (-2, -1, -2) \nonumber\\
{\rm chain}\ 3: &  &  (-2, -1, -2) \nonumber\\
{\rm chain}\ 4: &  & (-2, -1, -2)  \nonumber
\normalsize
\end{eqnarray}

\end{document}